\documentclass[
twocolumn, prl, nofootinbib,
superscriptaddress, 
 aps
]{revtex4-2}
\usepackage[T1]{fontenc}
\usepackage[utf8]{inputenc}
\usepackage{subfiles}
\usepackage{graphicx}%
\usepackage{multirow}%
\usepackage{amsmath,amssymb,amsfonts}%
\usepackage{amsthm}%
\usepackage{mathrsfs}%
\usepackage{xcolor}%
\usepackage{textcomp}%
\usepackage{booktabs}%
\usepackage{commath}
\usepackage{algorithm}%
\usepackage{algorithmicx}%
\usepackage{algpseudocode}%
\usepackage{listings}%
\usepackage{float}
\usepackage{matlab-prettifier}
\usepackage{tabularx}  
\usepackage{mathtools}
\usepackage{bibunits}
\defaultbibliographystyle{apsrev4-2}
\defaultbibliography{references} 

\begin{document}

\title{Mechanical instability generates monodisperse colloidosomes}

\author{Seungwoo Shin}
\affiliation{Department of Physics, University of California Santa Barbara, Santa Barbara, CA 93106, USA.}

\author{Federico Cao}
\affiliation{School of Engineering, Brown University, Providence, RI 02912, USA.}
\affiliation{Center for Fluid Mechanics, Brown University, Providence, RI 02912, USA.}

\author{Robert A. Pelcovits}
\affiliation{Department of Physics, Brown University, Providence, RI 02912, USA.}
\affiliation{Brown Theoretical Physics Center, Brown University, Providence, RI 02912, USA.}

\author{Thomas R. Powers}
\affiliation{School of Engineering, Brown University, Providence, RI 02912, USA.}
\affiliation{Center for Fluid Mechanics, Brown University, Providence, RI 02912, USA.}
\affiliation{Department of Physics, Brown University, Providence, RI 02912, USA.}
\affiliation{Brown Theoretical Physics Center, Brown University, Providence, RI 02912, USA.}

\makeatletter
\let\FM@theaffil\frontmatter@theaffiliation 
\gdef\frontmatter@theaffiliation{}          
\makeatother

\author{Zvonimir Dogic$^{1,\dagger}$}

\begin{abstract}
Formation and rupture of vesicles is a fundamental process underlying diverse phenomena in biology, materials science, and biomedical applications. Vesicles form when the area of a growing disk-like membrane exceeds a critical value at which the edge and bending energies balance each other. Observing such topological transitions in lipid bilayers is a challenge because of their nanoscale dimensions and rapid dynamics. We study a scaled-up model of colloidal membranes assembled from rod-shaped colloidal particles. The unique features of colloidal membranes enable the real-time visualization of spontaneous closure driven by instability relevant to all membrane-based materials. First-principles theory quantitatively predicts the instability point for vesicle formation and intermediate membrane conformations during the disk-to-vesicle transition. The instability generates monodisperse, selectively permeable colloidosomes with size controlled by gravity and membrane thickness, providing a scalable and programmable platform for diverse applications.
\end{abstract}

\maketitle

\begingroup
  \renewcommand\thefootnote{\fnsymbol{footnote}} 
  \footnotetext[2]{\href{mailto:zdogic@ucsb.edu}{zdogic@ucsb.edu}}
\endgroup

\begin{bibunit}
Lipid bilayers are a prototypical example of a two-dimensional fluid embedded in a three-dimensional space. Compared to conventional three-dimensional materials that grow in size without bounds, two-dimensional fluid membranes have an intrinsic size limit~\cite{helfrich_size_1974,fromherz_lipid-vesicle_1983,boal_topology_1992,lipowsky_budding_1992,baumgart_imaging_2003,hu_determining_2012,noguchi_cup--vesicle_2019,ding_shapes_2020}. Above a critical area, a flat disk-shaped membrane is mechanically unstable. It transitions into a topologically distinct three-dimensional edgeless capsule, a vesicle. Such transformations are driven by the competition between the edge and bending energy. At a critical size, the edge energy decrease associated with a diminishing perimeter of a closing vesicle compensates for the increase in the bending energy due to the membrane deviating from the flat state. The instability of two-dimensional membranes is of widespread importance. In biology, it underlies the formation of vesicles, which is the foundation for life-sustaining functions such as intracellular trafficking, transcellular communication, and signal transduction~\cite{mcmahon_membrane_2005,yanez-mo_biological_2015,shin_visualization_2018,wiklander_advances_2019,tavakoli_hemifusomes_2025}. In materials science, vesicle-like structures provide a robust and programmable platform for drug delivery, biochemical reactors, and synthetic analogs of cellular compartments~\cite{discher_polymersomes_1999,dinsmore_colloidosomes_2002,riske_electro-deformation_2005}. Conductive nanorods can also self-assemble into vesicle-like structures, enabling applications from photocatalysis and energy storage to light-management devices \cite{park_self-assembly_2004,baranov_assembly_2010,young_directional_2013,zanella_assembly_2010}. Rigorous tests of the mechanical instability require visualization of membrane conformation during the topological transition. Disk-like lipid bilayers become unstable on nanometer scales, making it challenging to visualize real-space dynamics~\cite{smith_electron_2024}. Therefore, probing the mechanical instability of two-dimensional membranes remains a challenge, despite its fundamental importance.

Micrometer-sized colloidal particles provide a powerful platform for visualizing real-space dynamics. Studies of such systems provided insight into diverse physical processes, including crystal nucleation, glass formation, and defect dynamics~\cite{van_blaaderen_real-space_1995,weeks_three-dimensional_2000,gasser_real-space_2001, schall_visualization_2004}. In a similar spirit, colloidal membranes, which are one-rod-length-thick fluid monolayers of aligned filamentous viruses, provide a scaled-up system that mimics numerous aspects of conventional lipid bilayers~\cite{barry_entropy_2010,sharma_hierarchical_2014,khanra_controlling_2022}. Submicron-thick colloidal membranes form two-dimensional flat disks. They can also form closed vesicles or colloidosomes through a gravity-assisted multi-step pathway that is distinct from the mechanical instability relevant to lipid bilayers~\cite{adkins_topology_2025}. Here, we identify conditions where disk-shaped colloidal membranes become unstable and form vesicles. The submicron-thick colloidal membranes enable real-space imaging as the vesicle formation takes place. The transition occurs when a growing membrane reaches a critical area, thus providing a robust mechanism for generating monodisperse colloidal vesicles or colloidosomes. The detailed understanding of the mechanical instability reveals how gravity and membrane mechanics control colloidosome size.

\subsection{Energetic stability of disks and vesicles}
To understand the relative stability of 2D flat disks and 3D vesicles, we write the free energy, consisting of the Helfrich energy and line tension energy,
\begin{equation}
E = \int \left[ \frac{\kappa}{2} (2K)^2 + \bar{\kappa} K_G \right]\, dA 
    +  \int \gamma ~dL,
\end{equation}
where $K$ and $K_G$ are the mean and Gaussian curvature respectively, $\kappa$ and $\bar{\kappa}$ are the bending modulus and Gaussian curvature modulus respectively, and $\gamma$ and $dL$ are the line tension and the line element along the membrane edge respectively~\cite{helfrich_elastic_1973}. The energy of a flat disk with radius \(R\) is proportional to its perimeter \(2\pi \gamma  R\). In comparison, an edgeless sphere-shaped vesicle only has the elastic energy required to bend the membrane away from its flat zero-curvature state. The energy cost to bend a spherical area element scales with the inverse radius squared. Integrating this energy density over the entire sphere yields a size-independent constant \(4\pi (2\kappa + \bar{\kappa})\). Thus, the energy to form a sphere is independent of its radius. Consequently, flat disks are energetically favorable for small areas, whereas closed vesicles are stable at larger sizes. Equating the two energies gives a critical area, 
\begin{equation}
A^* = 4\pi(2\kappa + \bar{\kappa})^2/\gamma^2,
\end{equation}
above which a closed vesicle is energetically favorable~\cite{boal_topology_1992}. For conventional lipid bilayers, typical values of the bending modulus \(\kappa \sim 10 \, k_\mathrm{B  }T\)~\cite{rawicz_effect_2000,dimova_hyperviscous_2002,dimova_recent_2014} and the line tension \(\gamma \sim 5\, k_\mathrm{B}T/\mathrm{nm}\)~\cite{baumgart_imaging_2003,tian_line_2007,elson_phase_2010}, yield a critical disk radius of $\sim 10 \, \mathrm{nm}$. This length scale poses a challenge for studying the instability of 2D lipid bilayer membranes with open edges.

\subsection*{\textbf{Visualizing the instability of 2D fluid disks}}
We study colloidal membranes as a scaled-up analog of lipid bilayers which are assembled from negatively charged monodisperse virus-like filaments (Fig.~1a)~\cite{barry_entropy_2010}. The attractive depletion interactions \cite{asakura_interaction_1954}, generated by non-adsorbing polymers, drive the lateral assembly of the particles into one-rod-length-thick membranes. The particle separation is determined by the competition between repulsive electrostatic interactions and depletion attraction. This balance yields an areal density of $\approx$\(30\%\)~\cite{balchunas_equation_2019}, which corresponds to a surface-to-surface filament spacing of $\approx$6 nm. Thus, colloidal membranes are nanoporous and permeable. Previous studies explored various aspects of 900-nm-thick colloidal membranes, which mainly adopted flat 2D disk-shaped morphologies or 1D twisted ribbons~\cite{gibaud_reconfigurable_2012,sharma_hierarchical_2014,gibaud_achiral_2017}.

\begin{figure*}[t]
\centering
\includegraphics[width=.8\linewidth]{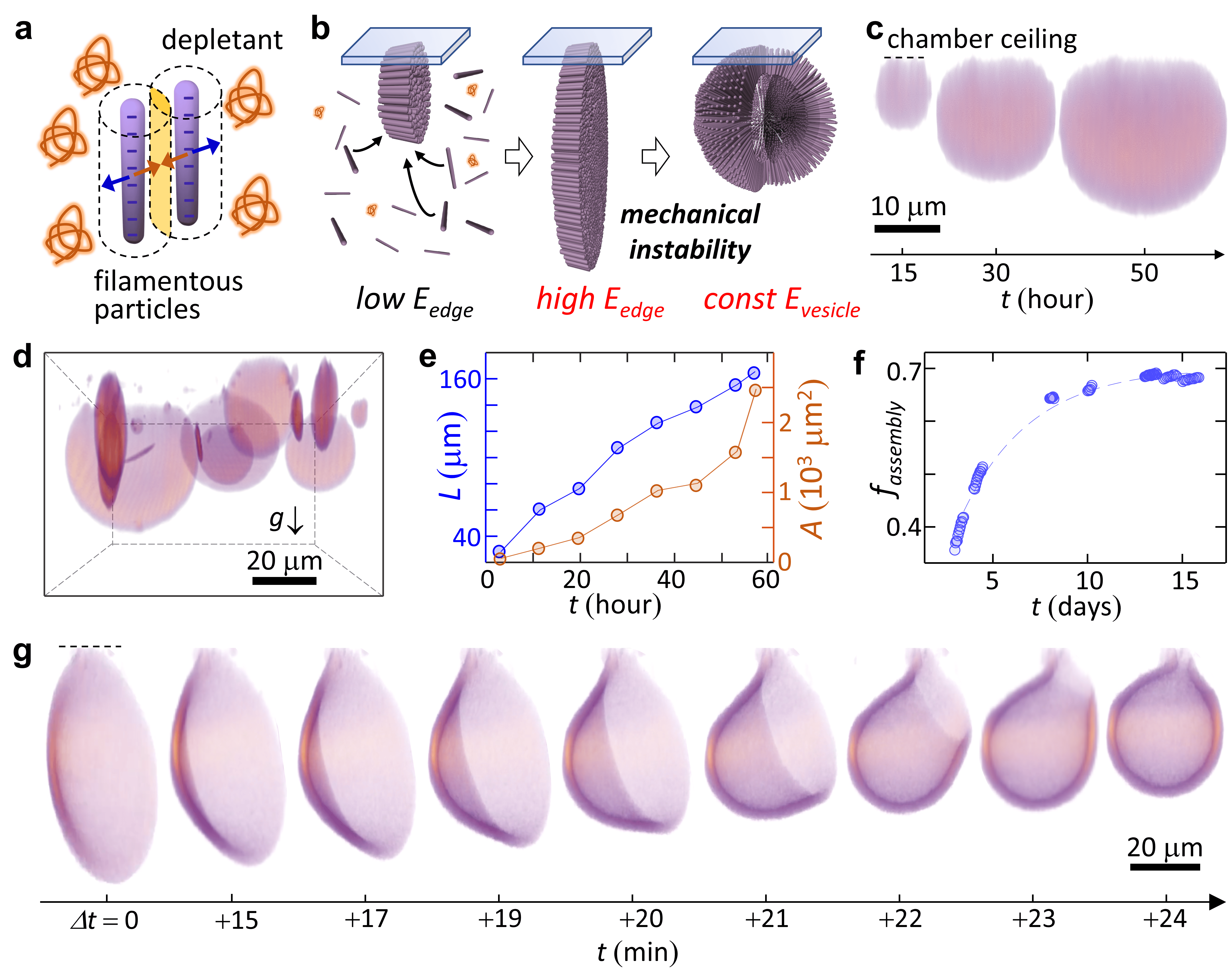}
\caption{\textbf{Vesicle formation from growing membranes.}\\
\textbf{a}, Rod-like particles assemble by the depletion attraction (orange arrows), which is balanced by electrostatic repulsion (blue arrows). \textbf{b}, A surface-anchored membrane continuously grows through the addition of individual rods. The edge energy of a growing membrane increases and eventually exceeds the size-independent bending energy of a closed vesicle, which triggers a mechanical instability that drives spontaneous closure. \textbf{c}, Continuous growth of a ceiling-anchored membrane lying in the plane of the page. \textbf{d}, Numerous polydisperse membranes anchored to the ceiling. \textbf{e}, Perimeter length \textit{L} and area \textit{A} over time during growth. \textbf{f}, Total fraction of particles incorporated into membranes \(f_\mathrm{assembly}\) over time. \textbf{g}, Transformation of a disk-like membrane into a closed vesicle takes about 30 minutes. Time is relative to the onset of instability.
}\label{fig1}
\end{figure*}

Disk-shaped colloidal membranes are stable below the critical area for vesicle formation. The critical area \(A^*\) is proportional to the bending modulus \(\kappa\) and inversely proportional to the line tension \(\gamma\) (Eq. (2)). In turn, these parameters are controlled by the membrane's thickness, $d$. In particular, \(\kappa \propto d^3\) and \(\gamma \propto d\)~\cite{szleifer_curvature_1988,rawicz_effect_2000}. Thus, reducing the membrane thickness reduces the critical area. Motivated by such considerations, we assembled membranes from 200-nm-long virus-like rods~\cite{nafisi_construction_2018}. This approach exploits the unique tunability of virus particles, whose tunable contour length controls the membrane thickness~\cite{dogic_development_2001}. 

A dilute suspension of rods was mixed with a depleting polymer. The ensuing assembly process was observed using confocal fluorescence microscopy. Shortly after sample preparation, small disk-like membranes were observed to anchor to the top surface (Fig.~1b, Fig.~E1). Such isolated disks grew over tens of hours (Fig.~1c), generating many vertically hanging membranes with different sizes (Fig.~1d). The perimeter length \(L\) and the area \(A\) of individual membranes increased continuously over time (Fig.~1e). Being anchored and unable to diffuse, membranes grew by recruiting particles from the background suspension (Fig.~E1). Fluorescence imaging estimated the fraction of filamentous particles incorporated into membranes (Fig.~1f). The growth kinetics plateaued, with $\approx70\%$ of the viruses incorporated into membranes after two weeks.

After reaching a critical area, we observed that a flat membrane suddenly developed a slight curvature (Fig.~1g, Fig.~E2). The curvature slowly increased, while the membrane edge gradually shortened, a process that lowered the line tension energy while increasing the bending energy. After 20 minutes, the closure sped up, forming a hemisphere-like structure that rapidly transformed into an almost closed vesicle. The closure transition took 25 minutes, revealing the membrane conformations throughout the instability process. Because of the size polydispersity of anchored membranes, vesicle closure events began to appear within two days of sample preparation, as individual membranes reached a critical area at different times. The spontaneous and continuous nature of the pathway suggests a monotonically decreasing energy landscape without barriers. These observations provide compelling experimental evidence for the instability of 2D fluid membranes and a unique opportunity to test theoretical models of vesicle closure.

\subsection*{\textbf{Monodisperse vesicle formation}}
The disk-to-vesicle closure transition was observed for hanging membranes (Fig.~2a). The vesicle diameter was narrowly distributed around \(36 \pm 3.2\,\mu\text{m}\) (mean ± s.d.; Fig.~2b). In comparison, a gravity inversion method yielded larger vesicles with a wider size distribution \(73.0 \pm 15.5\,\mu\text{m}\)~\cite{adkins_topology_2025}. Thus, the mechanical instability generated highly monodisperse vesicles.   
We compared the observed vesicle size with the critical area $A^*$  (Eq.~(2)). To realize this goal, we first measured the parameters of the total free energy (Eq.~(1)). We captured thermal fluctuations of flat disk-shaped membranes and closed vesicles (Fig.~E3, SI.~1--2). From these measurements, flickering spectroscopy yielded the bending modulus of \(\kappa = 1200 \pm 60\,k_\mathrm{B}T\), which is several orders of magnitude larger than the bending rigidity of conventional lipid bilayers~\cite{schneider_thermal_1984,gracia_effect_2010}. To measure the line tension, we modified the assembly protocol to generate disk-shaped membranes lying on the chamber floor (Fig.~E4). Such membranes exhibited visible edge fluctuations (Methods), whose analysis yielded a line tension of \(\gamma = 200 \pm 30\,k_\mathrm{B}T/\mu\text{m}\)~\cite{gibaud_reconfigurable_2012,jia_chiral_2017}. The Gaussian curvature modulus was estimated as \(\bar{\kappa} \approx 20\,k_\mathrm{B}T\) (SI.~3)~\cite{gibaud_achiral_2017}; since it is much smaller than the bending modulus, it can be neglected. The measured bending modulus and line tension predict a critical area of \(1840~\mu\mathrm{m}^2\), corresponding to a vesicle diameter of \(24.2\,\mu\mathrm{m}\). This value is significantly lower than the measured colloidosome diameter of \(36\pm3.2\,\mu \mathrm{m}\). This discrepancy motivated an analysis of the entire kinetic pathways of vesicle formation.

\begin{figure*}[t]
\centering
\includegraphics[width=.85\linewidth]{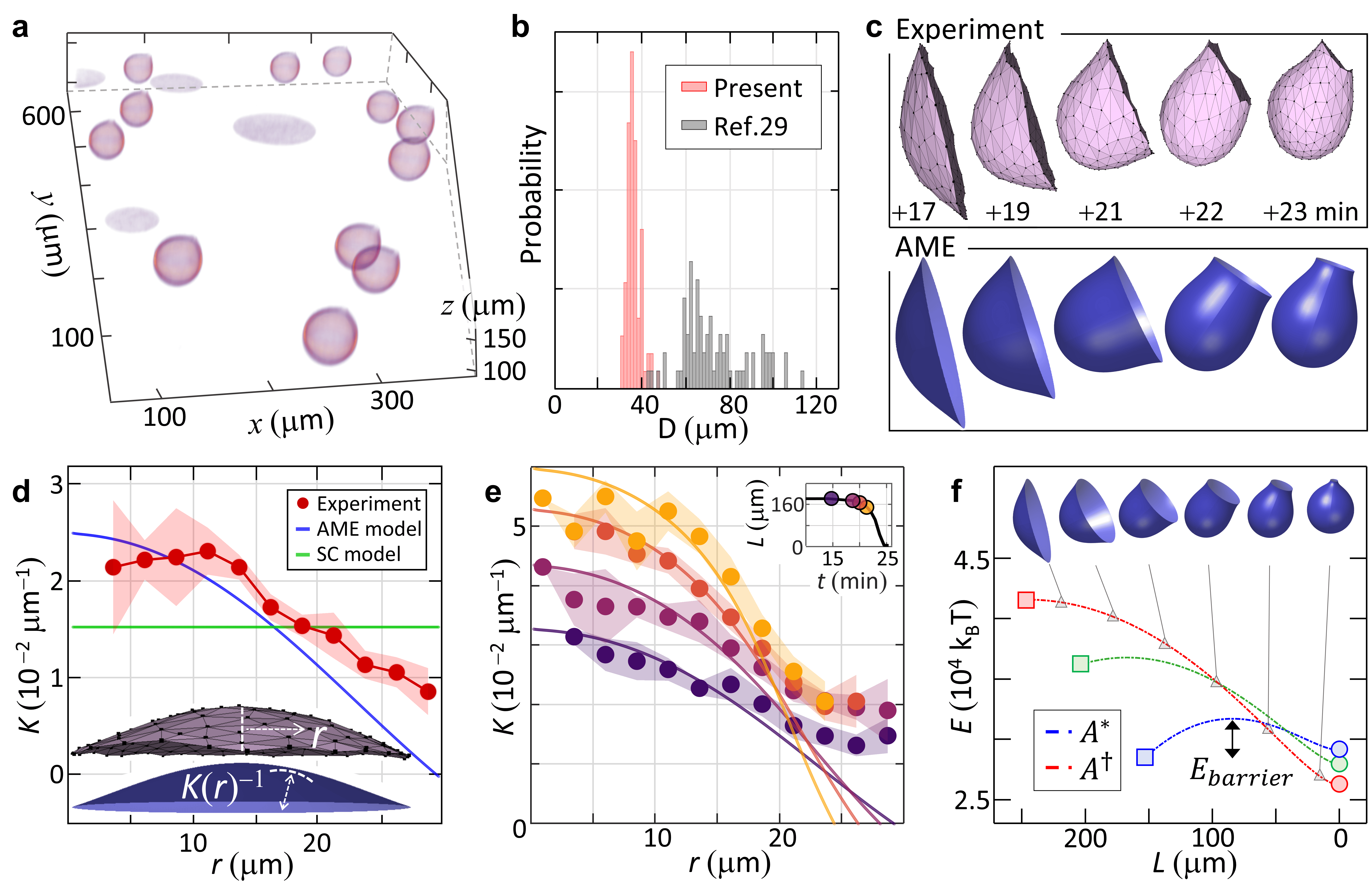}
\caption{\textbf{Kinetic pathways of vesicle formation}\\
\textbf{a}, Monodisperse vesicles and flattened membranes anchored to the ceiling. 
\textbf{b}, Distribution of vesicle diameters obtained via mechanical instability (red, \(36.1 \pm 3.2\,\mu\mathrm{m}\), \(N = 76\)) and gravity inversion method (gray, \(73.0 \pm 15.5\,\mu\mathrm{m}\), \(N = 73\)) \cite{adkins_topology_2025}. 
\textbf{c}, 3D meshes from imaged membrane (top) and axisymmetric minimum-energy (AME) model predictions (bottom) for the same area and perimeter. Experimental times are from the instability onset. 
\textbf{d}, Radial mean-curvature profiles \(K(r)\) from the experimental 3D mesh (red) and the AME prediction (blue) at the instability onset, compared with the spherical-cap (SC, green) model. Dots and shading denote the mean and 95\% confidence interval. Inset: corresponding 3D mesh and AME surface illustrating the definitions of the radial distance and mean curvature.
\textbf{e}, Time evolution of curvature profile \(K(r)\) during early closure, overlaid with AME model predictions (solid lines). The colors correspond to the times in the inset, which shows the perimeter over time.
\textbf{f}, Energy landscapes as a function of perimeter \(L\) for three areas \(A = 1840, 3260,\) and \(4790 \,\mu\mathrm{m}^2\) (blue, green, and red). Filled squares and circles denote flat disks and closed vesicles, respectively. The black arrow indicates the energy barrier. The top images show the AME model predictions corresponding to triangular markers along the curves. 
}
\label{fig2}
\end{figure*}

\subsection*{\textbf{Kinetic pathways of vesicle formation}}
The micron-scale size and minute-scale dynamics revealed the intermediate conformations of the disk-to-vesicle transition. To compare these shapes with theory, we reconstructed 3D meshes from confocal images, which provided the membrane area \(A\), the edge perimeter \(L\), and the curvature of the evolving membrane (Fig.~E5). We compared these measurements to two theoretical models. The spherical cap (SC) model approximates intermediate shapes as portions of a sphere with prescribed membrane area \(A\) and perimeter \(L\). It represents a simple, analytically tractable, and widely used description of intermediate states~\cite{fromherz_lipid-vesicle_1983,lipowsky_budding_1992,baumgart_imaging_2003,noguchi_cup--vesicle_2019}. The axisymmetric minimum energy (AME) model predicts intermediate shapes by minimizing the energy (Eq.~(1)) for given \(A\) and \(L\). It assumes axial symmetry and zero bending moment at the edge (SI.~4) \cite{boal_topology_1992,adkins_topology_2025,Julicher1994}. Because the gravitational contribution to the total free energy is small, the membrane shapes were estimated by accounting only for the bending and edge energy. Gravity was accounted for by setting the tilt angle for the ceiling-anchored membrane such that the center of mass was vertically aligned with the anchoring point. Comparison of intermediate shapes reveals that the AME model captures intermediate states more accurately than the SC model (Figs.~2c,~S1). 
Given the slow dynamics, we adopt a quasi-static approximation in order to compute the equilibrium shapes that minimize the free energy.

Both models assume axial symmetry and a constant membrane area. To test the assumptions, we monitored the morphological evolution during closure. In the early closure stages, the membrane conformations were symmetric around an axis, while the late-stage shapes, after 22 min, noticeably deviated from axial symmetry (Fig.~2c, Fig.~E6). We analyzed the initial stages, by extracting azimuthally-averaged radial profiles of the mean curvature $K(r)$ from the 3D meshes (Fig.~2d), where $r$ is the radial distance to the axis of the cylindrical symmetry. The SC model assumes constant curvature. In contrast, the AME model predicts a curvature that is maximum at the center ($r=0$), decreases with increasing $r$, and vanishes at the edge. A time series of measured early stage curvature profiles follows the AME predictions (Fig.~2e). The model explains the curvature profiles without adjustable parameters, as perimeter and area are extracted from meshes, and $\kappa$ and $\gamma$ are independently measured. We also monitored the membrane area throughout the closure process (Fig. E5c). The membrane area remained nearly constant during the early stages, consistent with the model assumption, but increased significantly during the late stages, likely due to the sweeping motion of the membrane edge that enhances the recruitment of background particles. Thus, the AME model quantitatively describes the initial stages of the closure pathway. 
   
With the quantitative descriptions of the intermediate conformations, we calculated the energy landscape associated with the disk-to-vesicle transition. The membrane shape and tilt angle were obtained within the AME framework. The gravitational contribution was given by \( \rho_\mathrm{a} A g z_{\mathrm{CM}} \), where \(\rho_\mathrm{a}\), $A$, \(g\), and  \( z_{\mathrm{CM}}\) denote the areal mass density, area, gravitational acceleration, and the center-of-mass height determined from the calculated shape (SI.~4). The effective gravitational term \(\rho_\mathrm{a} g\) was estimated to be \(0.04\,k_\mathrm{B} T/\mu\text{m}^3\) (SI.~5) \cite{gibaud_achiral_2017}. The energy landscape was calculated from the AME model as a function of the perimeter length $L$, for three areas \(A = 1840, 3260,\) and \(4790 \,\mu\mathrm{m}^2\) (Fig.~2f). For each fixed area curve, the largest \(L\) corresponds to a flat disk, and \(L=0\) corresponds to a closed vesicle. Smaller closed vesicles exhibit higher total energy than larger ones, due to gravitational contributions for ceiling-anchored vesicles. For the critical area, \(A^*\), the edge energy of a flat membrane equals the bending energy of a closed vesicle (Eq.~(2)), but a substantial barrier \(10^4\,k_\mathrm{B} T\) prevents vesicle closure (Fig.~2f, blue curve). Increasing the area decreases the energy barrier (Fig.~2f, green curve). Eventually, the barrier vanishes at \(A^\dagger = 4790 \,\mu\text{m}^2 = 2.6~A^*\) (Fig.~2f, red curve), corresponding to a predicted vesicle diameter of \(39.1\,\mu\text{m}\) that quantitatively agrees with the experimental measurement of \(36.1 \pm 3.2\,\mu\mathrm{m}\), without any adjustable parameters. 

\subsection*{\textbf{Gravity controlled vesicle size}}
Having established that mechanical instability yields monodisperse vesicles, we sought control of their size. In addition to the membranes hanging from the chamber ceiling, some membranes were anchored on the sample floor, growing upward from their edge (Fig.~3a). Intriguingly, the floor-anchored membranes formed vesicles earlier than the ceiling ones. The former had a diameter of \(32.1 \pm 1.9\,\mu\mathrm{m}\), whereas the latter were larger, at \(36.1 \pm 3.2\,\mu\mathrm{m}\) (Fig.~3b). The closure of ceiling-anchored membranes raises the center of mass, increasing the gravitational potential energy and the energy barrier, thus impeding vesicle formation. In contrast, closure of floor-anchored membranes lowers the center of mass, decreasing the gravitational potential energy, thereby assisting vesicle formation (Fig.~3c). Thus, gravity shifts the critical area, leading to distinct vesicle diameters under otherwise identical conditions. 

\begin{figure*}[t]
\centering
\includegraphics[width=.85\linewidth]{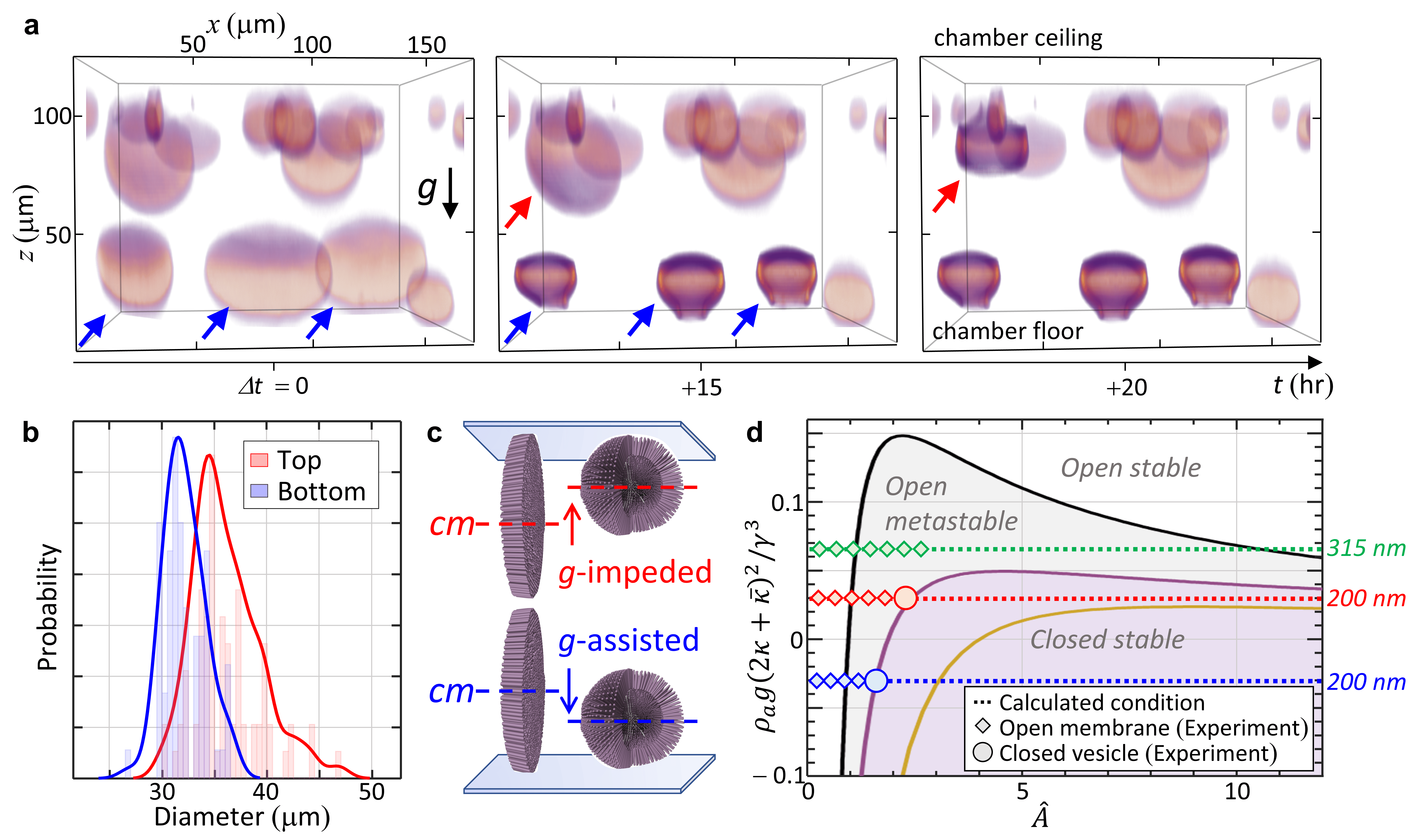}
\caption{\textbf{Gravity and membrane mechanics controls colloidosome size.\\
a}, Time lapse of floor (blue arrows) and ceiling (red arrows) anchored membranes. Floor-anchored membranes exhibit instability earlier in the growth cycle.
\textbf{b}, Size distributions for ceiling (red) and floor (blue) anchored membranes. Floor-formed vesicles have a smaller size (\(32.1 \pm 1.9\,\mu\mathrm{m}, N = 72\)). 
\textbf{c}, Due to the difference in the change in the vertical position of the center-of-mass, gravity impedes the formation of ceiling-anchored vesicles while assisting bottom-growing ones. 
\textbf{d}, Stability of the 3D vesicle versus 2D disks as a function of rescaled area \(\hat{A} = A/A^*\) and \(\rho_\mathrm{a} g (2\kappa + \bar{\kappa})^2 / \gamma^3\), calculated with the AME model. \textit{Open stable} indicates the regime where disk-shaped membranes are energetically favorable, \textit{Open metastable} indicates the regime where vesicles are energetically favored and disk-shaped membranes are metastable, and \textit{Closed stable} indicates the regime where the barrier for the transition from disk to vesicle disappears. The red and blue dashed lines correspond to ceiling and floor-anchored 200-nm membranes. The green dashed line represents 315-nm membranes anchored to the ceiling. The yellow curve is the SC model prediction for the onset of the mechanical instability.}
 \label{fig3}
\end{figure*}

To generalize these observations, we constructed a gravity-dependent stability diagram of the disk-to-vesicle transition in a dimensionless parameter space based on the AME model (Fig.~3d). The horizontal axis is rescaled by the critical area as \(\hat{A} = A/A^*\), while the vertical axis \(\rho_\mathrm{a} g (2\kappa + \bar{\kappa})^2 / \gamma^3\) incorporates the effects of gravity and membrane mechanics. The negative range of the $y$-axis corresponds to negative gravitational acceleration, which can be accomplished by inverting the sample. We identified three regimes: \textit{open stable} indicates the phase space where flat membranes are energetically favored; \textit{open metastable} indicates the regime where vesicles represent the global energy minimum but their formation is hindered by an energy barrier; and \textit{closed stable} indicates a phase space where disks are unstable and spontaneously transition into a vesicle without encountering a barrier. 

Vesicle closure for a given gravity and membrane thickness occurs when a horizontal line intersects the \textit{closed stable} regime (Fig. 3d, purple line). For 200-nm-thick membranes, ceiling- and floor-anchored configurations correspond to positive and negative values on the vertical $y$-axis. The intersections of these lines with the purple curve (\(\hat{A} = 2.60\) and \(1.77\)) coincide with the measured vesicle sizes (\(\hat{A} = 2.23\) and \(1.76\)). Thicker (\(d=315\) nm) membranes have modified mechanical properties (Fig.~S3, SI.~6). In this case, theory predicts stability of disk-shaped membranes for all areas, consistent with experimental observations (Fig.~3d, green line, Fig.~E7). Thus, the AME-based phase diagram agrees with experiments. Notably, both the black and purple curves decrease at large areas \(\hat{A}\) due to the increased mass of anchored membranes, implying that a drastic increase in membrane size can bypass the instability and suppressing vesicle closure. Our experimental results also show that the widely used SC model overestimates both the critical area and the energy barrier for spontaneous closure (Fig.~3d, yellow curve), as theoretically predicted \cite{boal_topology_1992}. These results establish that the vesicle diameter can be tuned by gravity and membrane mechanics.

\subsection*{\textbf{Complete closure and mass production of vesicles}}
To induce a complete closure of ceiling-anchored vesicles, we detached them using a short centrifugation step (Fig.~4a). The sample chamber was mounted in a horizontal rotor with the $xy$-image plane parallel to the rotor arm, which generated a 30~\(g\) lateral force for 30~s. In response, vesicles ruptured from the ceiling and sedimented toward the bottom with little shape change (Fig.~4b). The size distribution before detachment ($37.7 \pm 1.2~\mu\mathrm{m}$) and after sedimentation ($38.4 \pm 1.5~\mu\mathrm{m}$) remained identical. The ceiling-attached vesicles had an open pore at the anchor site. Upon sedimentation, the pore remained open, gradually diffusing across the vesicle surface (Fig.~4c). Unlike pores in lipid bilayer vesicles that reseal within milliseconds to seconds~\cite{riske_electro-deformation_2005,portet_new_2010}, pores in colloidal vesicles closed after 30--60 minutes, suggesting the presence of an energy barrier. Together with the nanoporous structure of colloidal membranes, the complete closure of monodisperse vesicles enables controlled encapsulation with nanoscale  permeability. 

\begin{figure*}
\centering
\includegraphics[width=.85\linewidth]{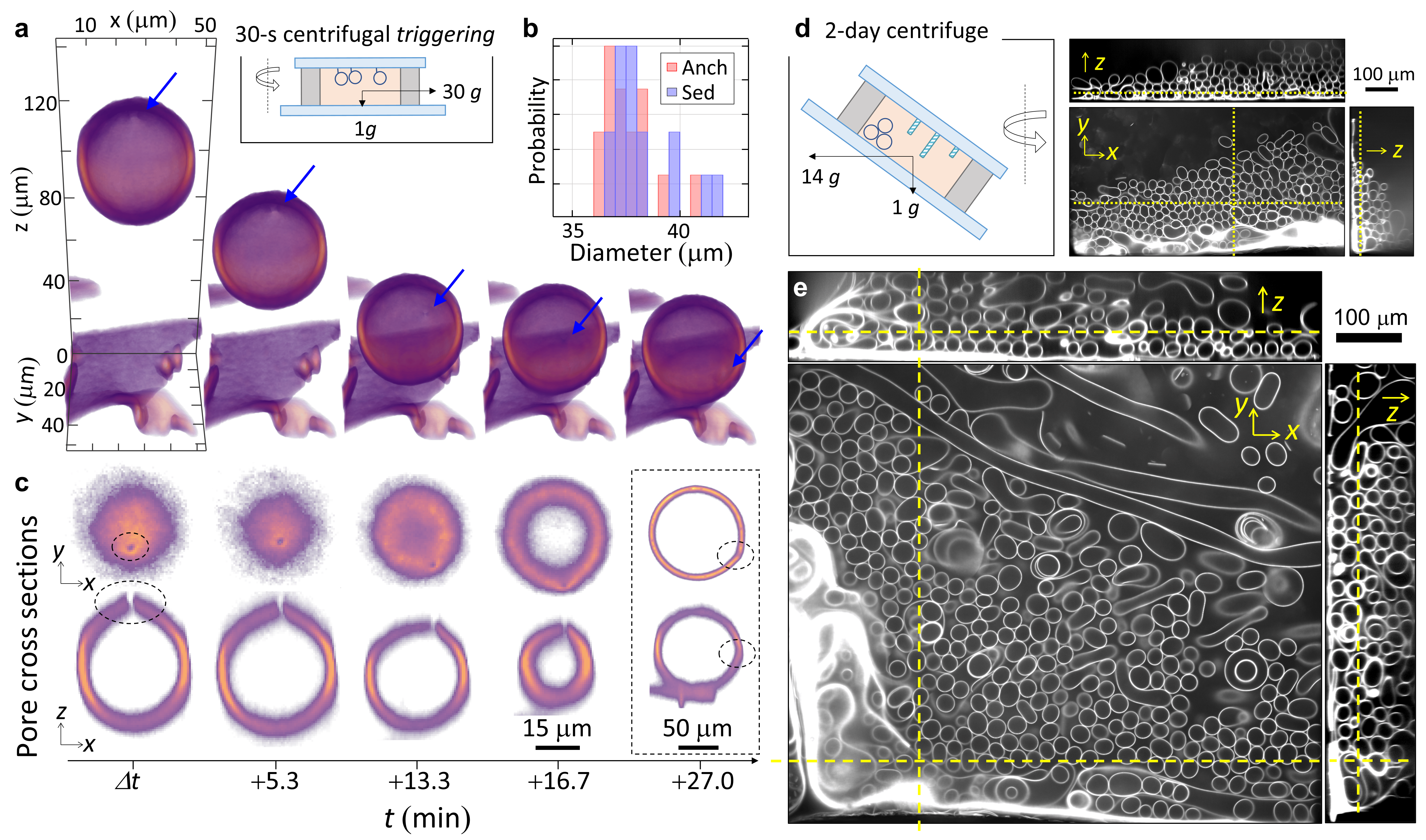}
\caption{\textbf{Complete closure and large-scale formation of vesicles\\
a}, Induced by a 30-s lateral centrifugal trigger, the ceiling-anchored vesicle detaches and sediments. Blue arrow marks the open pore, which diffuses across the vesicle surface and closes by 27 min. 
\textbf{b}, Size distributions for anchored (red, \(37.7 \pm 1.2\,\mu\mathrm{m}, N = 16\)) and sedimented (blue, \(38.4 \pm 1.5\,\mu\mathrm{m}, N = 16\)) vesicles.
\textbf{c}, Pore-centered cross-sectional views in the \textit{$xy$-} and \textit{$xz$-}planes. The dashed circles indicate the pore.
\textbf{d}, Vesicle formation over two days under an oblique centrifugal force. Cross-sectional slices in \textit{$xy$-}, \textit{$xz$-}, and \textit{$yz$-}planes reveal closed vesicles layered along the chamber wall. 
\textbf{e}, 3D cross-sectional views of the same sample from (\textbf{d}) after additional centrifugation along the $x$-axis. Dense vesicles accumulate along the chamber wall.
}
\end{figure*}

Next, we explored the influence of the oblique centrifugal field on membrane growth. The sample chamber was mounted in a horizontal rotor at a $45^{\circ}$ tilt, exposing the sample to a continuous $14~g$ force throughout the self-assembly process. Vesicles formed after two days of continuous centrifugation aligned parallel to the chamber edge, giving rise to dense stacks (Fig.~4d). To further concentrate them, we applied horizontal centrifugation at $14~g$ in a perpendicular direction, which led to their accumulation at the chamber corner (Fig.~4e). Most vesicles retained similar sizes and shapes, although some ellipsoidal, enlarged, and partially enveloping vesicles were also observed. Applying a centrifugal force along the sedimentation direction to ceiling-anchored vesicles caused them to deform into ellipsoids (SI.~7, Fig.~S4). We occasionally observed that a pore of a sedimented vesicle encountered another membrane edge, resulting in sudden merging and vesicle enlargement. In contrast, vesicles that landed on membranes without open-edge interactions remained intact. Thus, colloidal vesicles preserve their integrity during surface-to-surface contact, whereas open-edge contacts trigger abrupt merging~\cite{barry_entropy_2010,yang_self-assembly_2011}.

\subsection*{Discussion}
Our experiments revealed the spontaneous transition of open 2D disk-like membranes into closed 3D vesicles. Direct visualization of intermediate morphologies provided insight into the kinetic pathways of the transition and validated the AME model. The transition is driven by a mechanical instability, which occurs when the energetic barrier vanishes. The critical size for the disk-to-vesicle transition is determined by gravity and membrane mechanics, the latter being controlled by the length of virus-like particles (Fig.~3d)~\cite{dogic_development_2001,sharma_hierarchical_2014}. 

Anchoring membranes to the surface was critical for two reasons. First, surface anchoring fundamentally changed the energy landscape associated with vesicle formation. Notably, when equivalent membranes lay in the $xy$-plane of constant gravitational potential, they never became mechanically unstable (Figs.~E4b--E4c). For such configurations, we proved mathematically that there is always an energy barrier for vesicle formation (SI.~4). In this configuration, one can manually induce vesicle formation through gravity inversion---a complex multi-step process that has no direct analog in lipid vesicles~\cite{adkins_topology_2025}. In comparison, vertically oriented membranes exhibit the mechanical instability that spontaneously generates monodisperse vesicles. Second, free membranes can increase their size through continuous addition of isolated rods and membrane-membrane coalescence events~\cite{zakhary_imprintable_2014}, with the latter leading to discontinuous area growth. In comparison, surface-anchored membranes grow only through the gradual recruitment of particles from the background. Thus, they smoothly and continuously approach the critical instability area, which in turn ensures the formation of monodisperse capsules. In comparison to the colloidosomes presented in this work, conventional nanometer-sized liposomes exhibit polydisperse size distributions~\cite{hallett_determination_1991,lee_preparation_2001, maulucci_particle_2005,yewle_progressive_2016,huang_formation_2017}. This could be attributed to their lateral growth, which likely involves the discrete coalescence of membrane patches, which in turn leads to discontinuous area growth beyond the critical size. 

Our results have practical relevance. The complete closure, structural integrity, and scalable production underscore the potential of colloidosomes as selectively permeable encapsulation carriers. More broadly, the mechanical instability of fluid films provides a novel mechanism for assembly of monodisperse soft structures, one that is fundamentally different from existing technologies, including microfluidic techniques that generate monodisperse droplets by controlling the hydrodynamic instabilities, chemical synthesis of colloidal particles that require the control of nucleation and growth kinetics, or thermodynamic self-limited self-assembly methods~\cite{utada_monodisperse_2005,shum_microfluidic_2008,van_blaaderen_synthesis_1993,hagan_equilibrium_2021}

Our approach establishes a universal mechanism for mechanically driven vesicle formation that depends only on the membrane’s elastic properties, rather than on the molecular composition of its constituents. Although demonstrated here using biologically inspired, virus-like rods, the same mechanism can be applied to membranes self-assembled from conductive nanorods, underscoring its broad applicability across diverse material systems~\cite{park_self-assembly_2004,baranov_assembly_2010,young_directional_2013,zanella_assembly_2010}. More broadly, our work demonstrates that continuum deformations of both lipid bilayers and colloidal monolayers are described by the same model based on the Helfrich energy. Thus, colloidal membranes provide a powerful model system to visualize and quantitatively test diverse membrane-based biophysical processes that play essential roles in life science and materials development, including pore nucleation and closure, particle envelopment and membrane fusion, and fission~\cite{oglecka_oscillatory_2014,reynwar_aggregation_2007,kozlovsky_stalk_2002}.  

\subsection*{Acknowledgments}
This work was primarily supported by the National Science Foundation Grant 2308537. S.S. acknowledges the support of the Human Frontier Science Program (HFSP) postdoctoral fellowship [LT0003/2023-C]. F.C. acknowledges the support of the Natural Sciences and Engineering Research Council of Canada (NSERC), [567961-2022]. Cette recherche a été financée par le Conseil de recherches en sciences naturelles et en génie du Canada (CRSNG), [567961-2022]. R. A. P. and T. R. P. acknowledge support from National Science Foundation Grant CMMI-202009. This research was supported in part by grant NSF PHY-2309135 to the Kavli Institute for Theoretical Physics (KITP).  

\subsection*{Competing interests}
The authors declare no competing interests.

\section*{Methods}
\subsection*{Production of filamentous virus-like particles}
For our studies, we used rod-shaped phagemid particles in which DNA is packaged inside a protein capsid. Phagemids cannot infect bacteria and propagate alone. To produce 200-nm long phagemids, the plasmid pScaf1512.1 was co-transformed with the helper plasmid HP17\_K07 (Plasmid \#111402 and \#120346, Addgene)~\cite{nafisi_construction_2018} into \textit{E.~coli} cells (NEB 5-alpha F'\textit{I}$^q$ Competent \textit{E. coli}, Cat. No. C2992I, New England Biolabs). Transformed colonies were streaked onto LB agar plates containing 100~$\mu g$/mL ampicillin and 50~µg/mL kanamycin, and incubated at 37\textdegree C overnight. A single colony was used to inoculate 3~mL of 2$\times$YT starter culture supplemented with the same antibiotics. Cultures were grown at 30\textdegree C with shaking at 225~rpm until reaching an OD\textsubscript{600} of approximately 1.2.

The starter culture was transferred to a 250~mL Erlenmeyer flask containing 112~mL of small-scale growth medium, consisting of 100~mL of 2$\times$YT, 10~mL of phosphate buffer (7\% K$_2$HPO$_4$, 3\% NaH$_2$PO$_4$, pH~7.0; autoclaved), 1~mL of 50\% glucose, 0.5~mL of MgCl$_2$, and the same concentrations of antibiotics. After incubation to OD\textsubscript{600}~$\sim$1.8, the culture was scaled into twelve 2~L Erlenmeyer flasks, each containing 1.120~L of growth medium prepared by a tenfold volumetric scale-up of the small-scale growth medium while maintaining the same component ratios. Cultures were incubated at 30\textdegree C and 225~rpm until reaching OD\textsubscript{600}~$\sim$2.8.

Cells were chilled on ice and removed by sequential centrifugation. Low-speed centrifugation (4,200$\times g$, 15 minutes at 4\textdegree C) was followed by high-speed centrifugation (12,000$\times g$, 15 minutes at 4\textdegree C). The supernatant was collected, and phagemids were precipitated by adding polyethylene glycol (PEG, 8000 Da) and NaCl to final concentrations of 40~g/L and 30~g/L, respectively. The mixture was stirred for 1 hour at 4\textdegree C and centrifuged at 12,000$\times g$ for 30 minutes. The resulting pellets were resuspended in TE buffer. Bacterial debris was further removed by sequential centrifugation at 7,000$\times g$, 12,000$\times g$, and 45,000$\times g$. 

Residual DNA was digested with Benzonase\textsuperscript{\textregistered} Nuclease (purity $>$99\%, 10~KU, Cat. No. 70664, Sigma-Aldrich). Total concentrations of residual DNA and phagemids were first measured using a nanodrop spectrophotometer (NanoDrop One C, Thermo Fisher). Based on the measured DNA content, 2.5~$\mu$L of nuclease was added per 1~$\mu$g of DNA and incubated at 37\textdegree{}C for 30~minutes. The final purification was achieved by ultracentrifugation at 365,000$\times g$ for 4 hours at 4\textdegree C, after which phagemids were collected from the pellet. The pellet was resuspended in phagemid buffer (140~mM NaCl, 20~mM Tris-HCl, pH~8.0) and stored at 4\textdegree{}C.

In addition, 315-nm phagemids were produced by shortening the Litmus 28i plasmid (New England Biolabs) through the removal of non-essential regions. Specifically, a designated sequence including the \textit{LacZ}$\alpha$ site was deleted, reducing the total DNA length and resulting in a proportional decrease in the length of the assembled phage. The modified phagemid was transformed into a competent \textit{E.~coli} strain (NEB 5-alpha F'\textit{I}$^q$ Competent \textit{E. coli}, Cat. No. C2992I, New England Biolabs) and co-infected with the helper phage M13KO7, followed by amplification using the same protocols described above. The resulting phage particles were purified through sequential centrifugation steps. After purification, residual helper phages were removed by Dextran-based fractionation.

\subsection*{Phage labeling}
To fluorescently label the phagemids, an amine-reactive fluorophore (DyLight-NHS ester 550, Thermo Fisher) was conjugated to the primary amines on the major coat proteins\cite{lettinga_self-diffusion_2005}. Labeling was performed in phagemid buffer according to the manufacturer’s instructions. To minimize perturbation of membrane properties, the labeling ratio was kept low, targeting approximately 1\% of the coat proteins per phagemid. The labeled phagemids were stored at 4\textdegree{}C.

\subsection*{Dextran fractionation}
Low-polydispersity Dextran (500{,}000~Da, Sigma-Aldrich) was prepared by ethanol fractionation to improve membrane quality. Ethanol was slowly added to a 0.2\% (w/w) aqueous Dextran solution under vigorous stirring at 23\textdegree C. When the ethanol concentration reached 31\% (w/w), the precipitated Dextran was collected by centrifugation (20\, min at 17{,}000\,$g$, 23\textdegree C). Ethanol was subsequently added to increase the concentration to 32\% (w/w), and a second centrifugation step was performed to recover additional precipitates. The combined precipitates were freeze-dried and stored as a powder.

\subsection*{Surface coating of glass coverslips}
Glass coverslip (micro cover glass No.\,1.5, VWR International) surfaces were coated with polyacrylamide brushes. Most steps were followed from \cite{lau_condensation_2009}, with a modification of an acrylamide solution: 2\% w/v acrylamide solution was prepared by diluting 40\% acrylamide stock solution in DI water, degassed under vacuum for 1~hour, and polymerized by adding 0.07\% ammonium persulfate and 0.035\% N,N,N$'$,N$'$-Tetramethylethylenediamine. After polymerization, coverslips were stored immersed in the acrylamide solution and rinsed with DI water before use.

Most samples were prepared on coverslips coated with polyacrylamide brushes. For specific experiments, bovine serum albumin (BSA; Sigma-Aldrich) was additionally applied to the polyacrylamide-coated coverslips to further suppress surface adhesion. BSA powder was dissolved in phagemid buffer (140~mM NaCl, 20~mM Tris-HCl, pH~8.0) at a final concentration of 1~mg/mL. The BSA solution was sterilized using a 0.22~$\mu$m syringe filter (Millex polyethersulfone, SLGPR33RS, Sigma-Aldrich), and injected into the sample chamber assembled with polyacrylamide-coated coverslips. After a 1-hour incubation at room temperature, the BSA solution was removed, and the chamber was rinsed three times with buffer prior to sample injection.

\subsection*{Sample preparation}
Samples consisted of fluorescently labeled 200-nm phagemids and fractionated 500\,kDa Dextran, both suspended in buffer (140\,mM NaCl, 20\,mM Tris-HCl, pH~8.0). The powdered Dextran obtained from fractionation was dissolved in the buffer to a final concentration of 65\,mg/mL. Labeled phagemids were similarly diluted in the buffer to a working concentration of 1\,mg/mL. Prior to loading into sample chambers, the solution containing phagemids and Dextran was centrifuged using a tabletop centrifuge to remove irregular aggregates (15\,min, 2{,}000$\times g$, 21\textdegree C), and the supernatant was transferred to a sample chamber. 

Sample chambers were constructed by sandwiching surface-coated coverslips separated by parafilm spacers of approximately 150\,$\mu \text{m}$ thickness. Once the sample was loaded, the chamber was sealed using a UV-curable adhesive. In addition, the formation of vertical membranes on the chamber floor was suppressed by using coverslips coated with both polyacrylamide and BSA (Fig.~E4).

\subsection*{Fluorescence imaging}
Fluorescence imaging of membranes and vesicles was primarily conducted using a spinning disk confocal microscope (Crest X-Light V2) mounted on a Nikon Ti2 inverted microscope. Images were acquired with an sCMOS camera (Photometrics Prime 95B, 25~mm sensor). A 20$\times$ water-immersion objective lens (numerical aperture [NA] = 0.95) was used for most experiments. Unless otherwise noted, all data presented in this study were obtained using this configuration. For specific experiments, the following objective lenses were used instead: 20$\times$ air objective (NA = 0.75) for Fig.~3a and Fig.~E1a, 40$\times$ air objective (NA = 0.75) for Fig.~E2, and 60$\times$ water-immersion objective (NA = 1.27) for Fig.~E3a. The microscope was equipped with appropriate filter sets for the fluorophore used (DyLight 550; excitation peak: 562~nm, emission peak: 576~nm), and imaging was performed at room temperature. Z-stacks were collected when necessary to reconstruct three-dimensional morphologies.

To image membrane growth and the transition to vesicles, samples were imaged immediately after preparation. To observe vesicle sedimentation and closure, imaging was initiated immediately after the application of the lateral centrifugal trigger. 

Fig.~E3b was acquired using a wide-field epifluorescence microscope equipped with a 100$\times$ oil-immersion objective lens (NA = 1.3). An additional 1.5$\times$ magnification was applied. Images were acquired with an sCMOS camera (pco.edge 4.2 bi).

\par\vspace{15pt}
\noindent \textbf{Statistical analysis.} Data are presented as mean ± standard deviation (s.d.).

\putbib

\end{bibunit}

\pagebreak
\widetext
\setcounter{equation}{0}
\setcounter{figure}{0}
\setcounter{table}{0}
\setcounter{page}{1}
\makeatletter
\renewcommand{\theequation}{S\arabic{equation}}
\renewcommand{\thefigure}{S\arabic{figure}}
\renewcommand{\thefigure}{E\arabic{figure}}
\renewcommand{\bibnumfmt}[1]{[S#1]}
\renewcommand{\citenumfont}[1]{S#1}
\makeatother

\begin{bibunit}

\section*{Supplementary information}
\subsection*{Supplementary Figures}

\begin{figure}[h!]
\renewcommand{\thefigure}{E\arabic{figure}}%
\centering
\includegraphics[width=.8\textwidth]{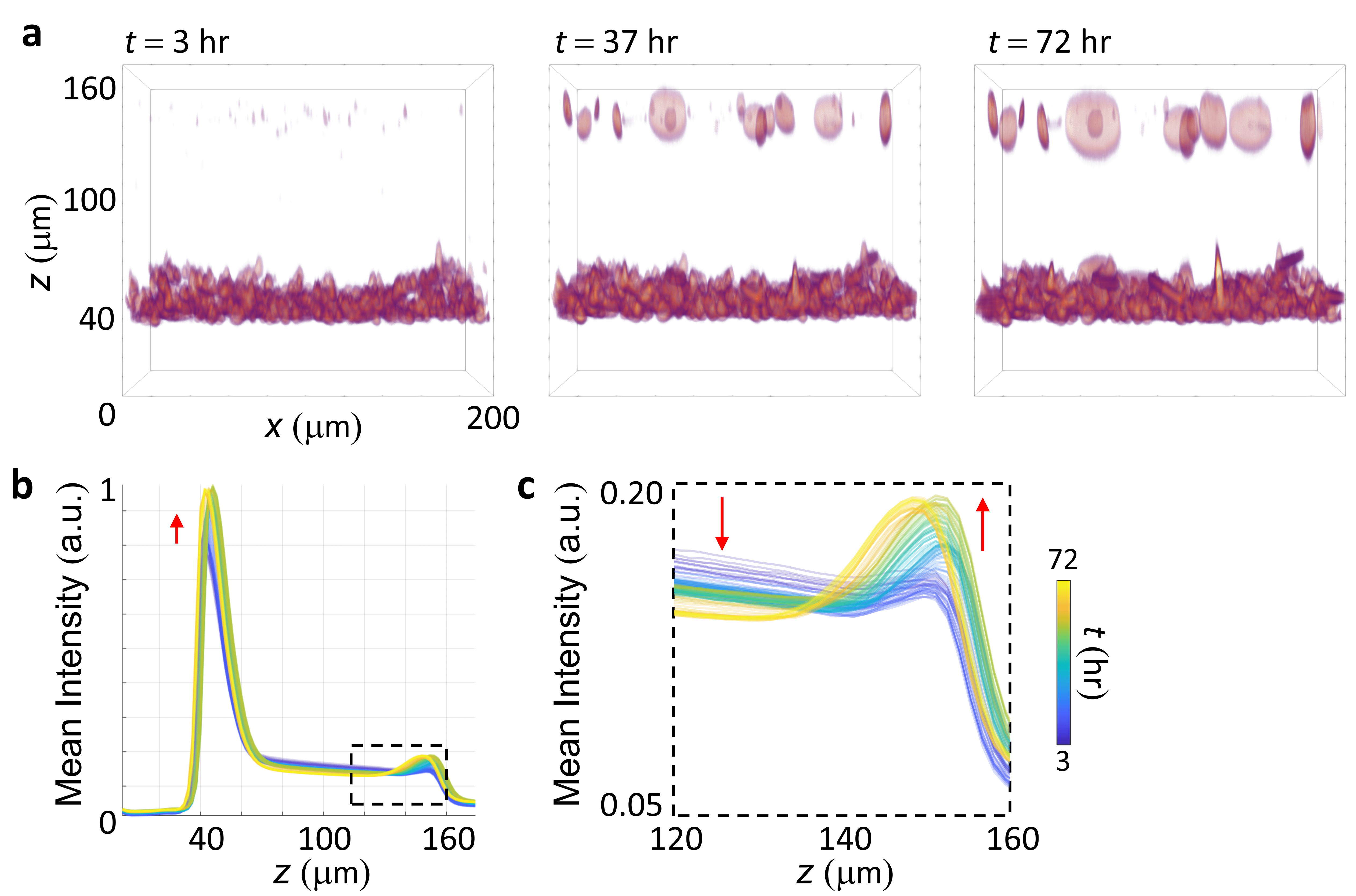}
\caption{\textbf{Time evolution of axial fluorescence intensity profiles.}
\textbf{a}, Time-lapse 3D imaging of the growth of membranes, starting 3 hours after sample preparation. 
\textbf{b}, Mean fluorescence intensity in the $xy$-plane at each axial $z$ position. The profiles are color-coded by time (3--72 hours). Two distinct peaks are observed near the chamber floor and ceiling, corresponding to membrane growth and vesicle formation at the surfaces. 
\textbf{c}, Magnified view of the dashed region in (\textbf{b}). Nonzero intensity is also maintained in the mid-axial region between the two surfaces, indicating the presence of isotropically dispersed particles. Over time, the mid-plane intensity gradually decreases, accompanied by a corresponding increase near the ceiling and floor planes, indicating continuous recruitment of dispersed particles into membranes.}
\end{figure}

\begin{figure}[h!]
\renewcommand{\thefigure}{E\arabic{figure}}%
\centering
\includegraphics[width=0.75\textwidth]{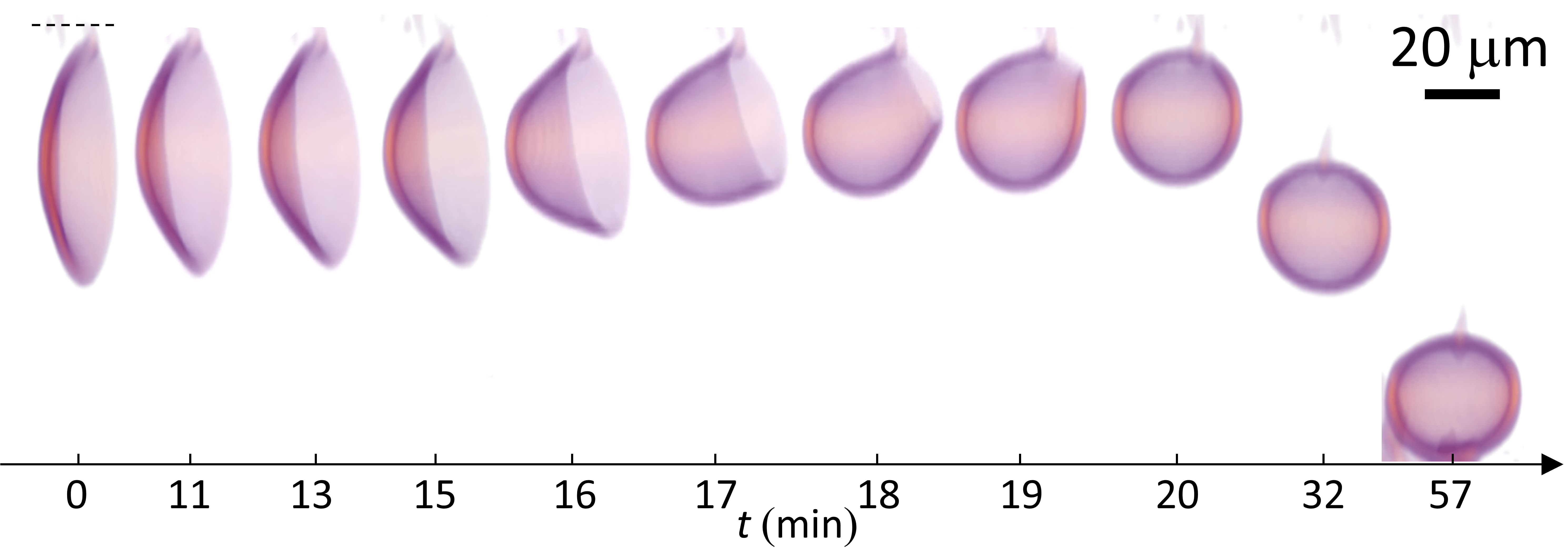}
\caption{\textbf{Time-lapse imaging of complete vesicle closure followed by sedimentation.}
Confocal fluorescence time-lapse images capturing the spontaneous transformation of a membrane into a fully closed vesicle, followed by its sedimentation onto the chamber floor. The dashed line indicates the chamber ceiling. While the majority of vesicles formed through self-assembly remain anchored near the ceiling as shown in Fig.~1g, a small subset of vesicles undergoes complete closure and subsequently sediments, as illustrated in this sequence.}
\end{figure}

\begin{figure}[h!]
\centering
\includegraphics[width=.9\textwidth]{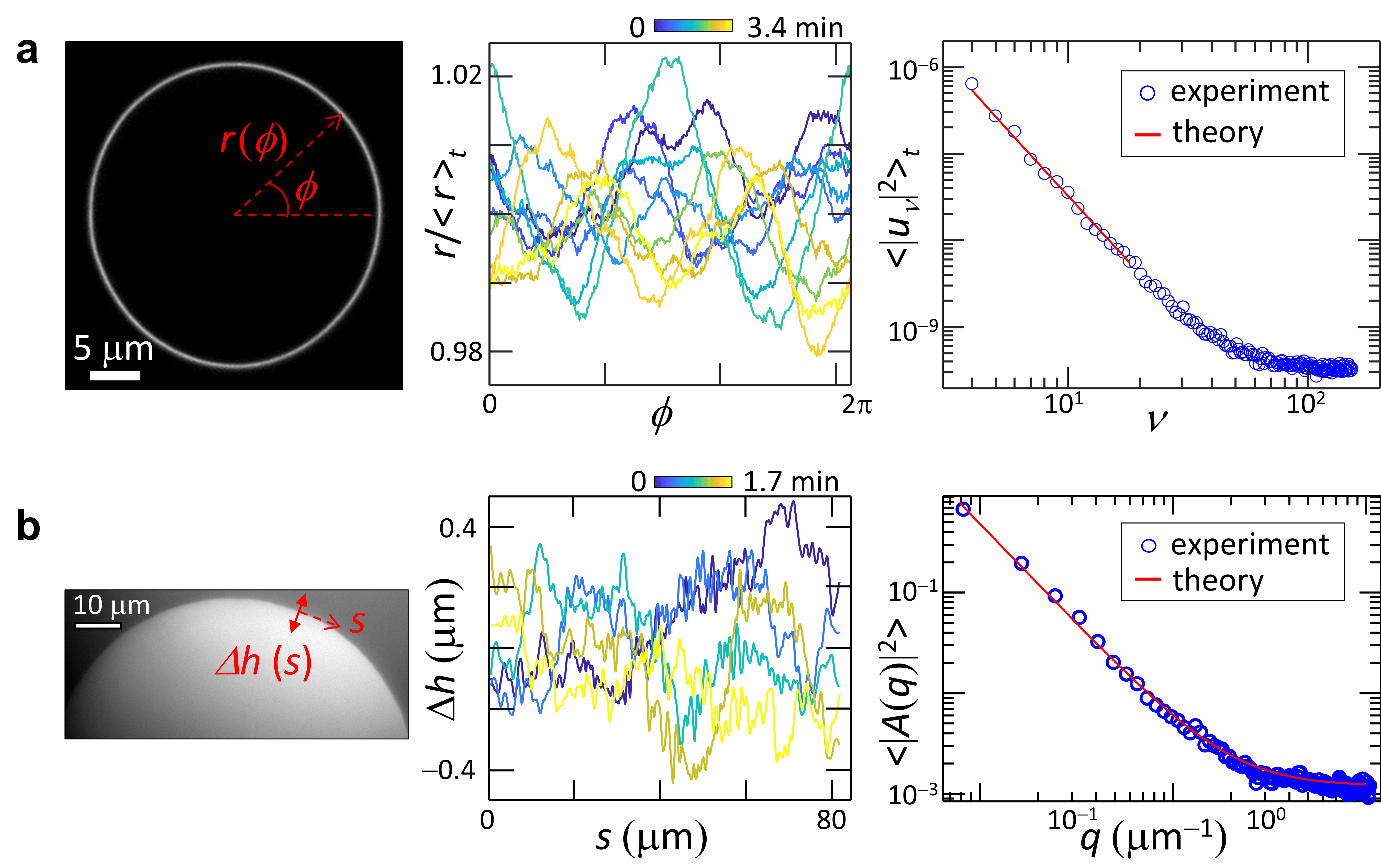}
\caption{\textbf{Measurement of membrane mechanical properties using flickering spectroscopy.}
\textbf{a}, Determination of the curvature modulus \( \kappa \) from thermal fluctuations of a closed vesicle.  
Left: A representative confocal fluorescence image of the equatorial cross-section of a closed vesicle. For each time frame, the radial distance \( r(\phi) \) from the center is measured as a function of the azimuthal angle \( \phi \).  
Middle: Temporal snapshots of normalized radial fluctuations, \( r(\phi)/\langle r \rangle_t \), shown for 10 representative frames sampled every 20 seconds.  
Right: Time-averaged mean square mode amplitudes \( \langle |u_\nu|^2 \rangle_t \) plotted against mode number \( \nu \), showing agreement between experimental data (blue circles) and theoretical prediction (red line) from SI.~1.
\textbf{b}, Determination of line tension from thermal fluctuations at the edge of a flat membrane.  
Left: Epi-fluorescence image of a flat membrane on the chamber floor, used to measure the edge profile. Displacement \( \Delta h(s) \) perpendicular to the edge contour is measured as a function of arc length \( s \).  
Middle: Five representative snapshots of \( \Delta h(s) \), sampled every 20 seconds.  
Right: Power spectral density \( \langle |A(q)|^2 \rangle \) plotted as a function of wave number \( q \), again showing agreement between experimental data and theory (see SI.~2).}
\end{figure}

\newpage

\begin{figure}[h!]
\centering
\includegraphics[width=.8\textwidth]{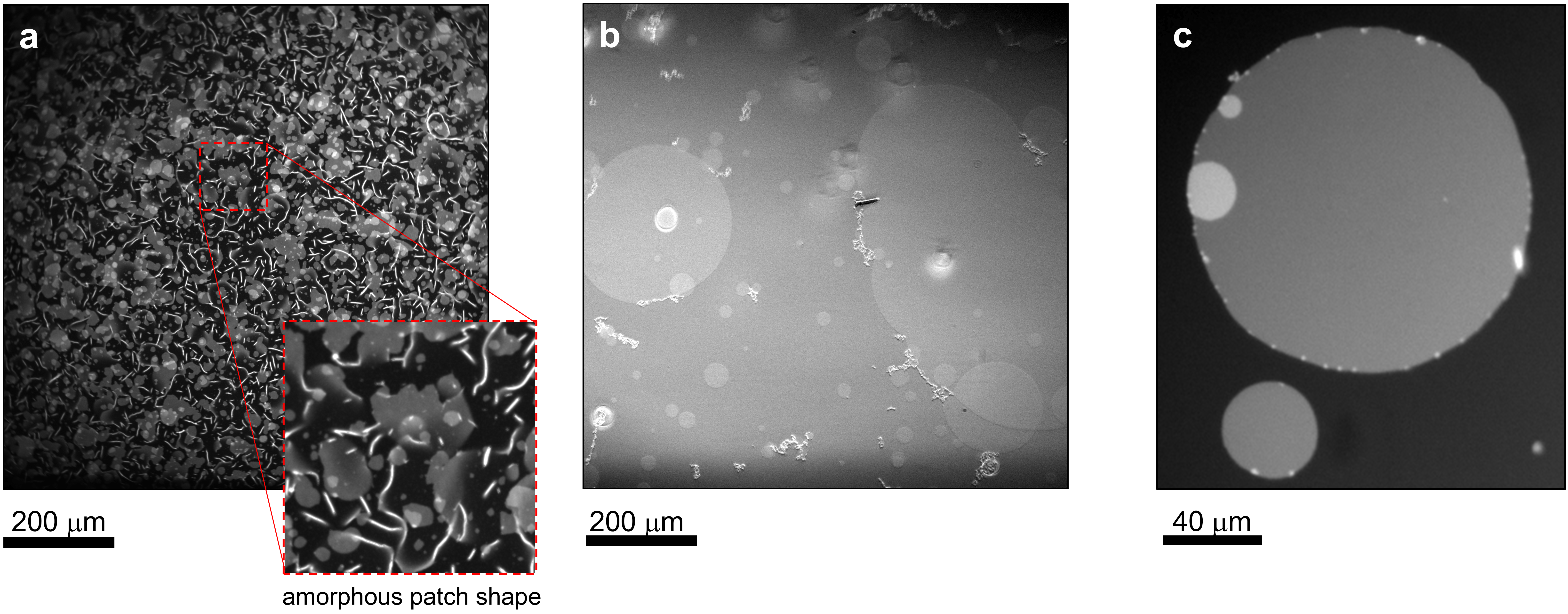}
\caption{\textbf{Additional surface coating to suppress the formation of vertical membranes attached to the chamber floor.} 
\textbf{a}, Confocal fluorescence image of the chamber floor after membrane growth on a coverslip coated with polyacrylamide. Both vertical membranes, observed through their cross-sectional profiles, and irregularly shaped amorphous patches are present on the chamber floor. These structures arise from membrane edges attaching to the polyacrylamide-coated surface. This condition prevents the formation of flat membranes lying horizontally on the chamber floor.
\textbf{b}, Differential interference contrast (DIC) image and \textbf{c}, epi-fluorescence image of the chamber floor when the coverslip surface is further coated with bovine serum albumin (BSA). In contrast to panel~a, vertical membrane formation is effectively suppressed. This suppression likely arises from BSA preventing membrane edges from adhering to the surface. Flat membranes are clearly visible lying parallel to the floor, appearing mostly circular due to minimized edge tension. Notably, even membranes whose area exceeds \( A^\dagger \), where the energy barrier vanishes, remain in the open, flat state, as discussed in the Discussion and SI.~4.}
\end{figure}

\begin{figure}[h!]
\centering
\includegraphics[width=0.77\textwidth]{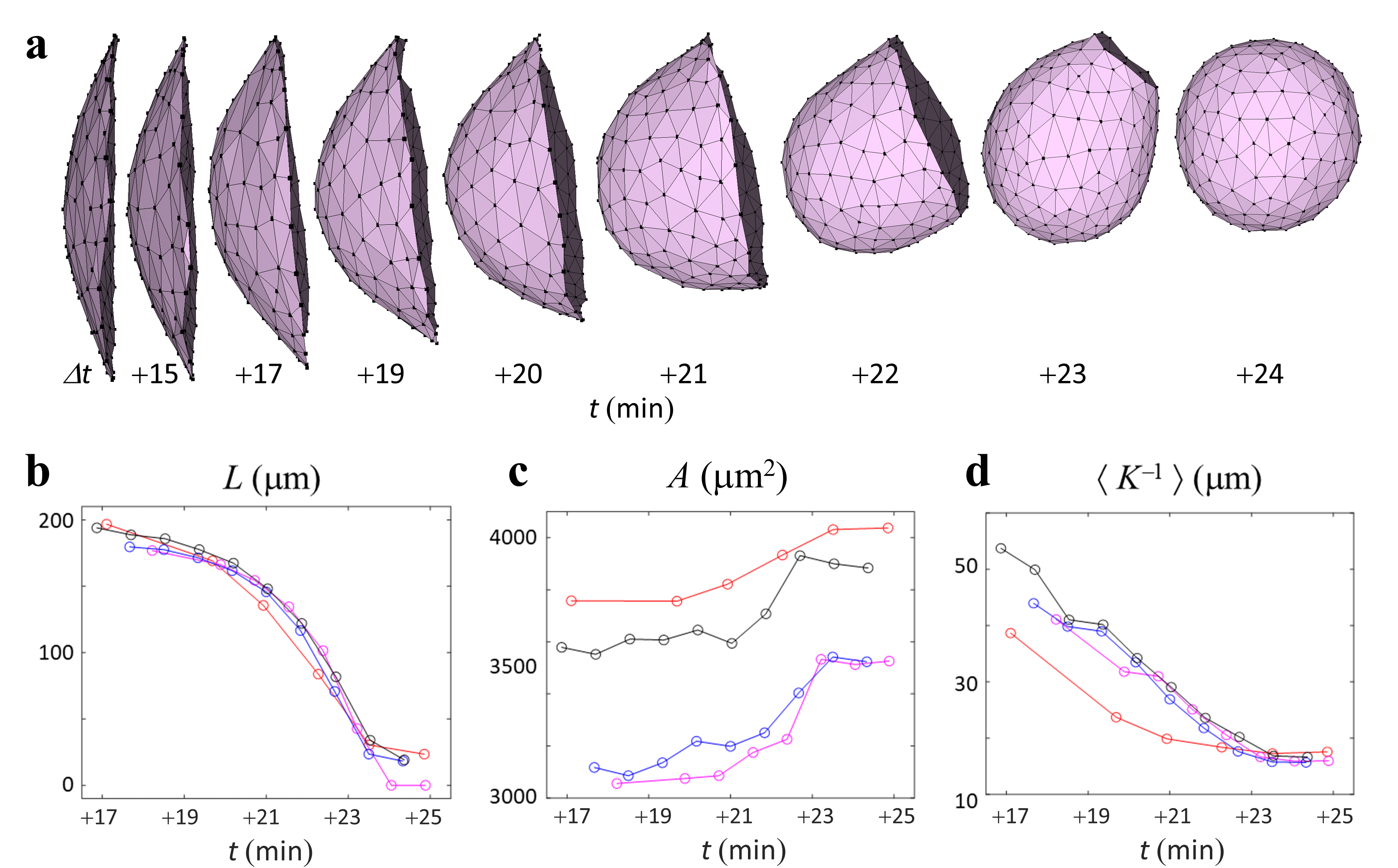}
\caption{\textbf{Time-resolved 3D quantification of the disk-to-vesicle transition.}
The slowed-down dynamics of the transition allowed direct 3D measurement of the morphological transformation between the open disk-like membrane and closed vesicle states.  
\textbf{a}, Representative time-lapse 3D mesh reconstructions capturing the progressive closure of a membrane shown in Fig.~E2. Time is shown relative to \(\Delta t\).
\textbf{b--d}, Time evolution of key morphological parameters extracted from the 3D meshes of four datasets:  
\textbf{b}, perimeter length \( L \),  
\textbf{c}, surface area \( A \), and  
\textbf{d}, spatial average of the inverse mean curvature, \( \langle K^{-1} \rangle \), over the vesicle surface. \\
Interestingly, although the four membranes have different areas, their perimeters evolve in a very similar manner over time. Another notable feature is that the membrane area increases by approximately 10\% during the late stages of the vesicle closure process. The magenta curve corresponds to the vesicle shown in (a) which fully closed and subsequently sedimented. These quantitative metrics characterize the shape evolution throughout the closure process.}
\end{figure}

\begin{figure}[h!]
\centering
\includegraphics[width=0.85\textwidth]{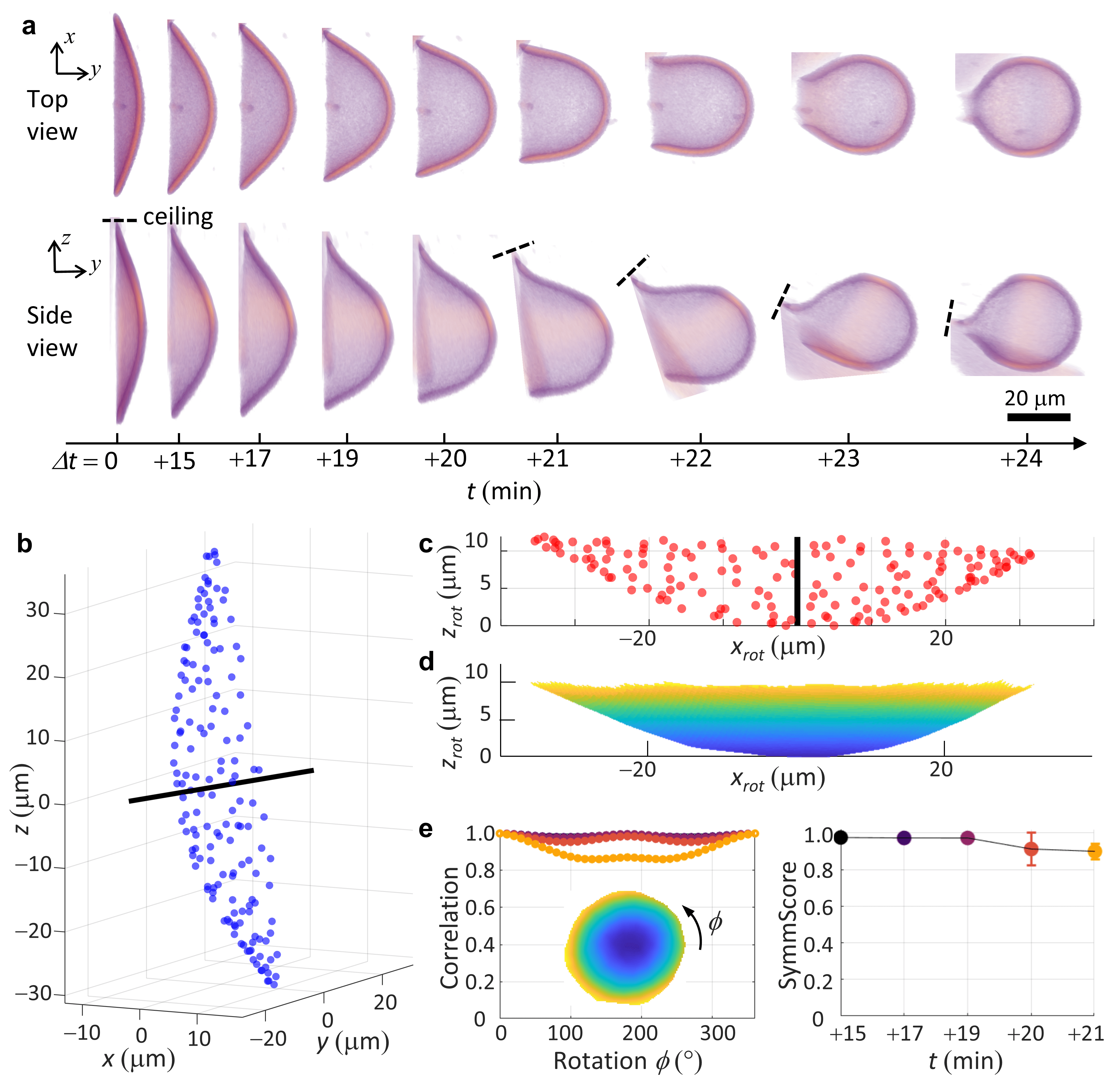}
\caption{\textbf{Quantification of axisymmetry during disk-to-vesicle transition.}
\textbf{a}, Time-lapse 3D confocal images capturing the disk-to-vesicle closure transition shown in Fig.~1g. Top and side views visualize the evolution of membrane shape to assess axisymmetry. Dashed lines in the side views indicate the chamber ceiling. Depending on the viewing angle and time, nearby membranes occasionally appear overlapped. The side-view images are tilted to best illustrate the degree of axisymmetry at each time point. 
\textbf{b}, 3D mesh vertices obtained from an observed membrane. 
\textbf{c}, Cylindrically symmetric axis (black line) identified by principal component analysis and aligned with the \(z\)-axis. 
\textbf{d}, Interpolated vertices and reconstructed surface. 
\textbf{e}, Left: Correlation function \( \mathrm{corr}(\phi) \) calculated by rotating the reconstructed surface around the \(z_{\mathrm{rot}}\)-axis by an angle \(\phi\). Colors correspond to the time points shown on the right. Right: Symmetry score, defined as \(1 - \mathrm{std}(\mathrm{corr}(\phi))/\mathrm{mean}(\mathrm{corr}(\phi))\), obtained from three data sets of closure dynamics. Time points correspond to those in (\textbf{a}). Membranes are highly axisymmetric in the early stages of the disk-to-vesicle transition, and symmetry gradually decreases as closure progresses.
}
\end{figure}
\newpage

\begin{figure}[h!]
\centering
\includegraphics[width=0.5\textwidth]{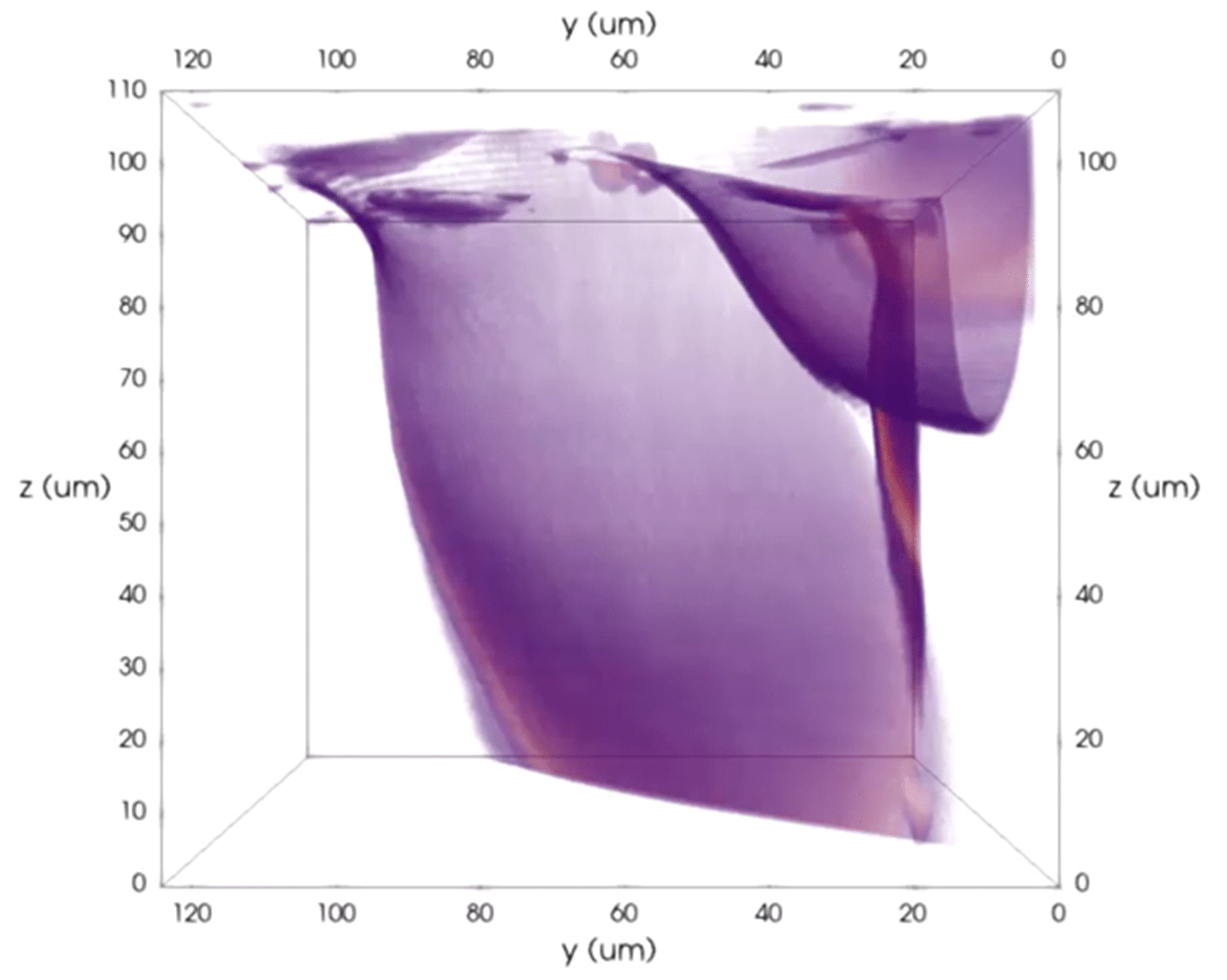}
\caption{\textbf{Anchored membrane with increased thickness.}
Volumetric 3D rendering of a 315-nm thick membrane anchored to the chamber ceiling. Unlike thinner membranes that undergo spontaneous vesiculation, this thicker membrane remains flat and extended under the same gravitational conditions, consistent with theoretical predictions that no critical size exists in this regime. The stable, hanging configuration further highlights the mechanical robustness of the system and confirms the role of membrane thickness in controlling vesicle formation.}
\end{figure}

\newpage
\clearpage
\setcounter{figure}{0}

\subsection*{Supplementary Information 1: Measurement of bending modulus }

We used flickering spectroscopy to measure the bending modulus of 200-nm-thick membranes $\kappa$ \cite{mutz_bending_1990,faizi_curvature_2024}. We recorded time-lapse confocal images of the equatorial cross-sections of closed vesicles (Fig.~E3a). The vesicle contour was determined from the fluorescence intensity profile of each cross-sectional image. First, the vesicle center was identified from the centroid of the overall intensity distribution. Then, intensity profiles were extracted radially from the center at evenly spaced azimuthal angles (\( \phi \)). For each angular direction, the intensity along the radial line was interpolated and fitted to a Gaussian function to accurately locate the peak corresponding to the vesicle edge. This fitting procedure provided subpixel precision for the radial distance \( r(\phi, t) \) from the center as a function of the azimuthal angle \( \phi \). A sequence of \( r(\phi, t) \) profiles was measured at one-second intervals. The fluctuations were decomposed into Fourier modes \( u_\nu(t) \) via

\[
u_\nu(t) = \frac{1}{2\pi R} \int_0^{2\pi} r(\phi, t) e^{-i \nu \phi} \, d\phi = \sum_{\ell = \nu}^{\ell_{\mathrm{max}}} f_{\ell \nu}(t) n_{\ell \nu} P_{\ell \nu}(0),
\]
where \( R \) is the average radial distance, \( f_{\ell \nu}(t) \) are the amplitudes of the spherical harmonic modes, \( n_{\ell \nu} \) is a normalization constant, and \( P_{\ell \nu}(0) \) is the associated Legendre polynomial evaluated at the equator \cite{faizi_curvature_2024}. This expression follows from the projection of \( r(\phi, t) \) onto spherical harmonics, where the standard spherical harmonic functions are defined as

\[
Y_{\ell m}(\theta, \phi) = n_{\ell m} P_{\ell m}(\cos \theta) e^{i m \phi}, \quad
n_{\ell m} = \sqrt{\frac{(2\ell + 1)(\ell - m)!}{4\pi(\ell + m)!}}.
\]
Since we focus on equatorial cross-sections, \( \theta = \pi/2 \), leading to the simplification \( P_{\ell m}(\cos \theta) = P_{\ell m}(0) \).

The time-averaged mean square amplitude \( \langle |u_\nu|^2 \rangle \) was compared to the theoretical spectrum,

\[
\langle |u_\nu|^2 \rangle = k_\mathrm{B} T \sum_{\ell = \nu}^{\ell_{\mathrm{max}}} \left[ (\ell + 2)(\ell - 1)(\ell(\ell + 1)\kappa + \sigma R^2) \right]^{-1} n_{\ell \nu}^2 |P_{\ell \nu}(0)|^2,
\]
where \( \kappa \) is the bending modulus and \( \sigma \) is the surface tension. 

Fitting the experimental data to this model yielded curvature modulus \( \kappa = 1200 \pm 60 \, k_\mathrm{B} T \) and surface tension \( \sigma = 0 \pm 16 \, k_\mathrm{B} T \) (Fig.~E3a). Fluid membranes adopt an energetically optimal area and resist stretching or compression. The vanishing surface tension is consistent with theoretical expectations for fluid membranes in the absence of external forces \cite{peliti_effects_1985,evans_entropy-driven_1990,lipowsky_conformation_1991}. 

\subsection*{Supplementary Information 2: Measurement of edge tension }

To determine the edge tension we recorded the in-plane thermal fluctuations of flat membrane lying on the chamber bottom. Flat membranes were imaged using epi-fluorescence microscopy (Methods), and their edge contours were extracted from the intensity profiles. For each membrane, we defined the arc length coordinate \( s\) along the membrane edge, and the vertical edge profile \( h(s) \) (Fig.~E3b). We computed the first derivative of the images to locate the position of the steepest increase, corresponding to the membrane edge. 
The position of the maximum gradient was refined to subpixel accuracy by fitting a quadratic function over a narrow window (typically 5 pixels wide) centered around the peak. The vertex of the fitted parabola was analytically computed and taken as the subpixel-resolved edge position.

Temporal fluctuations of the edge were quantified by tracking the local height \( h(s, t) \), measured normal to the membrane edge as a function of arc length \( s \) and time \( t \). The deviation from the time-averaged position, \( \Delta h(s, t) = h(s, t) - \langle h(s, t) \rangle_t \), represents the local edge displacement. The fluctuations \( \Delta h(s, t) \) were sampled at one-second intervals. These resulting fluctuations spectrum can be fitted to the following expression:
\[
\langle |A(q)|^2 \rangle = \langle \nu^2 \rangle + \frac{k_\mathrm{B} T}{\gamma q^2 + \kappa_B q^4},
\]
where \( \gamma \) is the edge tension, \( \kappa_B \) is the edge bending stiffness, and \( \langle \nu^2 \rangle \) accounts for baseline noise or instrumental uncertainty \cite{gibaud_reconfigurable_2012,jia_chiral_2017}. From four data sets, we extracted a edge tension of \( \gamma = 200 \pm 30 \, k_\mathrm{B} T/\mu\mathrm{m} \) (Fig.~E3b). The corresponding edge bending stiffness is \( \kappa_B = 1.4 \pm 2.2 \, k_\mathrm{B} T \), indicating that the edge bending contribution is negligible compared to the total edge energy.

\subsection*{Supplementary Information 3: Estimate of the Gaussian curvature modulus}

Gaussian curvature modulus $\bar{\kappa}$ can be approximated by
\(\bar{\kappa} \approx \frac{1}{6} n l^2 R_g \, k_\mathrm{B} T\) where $n$ is the depletant number density, and $R_g$ is the radius of gyration of the depletant~\cite{gibaud_achiral_2017}. For our system, we use $l = 206~\mathrm{nm}$, a polymer concentration of $n = 65~\mathrm{mg/mL}$, and a radius of gyration $R_g = 30~\mathrm{nm}$ corresponding to 500~kDa dextran \cite{senti_viscosity_1955}. Substituting these values yields $\bar{\kappa} \approx 20~k_\mathrm{B} T$.

\subsection*{Supplementary Information 4: Theoretical Models}

We present the equilibrium shape analysis for a membrane attached by a single edge point to the chamber ceiling. Then, we discuss the analysis of two other experimental conditions: (1) a membrane resting by a single point on its edge at the chamber floor, and (2) a membrane with its entire circumference supported by the chamber floor.

The equilibrium shapes are determined by minimizing the total free energy, consisting of a sum of the Helfrich energy $(E_{\mathrm{Helfrich}})$, gravitational energy $(E_{\mathrm{grav}})$, and edge tension energy $(E_{\mathrm{edge}})$
\begin{equation}
E = \underbrace{\int_S \left[ \frac{\kappa}{2} (2K)^2 + \bar{\kappa} K_G   \right] \, dS}_{E_{\text{Helfrich}}} +  \underbrace{\int_S \rho_\mathrm{a} g z \, dS}_{E_{\text{grav}}} 
+ \underbrace{\gamma \int_{\partial S} dL}_{E_{\mathrm{edge}}} +  \underbrace{\int_S \mu \, dS}_{\text{area constraint}}, \label{eq:Helfrich}
\end{equation}
where $K$ and $K_G$ are the mean and Gaussian curvatures, $\kappa$ and $\bar{\kappa}$ are the bending stiffness and Gaussian curvature modulus, $\rho_\mathrm{a}$ is the areal mass density, $g$ is the gravitational acceleration, $\mu$ is a Lagrange multiplier to enforce fixed surface area, $\gamma$ is the edge tension along the membrane boundary $\partial S$, and $z$ denotes the height of a surface element. 

Membrane shapes are axisymmetric during the early stages of the  transition (Fig.~E6). Therefore, our models assume such symmetry. To assess the relevance of gravity, we compare the gravitational energy to the total membrane bend and edge tension energy. For a disk the characteristic length is $R = (A/\pi)^{1/2}$; for a sphere, $R = (A/(4\pi))^{1/2}$. We introduce a Bond number ($\text{Bo}_{\gamma}$) based on edge tension and Bond number ($\text{Bo}_{\kappa}$) based on bending stiffness \cite{RAPP2017243}
\begin{align}
    \text{Bo}_{\gamma} &= \dfrac{\rho_\mathrm{a} g R^2}{\gamma} \label{eq:Bondgamma} \\
    \text{Bo}_{\kappa} &= \dfrac{\rho_\mathrm{a} g R^3}{\kappa}. \label{eq:Bondkappa}
\end{align}
Both numbers are $10^{-2}$--$10^{-1}$, indicating the minor role of gravity. Thus, consistent with the experimental observations, we assume a circular edge. Due to slow transition dynamics, we adopt a quasi-static approximation. Accordingly, for fixed area and edge length, we determine shapes that minimize the Helfrich and edge tension energy, neglecting gravity (Eq.~\eqref{eq:Helfrich}). Gravity, however, does influence the vesicle distributions since it contributes to the total free energy (Eq.~\eqref{eq:Helfrich}). Specifically, the floor bound membranes formed smaller vesicle than the ceiling ones (Fig.~\ref{fig3}). In both cases, the membranes tilt from the vertical position, shifting the center of mass height, thereby modifying the gravitational energy along the transition pathway (Fig.~E2). 

\begin{figure}[h!]
\renewcommand{\thefigure}{S\arabic{figure}}%
\centering
\includegraphics[width=0.45\textwidth]{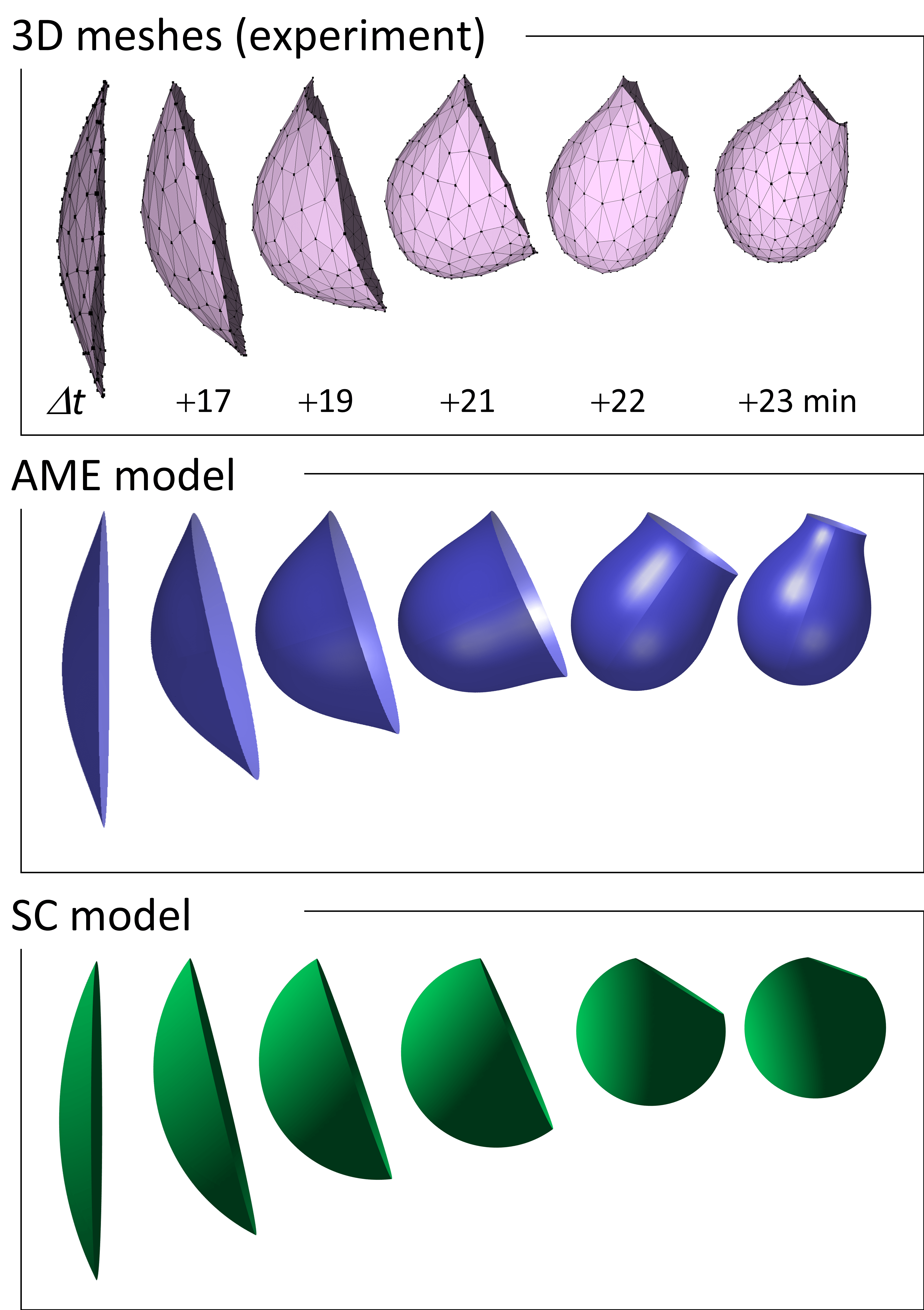}
\caption{\textbf{Comparison of morphologies during vesicle closure} \\
Top: reconstructed 3D meshes from experimental images, middle: predicted shapes based on the AME model, bottom: predicted shapes based on the SC model. 
}\label{figS2}
\end{figure}

Intermediate membrane conformations are modeled with two approaches. First, in the spherical cap (SC) approach we construct a simple, analytical solution by approximating intermediate conformations as spherical caps. Second, the axisymmetric minimum energy (AME) model reduces the full variational problem to the Euler-Lagrange equations, which are numerically solved to obtain the energy minimizing shapes. The simpler analytically tractable SC model provides a rough approximation of the experimental shapes (Fig.~\ref{figS2}), while the full numerical solution provided by the AME model more accurately represents the experiments.

\subsubsection*{Spherical Cap (SC) model}
We simplify the axisymmetric problem by assuming that the intermediate shapes are spherical caps. The two limiting cases are the flat disk corresponding to the initial configuration, and the closed sphere corresponding to the final configuration.  The gravitational potential is zero at the chamber ceiling, with downward $z$ being negative. 

A closed sphere with area $A$ and radius $(A/(4\pi))^{1/2}$ attached to the chamber ceiling has no edge energy ($E_{\mathrm{edge}}^{\mathrm{sph}} = 0$), while the Helfrich and gravitational energy are
\begin{align}
    E_{\mathrm{Helfrich}}^{\mathrm{sph}} &= 4\pi (2\kappa + \bar{\kappa}) \\
    E_{\mathrm{grav}}^{\mathrm{sph}} &= -\rho_\mathrm{a} g A \left(\dfrac{A}{4\pi}\right)^{1/2}.
\end{align}
A flat disk with area $A$ and radius $ (A/\pi)^{1/2}$ attached to the chamber ceiling by a point on its circumference has zero Helfrich energy ($E^{\mathrm{disk}}_{\mathrm{Helfrich}} = 0$), while the edge tension energy and gravitational energy are
\begin{align}
 E_{\mathrm{line}}^{\mathrm{disk}} &= 2\pi \gamma\left( \dfrac{A}{\pi}\right)^{1/2}\\
E_{\mathrm{grav}}^{\mathrm{disk}} &= -\rho_\mathrm{a} g A \left(\dfrac{A}{\pi}\right)^{1/2}.
\end{align}
The disk has no bending cost, but incurs an edge tension penalty proportional to its perimeter. The co-existence boundary (black line in Fig.~\ref{fig3}d) is determined by equating total disk and vesicle energies 
\begin{align}
    E_{\mathrm{Helfrich}}^{\mathrm{sph}} + E_{\mathrm{grav}}^{\mathrm{sph}} &=  E_{\mathrm{line}}^{\mathrm{disk}} + E_{\mathrm{grav}}^{\mathrm{disk}}\label{eq:coexistence}.
\end{align}
For membranes with area $A > A^* = 4\pi(2\kappa + \bar{\kappa})^2/\gamma^2$, disks can be metastable due to an energy barrier. This barrier arises because of the competition between the bending energy, which penalizes curvature and resists closure, and the edge tension which favors perimeter reduction and drives closure. Reducing the disk radius decreases the edge energy; however, at a fixed area, the disk must bend, which increases the bend energy. For small membranes, the increase in bend energy is larger than the decrease in edge tension energy, so the flat disk state represents a local minimum. For larger membranes areas, the reduction in the perimeter lowers the edge tension energy more than it raises the bend energy, lowering the barrier. For large enough areas the barrier vanishes and the disk is unstable.

\begin{figure}[!h]
\renewcommand{\thefigure}{S\arabic{figure}}%
\centering
\includegraphics[width=\textwidth]{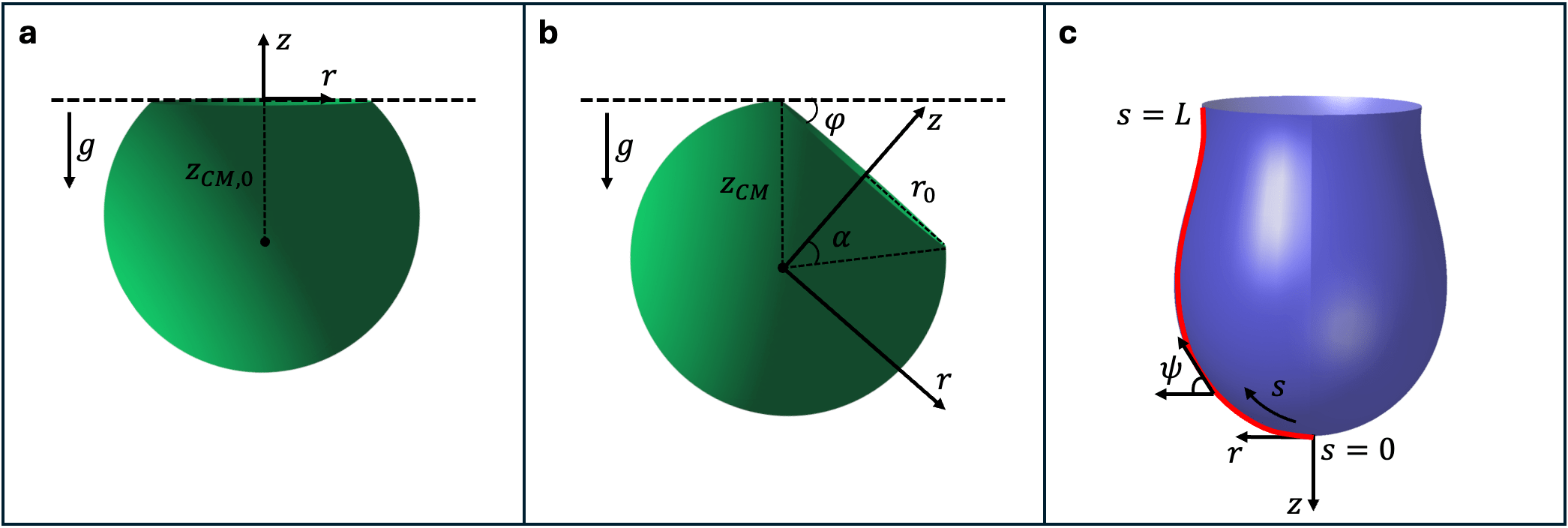}
\caption{\textbf{Schematic of geometric parameters in the SC and AME models } \\
\textbf{a}, A spherical cap whose base is attached parallel to the ceiling with the base center at the origin. The center of mass is located at $(0, z_{\mathrm{CM,0}})$. 
\textbf{b}, 
A spherical cap with curvature radius $R$ and rim radius $r_0$ attached by an edge to the ceiling. $\varphi$ denotes the tilt angle so that the center of mass $z_{\mathrm{CM}}$ lies directly below the attachment point. $\alpha$ denotes the polar angle where $\alpha \rightarrow 0$ is a closed sphere and $\alpha \rightarrow \pi$ is a flat disk. The zero point of the gravitational potential is taken to be the ceiling, with negative downward. This configuration is achieved from \textbf{a} by a translation and rotation, as discussed by following Eq.~\eqref{eq:zcm0}--\eqref{eq:zcm}. \textbf{c}, A shape solution of the AME model in the absence of gravity, with $\psi$ denoting the tangent angle to the meridional curve (red) for the axisymmetric shape, and $s$ is the arclength parameter, $s \in [0,L]$. }
\label{figscheme}
\end{figure}

To quantify how the energy barrier depends on the weight difference and surface area, we compute the energies of the disk and cap. A hollow spherical cap with surface area $A$, curvature radius $R$ and rim radius $r_0 = \left({{A}/{\pi} - A^2/(2 \pi R )^2}\right)^{1/2}$ has the bend and edge energy 
\begin{align}
    E_{\mathrm{Helfrich}}^{\mathrm{cap}} &= \dfrac{2\kappa + \bar{\kappa}}{R^2} A \label{eq:bendEnergy} \\ 
    E_{\mathrm{line}}^{\mathrm{cap}} &= 2\pi \gamma r_0. \label{eq:edgeEnergy} 
\end{align}
These energies interpolate between the limiting cases of the disk and closed vesicle, incurring both bend and edge tension penalties. 
Experiments show that the membrane tilts during the transitions. To account for this, we minimize the gravitational energy of a tilted spherical cap. The stable equilibrium configuration occurs when the cap's center of mass is directly below the suspension point. We first determine the required tilt angle $\varphi$, and then the center of mass height in this tilted configuration. We begin by considering an untilted spherical cap whose base is attached parallel to the ceiling with the base center at the origin (Fig.~\ref{figscheme}a). Its center of mass is located at
\begin{align}
    \begin{bmatrix}
        r_{\mathrm{CM},0} \\ z_{\mathrm{CM},0} 
    \end{bmatrix}
    &= 
    \begin{bmatrix}
         0 \\ -\dfrac{A}{4\pi R}
    \end{bmatrix}. \label{eq:zcm0}
\end{align} 
The following translation by the edge radius $r_0$ and rotation by $\varphi$ about the suspension point yields the center of mass location in the tilted configuration
\begin{align}
    \begin{bmatrix}
        r_{\mathrm{CM}} \\ {z}_{\mathrm{CM}}
    \end{bmatrix} &= 
    {\begin{bmatrix}
        \cos \varphi & \sin\varphi \\
        -\sin \varphi & \cos \varphi 
    \end{bmatrix}}\left(
    \begin{bmatrix}
        0\\
        z_{\mathrm{CM},0} \\ 
    \end{bmatrix} + \begin{bmatrix}
         r_0 \\ 0\\
    \end{bmatrix}\right).
\end{align}
Requiring the center of mass to lie directly below suspension means $r_{\mathrm{CM}} = 0$, or equivalently,
\begin{align}
   \tan\varphi &= - \dfrac{r_0}{z_{\mathrm{CM},0}}. 
\end{align}
Solving for $\varphi$, the height of the center of mass in the tilted configuration with respect to the suspension is
\begin{align}
    z_{\mathrm{CM}} =-r_0\sin\varphi + z_{\mathrm{CM,0}}\cos\varphi. \label{eq:zcm}
\end{align}
Finally, the gravitational potential energy for the spherical cap in its tilted configuration is given in terms of $z_{\mathrm{CM}}$ (Fig.~\ref{figscheme}b), 
\begin{align}
    E_{\mathrm{grav}}^{\mathrm{cap}} &= \rho_\mathrm{a} g A z_{\mathrm{CM}}.
\end{align}
With known energy contributions, we formulate the metastability condition. The membrane area $A$ and weight difference per area $\rho_{\mathrm{a}}g$ is fixed. We denote the difference between the cap and disk energy as $\Delta E(R; A, \rho_{\mathrm{a}}g)$. The admissible curvature radii are $R \in [R_{\mathrm{min}}, \infty)$, where $R_{\mathrm{min}} =  \left({A}/{4\pi}\right)^{1/2}$. For convenience, set $\epsilon = 1/R$. Then, the boundary of metastability is given by the locus of parameters satisfying
\begin{align}
    \dfrac{\partial}{\partial \epsilon} \Delta E(\epsilon;A,\rho_{\mathrm{a}}g)\bigg|_{\epsilon=\epsilon^*} = 0, \qquad
   \frac{\partial^2}{\partial \epsilon^2} \Delta E(\epsilon;A, \rho_{\mathrm{a}}g)\bigg|_{\epsilon=\epsilon^*} = 0 \label{eq:meta_conds}
\end{align}
for some $\epsilon^* \in (0, 1/R_{\mathrm{min}}]$. 
Since the energy barrier lies between the disk and shallow cap configurations, we derive an analytical expression for the metastability curve by expanding the non-dimensional energy difference about the flat disk limit ($\epsilon \rightarrow 0$). Applying the metastability conditions (Eq.~\eqref{eq:meta_conds}) to this expansion, we find 
\begin{align}
    \hat{\rho} &= \dfrac{2}{3} \dfrac{\hat{A}^{3/2} - 2\hat{A}}{\hat{A}^{5/2}}, \label{eq:metastability_hanging}
\end{align}
where $\hat{A} = A/A^*$ and $\hat{\rho} = \rho_\mathrm{a} g (2\kappa + \bar{\kappa})^2/\gamma^3$ are the dimensionless area and dimensionless membrane areal weight respectively. We use $A^* = \pi(R^*)^2$ and $R^* = 2(2\kappa + \bar{\kappa})/\gamma$ to denote the characteristic area and length. The metastability curve (Eq.~\eqref{eq:metastability_hanging}) represents the locus of parameters where the energy barrier vanishes (Fig.~\ref{fig3}d orange). 

\subsubsection*{Axisymmetric Minimum Energy (AME) model}
For a more accurate determination of the membrane shapes, we solve the full variational problem assuming axisymmetry.
An axisymmetric surface obtained by revolving a meridional curve $(r(s),z(s))$ about the $z$-axis is parametrized by  
\begin{align}
\vec{x}(s,\theta) = ( r(s) \cos\theta , r(s)\sin\theta, z(s)),
\end{align}
where $\theta \in [0, 2\pi)$ is the azimuthal angle and $s$ is the arclength coordinate in the meridional plane such that $r'(s)^2+z'(s)^2=1$. Since $(r',z')$ is the unit tangent, we define the tangent angle $\psi(s)$ such that  
\begin{equation}
    r'(s) = \cos \psi(s), \qquad z'(s) = -\sin \psi(s) \label{eq:geo}
\end{equation}
where the negative sign in Eq.~\eqref{eq:geo} is chosen so that $z$ decreases as $s$ increases. 

With this parametrization, the total free energy becomes 
\begin{multline}
    \dfrac{E}{2\pi} = \int  \left[\dfrac{\kappa}{2} \left( \psi' + \dfrac{\sin \psi}{r} \right)^2 + \bar{\kappa}\psi' \dfrac{\sin \psi}{r} \right] r\, ds + \int \rho_\mathrm{a} g z r \, ds  \\  + \int \left[ q(r' - \cos \psi) + \eta (z' + \sin \psi )\, \right] ds + \dfrac{\gamma}{2\pi} \int dL + \int \mu r \, ds, \label{eq:FreeEnergyAxi}
\end{multline}
where $q(s)$ and $\eta(s)$ are Lagrange multipliers that correspond to the radial and vertical forces respectively and enforce the geometric constraints~(Eq.~\eqref{eq:geo}), and $\psi'$ is the derivative of the tangent angle with respect to the arclength $s$. 
Since the Bond numbers are small, gravity does not significantly affect the shape. Hence, we discard it in Eq.~\eqref{eq:FreeEnergyAxi} for simplicity. To determine the equilibrium shapes, the variational problem is converted into a system of boundary value problems (BVPs) \cite{Julicher1994, adkins_topology_2025}
\begin{align}
    \dfrac{d\psi}{dT} &= L \psi' \label{eq:BVP1} \\
    \dfrac{d \psi'}{dT} &= - L \dfrac{\cos \psi \psi'}{r} + L \dfrac{\sin \psi \cos \psi}{r^2} + L \dfrac{q}{r\kappa} \sin \psi + L \dfrac{\eta }{r\kappa} \cos \psi \\ 
    \dfrac{dr}{dT} &= L \cos \psi \\
    \dfrac{dz}{dT} &= -L \sin \psi \\
    \dfrac{dq}{dT} &= L \dfrac{\kappa}{2} \psi'^2 - L \dfrac{\kappa}{2} \dfrac{\sin^2 \psi}{r^2} + L \mu \\
    \dfrac{dA}{dT} &= 2\pi r L \\
    \dfrac{d\mu}{dT} &= 0 \\
    \dfrac{d\eta}{dT} &= 0 \\
    \dfrac{dL}{dT} &= 0, \label{eq:BVP9}
\end{align}
where $L$ is the total arclength of the curve and $T = s/L \in [0,1]$ is the reduced arclength. The arclength is measured starting from the bottom of the membrane, at $z = 0$, increasing along the clockwise direction (Fig.~\ref{figscheme}c). 

We impose the following boundary conditions (BCs) at $T = 0$: zero radius, zero integrated area, and zero tangent angle
\begin{align}
    z(0) = 0, r(0) = 0, A(0) = 0, \psi(0) = 0. 
\end{align}
At the rim $(T=1)$, the BCs consist of fixed rim radius $r_1$, fixed integrated surface area, and zero bending moment at the rim,
\begin{align}
    r(1) = r_1, A(1) = A, \psi_s(1) = -\dfrac{\sin \psi(1)}{r(1)}.
\end{align}
Finally, we impose zero force at $s = 0$
\begin{align}
    \eta(0) = 0, q(0) = 0.
\end{align} 
These boundary value problems are solved numerically in MATLAB using  function \lstinline[style=Matlab-editor]{bvp4c}. 

We quantify the dependence of energy barrier on the membrane area and areal weight in the AME model. Let $S_0(A)$ be the family of equilibrium shapes at zero gravity for fixed area $A$, obtained by solving the BVPs (Eq.~\eqref{eq:BVP1}--\eqref{eq:BVP9}).  Since the Bond numbers are small (Eq.~\eqref{eq:Bondgamma}--\eqref{eq:Bondkappa}), gravity weakly perturbs  the zero gravity shape solutions $S_0$. By up-down symmetry, flipping the sign of gravity does not alter the shape of a membrane whether attached by a point on its rim to the ceiling or the floor. To first order, the shape is gravity independent: the shape correction is $S(\mathrm{Bo}) = S_0 + O(\mathrm{Bo}^2)$. Consequently, to first order in $\mathrm{Bo}$, we can evaluate the energy based on $S_0$. The center of mass height is $z_{\mathrm{CM}}(S)$ and the tilt angle is $\varphi$  (Fig.~\ref{figscheme}). Minimizing over $\varphi$ so that the center of mass lies directly below the suspension point (zero torque condition), the total free energy is 
\begin{align}
    E^{\mathrm{AME}}(S,\mathrm{Bo}) &= E_{\mathrm{Helfrich}}(S_0) + E_{\mathrm{line}}(S_0) + \rho_\mathrm{a} g A z_{\mathrm{CM}}(S_0) + O(\mathrm{Bo}^2). 
\end{align}
The gravitational energy is determined by numerically calculating $z_\mathrm{CM}(S_0)$ in the tilted configuration. We use the method of a translation and rotation described for the spherical cap, to compute the tilt angle $\varphi$. The metastability curve is computed numerically by a parameter sweep over the membrane areal weight $\rho_\mathrm{a} g$ and surface area $A$. For any $(A, \rho_\mathrm{a} g)$, we compute the energy difference $\Delta E = E^{\mathrm{AME}} - E^{\mathrm{disk}}$ along the family of shape solutions connecting disk-like structures to vesicles, indexed by the rim radius. A barrier exists when $\Delta E > 0$. Therefore, for each $A$, we seek the smallest $\rho_{\mathrm{a}}g$ such that $\max \Delta E = 0$. The locus of parameters $(\rho_{\mathrm{a}} g, A)$ defines the metastability curve (Fig.~\ref{fig3}d, purple), representing the set of solutions where the energy barrier vanishes. 

\subsection*{Membranes on the chamber floor}
Experiments show that membranes supported by an edge point at the floor also form closed vesicles (Fig.~\ref{fig3}a). The geometrical setup is identical to Fig.~\ref{figscheme}b except that gravity is inverted, with the potential increasing with height. Accordingly, this case is obtained from our existing framework with $\rho_\mathrm{a} g\mapsto -\rho_{\mathrm{a}} g$. The critical membrane floor and ceiling attached membranes correspond to the intersections of the metastability curve (purple) with the blue and red dashed lines (Fig.~\ref{fig3}d). The model captures the observations that floor-bound membranes form smaller vesicles.

For a membrane whose entire circumference lies in the plane of  the chamber floor, the membrane-to-vesicle transformation is suppressed; there is always an energy barrier. Unlike the single-point attachment cases, the axis of symmetry is aligned with gravity and the membrane does not tilt. We prove the existence of the local barrier for both the SC and AME models. The floor is the zero of gravitational potential with upward positive. The center of mass for a dome spherical cap (apex at $z$ non-zero) is the same for an inverted dome spherical cap (apex at $z=0$). Without loss of generality, we assume the spherical caps are domes. The spherical cap gravitational energy is
\begin{align}
E_{\mathrm{grav}}^{\mathrm{cap}} = \rho_{\mathrm{a}}g \dfrac{A^2}{4\pi R}.
\end{align}
The bend and edge tension energies are as before. The disk has zero gravitational energy. Hence, the cap-disk energy difference is given by
\begin{align}
    \Delta E &= \dfrac{\rho_\mathrm{a} g A^2}{4\pi R} + \dfrac{(2\kappa+\bar{\kappa})A}{R^2} + 2\pi \gamma \left(r_0 - r_\mathrm{d} \right),
\end{align}
where $r_0= \left({{A}/{\pi} - ({A}^2/{(2\pi R}})^2\right)^{1/2}$ is the rim radius and $r_\mathrm{d} = \left({A}/{\pi}\right)^{1/2}$ is the disk radius. 

\begin{proof}(Energy barrier for SC model) By concavity, we have the identity $\sqrt{y-x} - \sqrt{y} \geq - {x}/\sqrt{y}$, for all $0 \leq x \leq y$. With $y = {A}/{\pi}, x = ({A}/\left({2\pi R}\right))^2$, the edge tension term is bounded below, 
\begin{align*}
    2\pi \gamma (r_0 - r_\mathrm{d}) \geq - \dfrac{\gamma A^{3/2}}{2\pi^{1/2}R^2}.
\end{align*}
Hence, $\Delta E$ has the following lower bound, 
\begin{align*}
    \Delta E \geq \dfrac{c}{ R} +  \dfrac{c_{\mathrm{lower}}}{R^2},
\end{align*}
where $c = {\rho_\mathrm{a} g A^2}/{(4\pi)}, 
    c_{\mathrm{lower}} = (2\kappa+\bar{\kappa})A - {\gamma A^{3/2}}/(2\pi^{1/2}).$
The minimum admissible radius is $R_{\mathrm{min}} = \left({A}/{(4\pi)}\right)^{1/2}$. Choosing $R_c = \max\{R_{\text{min}}, 2\abs{c_\mathrm{lower}}/c\}$, we find that for all $R \geq R_c$
\begin{align*}  
 \Delta E \geq \dfrac{c}{R} -\abs{\dfrac{c_{\mathrm{lower}}}{R^2}} \geq \dfrac{c}{2R} > 0.
\end{align*}
\end{proof}
For the AME model, there is also a local energy barrier. 
\begin{proof}(Energy barrier for AME model) Let $\vec{X}(r,\theta) = (r\cos\theta, r\sin\theta, h(r))$ be a parametrization of an axisymmetric surface, where $\theta \in [0,2\pi)$, $r \in [0, R]$, and $R$ is the outer radius. We assume the deflection is small, $\abs{h'} \ll 1$, sufficient regularity at the axis, $h'(0) = 0$ and upward deflection $h > 0$. Let $R_0$ be the radius of the flat disk. Starting with the total free energy (Eq.~\eqref{eq:Helfrich}) and the edge tension energy of the disk,  the energy difference for small deflections is, to leading order,  
\begin{align*}
   \Delta E = 2\pi \int_0^{R_0} \dfrac{\kappa}{2} (\Delta_r h)^2 \, r dr + 2\pi \rho_{\mathrm{a}} g \int_0^{R_0} h r \, dr - \dfrac{\pi \gamma}{R_0} \int_0^{R_0} h'^2 r\, dr,
\end{align*}
where we used the constraint of fixed area to solve for $R$ in terms of $R_0$, $R =  R_0 - \displaystyle \dfrac{1}{2R_0}\int_0^{R_0} h'^2 r \, dr$. The first and second terms are non-negative, since $h, r \geq 0$. Define $w(r) = h' r$ with $w(0) = 0$. Then, $w'(r) = h''r + h' = r \Delta_r h$. We can express the third term using $w$ so that it can be controlled by the first term via the radial component of the 2D Laplacian. Therefore in terms of $w$, we have 
\begin{align*}
   \int_0^{R_0} h'^2 r \,dr = \int_0^{R_0} \dfrac{w^2(r)}{r} \, dr.
\end{align*}
Let $s \in [0, r]$. Using the Cauchy-Schwarz inequality,
\begin{align*}
    w^2(r) = \left(\int_0^r w'(s) \, ds \right)^2
    &\leq \left( \int_0^r s \,ds \right) \left(\int_0^r (\Delta_s h)^2 s \, ds \right) 
    = \dfrac{r^2}{2} \int_0^r (\Delta_s h)^2 s \,ds. 
\end{align*}
Dividing by $r$ and integrating from $0$ to $R_0$, since the integrand is non-negative and integrable, we can swap the order of integration. A short calculation then shows 
\begin{align*}
    \int_0^{R_0} \dfrac{w^2(r)}{r} \,dr &\leq \int_0^{R_0} \dfrac{r}{2} \int_0^r (\Delta_s h)^2 s \, ds \, dr \\
    &\leq \dfrac{1}{2} \int_0^{R_0}  (\Delta_s h)^2 s \left(\int_s^{R_0} r \, dr\right) \, ds \\
    &\leq \dfrac{R_0^2}{4} \int_0^{R_0} (\Delta_r h)^2 r \,dr.
\end{align*}
Therefore, we find that the difference in the energy is 
\begin{align*}
    \Delta E &\geq \left( \pi \kappa - {\pi \gamma}\dfrac{R_0}{4} \right) \int_0^{R_0} (\Delta_r h)^2 r\, dr + 2\pi \rho_{\mathrm{a}}g \int_0^{R_0} hr \, dr. 
\end{align*}
If $\gamma$ is sufficiently large, the first term may be negative; however, it can be controlled by expressing the deformation in terms of a small amplitude $\epsilon$. If we define $h(r) = \epsilon H(r)$, then $\Delta E$ will remain positive for any $0 < \epsilon \leq \epsilon^*$ where
\begin{align*}
    \epsilon^* = \dfrac{{c_2}}{2\abs{c_1}}, \qquad
    c_1 = \left( \pi \kappa - \dfrac{\pi \gamma}{R_0} \dfrac{R_0^2}{4} \right) \int_0^{R_0} (\Delta_r H)^2 r\, dr, \qquad c_2 = 2\pi \rho_{\mathrm{a}}g \int_0^{R_0} Hr \, dr,
\end{align*}
since we have for sufficiently large $\gamma $ (equiv. $c_1 < 0$), 
\begin{align*}
    \Delta E \geq  c_1\epsilon^2 + c_2\epsilon = (c_2 \epsilon - \abs{c_1} \epsilon^2)  \geq \epsilon(c_2 - \abs{c_1}\epsilon^*) = \epsilon \left(c_2 - \abs{c_1} \dfrac{c_2}{2\abs{c_1}}\right) = \epsilon \dfrac{c_2}{2} > 0.
\end{align*}
\end{proof}

\subsection*{Supplementary Information 5: Estimate of membrane density}
To estimate the areal mass density of 200-nm-thick membranes, we used the reported mass density difference between colloidal membranes and the surrounding medium, given as 0.091~g/cm\(^3\) \cite{balchunas_equation_2019,adkins_topology_2025}. Multiplying this value by the membrane thickness yields the areal density:
\[
\rho_{\mathrm{a}} = 0.091~\mathrm{g/cm}^3 \times 200~\mathrm{nm} = 1.82 \times 10^{-5}~\mathrm{kg/m^2}.
\]
Assuming a gravitational field \(g = 9.8~\mathrm{m/s^2}\), the effective weight per area becomes:
\[
\rho_\mathrm{a} g = 1.78 \times 10^{-4}~\mathrm{J/m^3}.
\]
To ensure consistency with other physical quantities, the effective weight per area \(\rho_\mathrm{a} g\) can be expressed in thermal energy units \(0.043~k_\mathrm{B} T/\mu\mathrm{m}^3\).

\subsection*{Supplementary Information 6: Properties of 315-nm-thick membranes}

To estimate the properties of 315-nm-thick colloidal membranes, we extrapolated from measurements of 200-nm and 385-nm membranes (Fig.~S3). Flickering spectroscopy analysis yielded bending moduli of \( 1200 \pm 60~k_\mathrm{B} T \) and \( 11000 \pm 1000~k_\mathrm{B} T \) for 200-nm and 385-nm membranes \cite{adkins_topology_2025}. Assuming that the bending modulus \( \kappa \) scales with the cube of membrane thickness \cite{szleifer_curvature_1988,rawicz_effect_2000}, the estimate that for 315-nm-thick membranes $\kappa$ is \( 5800~k_\mathrm{B} T \) (Fig.~S3a).

For edge tension, the measured values were \( 200 \pm 30~k_\mathrm{B} T/\mu\mathrm{m} \) and \( 700 \pm 40~k_\mathrm{B} T/\mu\mathrm{m} \) for 200-nm and 385-nm membranes \cite{adkins_topology_2025}. The edge tension should be linearly proportional to the membrane thickness. From here, a linear fit estimates the value of \(500~k_\mathrm{B} T/\mu\mathrm{m} \) for 315-nm-thick membranes (Fig.~S3b).
\begin{figure}[h!]
\renewcommand{\thefigure}{S\arabic{figure}}%
\centering
\includegraphics[width=0.75\textwidth]{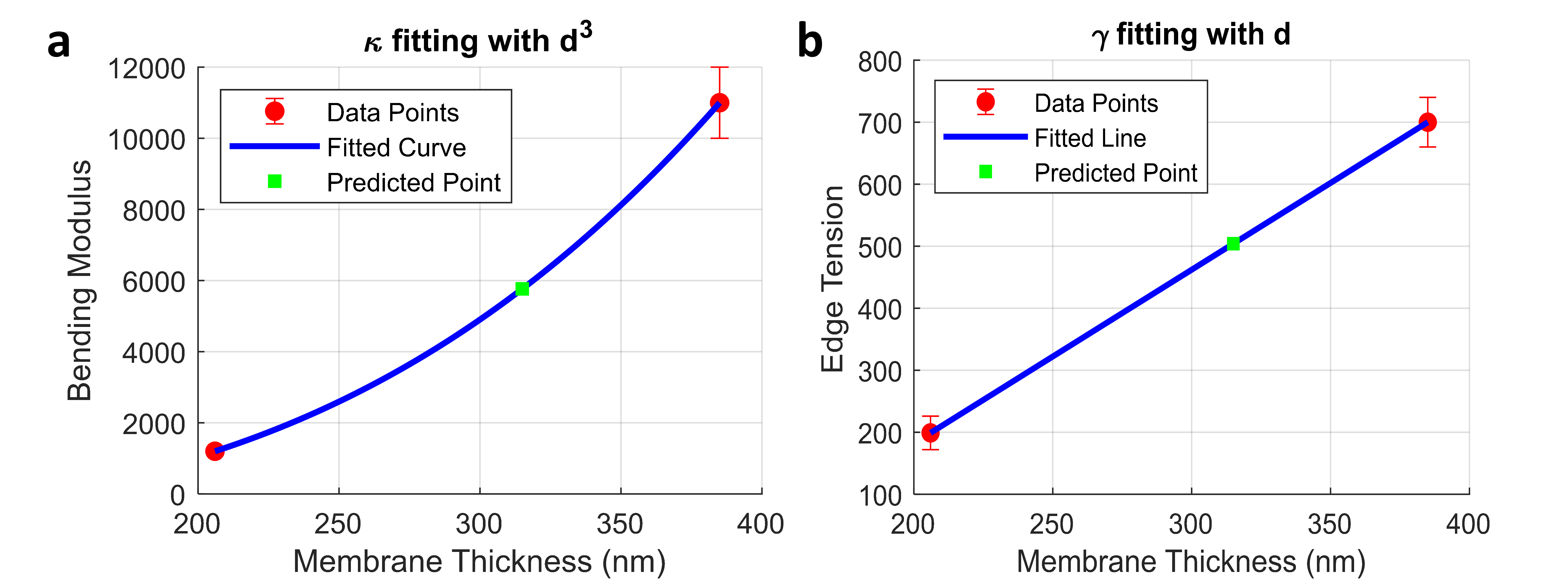}
\caption{\textbf{Estimating properties of 315-nm-thick membranes.} \\
\textbf{a,} Bending modulus as a function of membrane thickness. The red circles represent experimental data from 200-nm and 385-nm membranes. The blue curve is a cubic fit, and the green square indicates the predicted value for 315-nm-thick membranes.  
\textbf{b,} edge tension versus membrane thickness. The red circles show measured values for 200-nm and 385-nm membranes. A linear fit (blue line) is used to estimate the edge tension for 315-nm-thick membranes (green square).} \label{figS3}
\end{figure}
\subsection*{Supplementary Information 7: Vesicle deformation under centrifugation}
Contrary to the lateral centrifugal force (Fig.~4a), applying a centrifugal force along the sedimentation direction caused the sedimented vesicles to deform into ellipsoids and lose their monodispersity. To ensure that the force was applied precisely perpendicular to the chamber floor, we placed the sample chamber flat on the swing bucket rotor and centrifuged it at 60 g for 1 min. Thirty minutes after centrifugation, the vesicles remained at the chamber floor, and their morphology was monitored by epi-fluorescence imaging. The 3D cross-sectional slices revealed that the sedimented vesicles lost their monodispersity and were deformed into ellipsoidal shapes. The crosshair in Fig.~S4 highlights one representative vesicle with a pronounced ellipsoidal geometry, while other vesicles with either ellipsoidal or spherical shapes can also be observed in the $xy$-plane.
\begin{figure}[h!]
\centering
\includegraphics[width=0.4\textwidth]{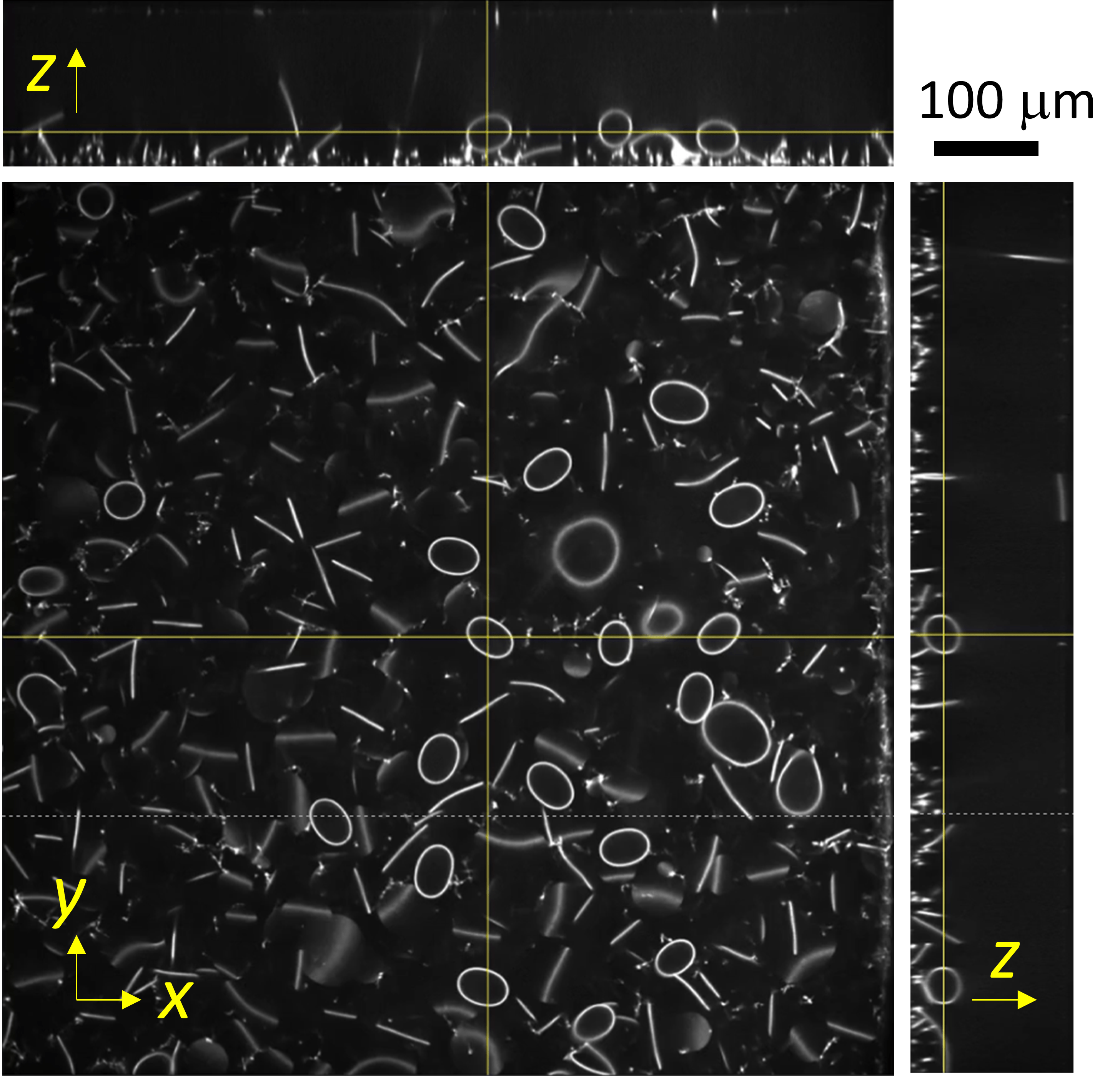}
\caption{Cross-section slices of 3D epi-fluorescence image after applying centrifugal force along the sedimentation direction (60 g, 1 min).}
\end{figure}

\newpage
\putbib
\end{bibunit}


\begin{thebibliography}{0}%
\makeatletter
\providecommand \@ifxundefined [1]{%
 \@ifx{#1\undefined}
}%
\providecommand \@ifnum [1]{%
 \ifnum #1\expandafter \@firstoftwo
 \else \expandafter \@secondoftwo
 \fi
}%
\providecommand \@ifx [1]{%
 \ifx #1\expandafter \@firstoftwo
 \else \expandafter \@secondoftwo
 \fi
}%
\providecommand \natexlab [1]{#1}%
\providecommand \enquote  [1]{``#1''}%
\providecommand \bibnamefont  [1]{#1}%
\providecommand \bibfnamefont [1]{#1}%
\providecommand \citenamefont [1]{#1}%
\providecommand \href@noop [0]{\@secondoftwo}%
\providecommand \href [0]{\begingroup \@sanitize@url \@href}%
\providecommand \@href[1]{\@@startlink{#1}\@@href}%
\providecommand \@@href[1]{\endgroup#1\@@endlink}%
\providecommand \@sanitize@url [0]{\catcode `\\12\catcode `\$12\catcode `\&12\catcode `\#12\catcode `\^12\catcode `\_12\catcode `\%12\relax}%
\providecommand \@@startlink[1]{}%
\providecommand \@@endlink[0]{}%
\providecommand \url  [0]{\begingroup\@sanitize@url \@url }%
\providecommand \@url [1]{\endgroup\@href {#1}{\urlprefix }}%
\providecommand \urlprefix  [0]{URL }%
\providecommand \Eprint [0]{\href }%
\providecommand \doibase [0]{https://doi.org/}%
\providecommand \selectlanguage [0]{\@gobble}%
\providecommand \bibinfo  [0]{\@secondoftwo}%
\providecommand \bibfield  [0]{\@secondoftwo}%
\providecommand \translation [1]{[#1]}%
\providecommand \BibitemOpen [0]{}%
\providecommand \bibitemStop [0]{}%
\providecommand \bibitemNoStop [0]{.\EOS\space}%
\providecommand \EOS [0]{\spacefactor3000\relax}%
\providecommand \BibitemShut  [1]{\csname bibitem#1\endcsname}%
\let\auto@bib@innerbib\@empty
\end{thebibliography}%


\begin{thebibliography}{63}%
\makeatletter
\providecommand \@ifxundefined [1]{%
 \@ifx{#1\undefined}
}%
\providecommand \@ifnum [1]{%
 \ifnum #1\expandafter \@firstoftwo
 \else \expandafter \@secondoftwo
 \fi
}%
\providecommand \@ifx [1]{%
 \ifx #1\expandafter \@firstoftwo
 \else \expandafter \@secondoftwo
 \fi
}%
\providecommand \natexlab [1]{#1}%
\providecommand \enquote  [1]{``#1''}%
\providecommand \bibnamefont  [1]{#1}%
\providecommand \bibfnamefont [1]{#1}%
\providecommand \citenamefont [1]{#1}%
\providecommand \href@noop [0]{\@secondoftwo}%
\providecommand \href [0]{\begingroup \@sanitize@url \@href}%
\providecommand \@href[1]{\@@startlink{#1}\@@href}%
\providecommand \@@href[1]{\endgroup#1\@@endlink}%
\providecommand \@sanitize@url [0]{\catcode `\\12\catcode `\$12\catcode `\&12\catcode `\#12\catcode `\^12\catcode `\_12\catcode `\%12\relax}%
\providecommand \@@startlink[1]{}%
\providecommand \@@endlink[0]{}%
\providecommand \url  [0]{\begingroup\@sanitize@url \@url }%
\providecommand \@url [1]{\endgroup\@href {#1}{\urlprefix }}%
\providecommand \urlprefix  [0]{URL }%
\providecommand \Eprint [0]{\href }%
\providecommand \doibase [0]{https://doi.org/}%
\providecommand \selectlanguage [0]{\@gobble}%
\providecommand \bibinfo  [0]{\@secondoftwo}%
\providecommand \bibfield  [0]{\@secondoftwo}%
\providecommand \translation [1]{[#1]}%
\providecommand \BibitemOpen [0]{}%
\providecommand \bibitemStop [0]{}%
\providecommand \bibitemNoStop [0]{.\EOS\space}%
\providecommand \EOS [0]{\spacefactor3000\relax}%
\providecommand \BibitemShut  [1]{\csname bibitem#1\endcsname}%
\let\auto@bib@innerbib\@empty
\bibitem [{\citenamefont {Helfrich}(1974)}]{helfrich_size_1974}%
  \BibitemOpen
  \bibfield  {author} {\bibinfo {author} {\bibfnamefont {W.}~\bibnamefont {Helfrich}},\ }\href {https://doi.org/10.1016/0375-9601(74)90899-8} {\bibfield  {journal} {\bibinfo  {journal} {Phys. Lett. A}\ }\textbf {\bibinfo {volume} {50}},\ \bibinfo {pages} {115} (\bibinfo {year} {1974})}\BibitemShut {NoStop}%
\bibitem [{\citenamefont {Fromherz}(1983)}]{fromherz_lipid-vesicle_1983}%
  \BibitemOpen
  \bibfield  {author} {\bibinfo {author} {\bibfnamefont {P.}~\bibnamefont {Fromherz}},\ }\href {https://doi.org/10.1016/0009-2614(83)87083-3} {\bibfield  {journal} {\bibinfo  {journal} {Chem. Phys. Lett.}\ }\textbf {\bibinfo {volume} {94}},\ \bibinfo {pages} {259} (\bibinfo {year} {1983})}\BibitemShut {NoStop}%
\bibitem [{\citenamefont {Boal}\ and\ \citenamefont {Rao}(1992)}]{boal_topology_1992}%
  \BibitemOpen
  \bibfield  {author} {\bibinfo {author} {\bibfnamefont {D.~H.}\ \bibnamefont {Boal}}\ and\ \bibinfo {author} {\bibfnamefont {M.}~\bibnamefont {Rao}},\ }\href {https://doi.org/10.1103/PhysRevA.46.3037} {\bibfield  {journal} {\bibinfo  {journal} {Phys. Rev. A}\ }\textbf {\bibinfo {volume} {46}},\ \bibinfo {pages} {3037} (\bibinfo {year} {1992})}\BibitemShut {NoStop}%
\bibitem [{\citenamefont {Lipowsky}(1992)}]{lipowsky_budding_1992}%
  \BibitemOpen
  \bibfield  {author} {\bibinfo {author} {\bibfnamefont {R.}~\bibnamefont {Lipowsky}},\ }\href {https://doi.org/10.1051/jp2:1992238} {\bibfield  {journal} {\bibinfo  {journal} {J. Phys. II}\ }\textbf {\bibinfo {volume} {2}},\ \bibinfo {pages} {1825} (\bibinfo {year} {1992})}\BibitemShut {NoStop}%
\bibitem [{\citenamefont {Baumgart}\ \emph {et~al.}(2003)\citenamefont {Baumgart}, \citenamefont {Hess},\ and\ \citenamefont {Webb}}]{baumgart_imaging_2003}%
  \BibitemOpen
  \bibfield  {author} {\bibinfo {author} {\bibfnamefont {T.}~\bibnamefont {Baumgart}}, \bibinfo {author} {\bibfnamefont {S.~T.}\ \bibnamefont {Hess}},\ and\ \bibinfo {author} {\bibfnamefont {W.~W.}\ \bibnamefont {Webb}},\ }\href {https://doi.org/10.1038/nature02013} {\bibfield  {journal} {\bibinfo  {journal} {Nature}\ }\textbf {\bibinfo {volume} {425}},\ \bibinfo {pages} {821} (\bibinfo {year} {2003})}\BibitemShut {NoStop}%
\bibitem [{\citenamefont {Hu}\ \emph {et~al.}(2012)\citenamefont {Hu}, \citenamefont {Briguglio},\ and\ \citenamefont {Deserno}}]{hu_determining_2012}%
  \BibitemOpen
  \bibfield  {author} {\bibinfo {author} {\bibfnamefont {M.}~\bibnamefont {Hu}}, \bibinfo {author} {\bibfnamefont {J.~J.}\ \bibnamefont {Briguglio}},\ and\ \bibinfo {author} {\bibfnamefont {M.}~\bibnamefont {Deserno}},\ }\href {https://doi.org/10.1016/j.bpj.2012.02.013} {\bibfield  {journal} {\bibinfo  {journal} {Biophys. J.}\ }\textbf {\bibinfo {volume} {102}},\ \bibinfo {pages} {1403} (\bibinfo {year} {2012})}\BibitemShut {NoStop}%
\bibitem [{\citenamefont {Noguchi}(2019)}]{noguchi_cup--vesicle_2019}%
  \BibitemOpen
  \bibfield  {author} {\bibinfo {author} {\bibfnamefont {H.}~\bibnamefont {Noguchi}},\ }\href {https://doi.org/10.1063/1.5113646} {\bibfield  {journal} {\bibinfo  {journal} {J. Chem. Phys.}\ }\textbf {\bibinfo {volume} {151}},\ \bibinfo {pages} {094903} (\bibinfo {year} {2019})}\BibitemShut {NoStop}%
\bibitem [{\citenamefont {Ding}\ \emph {et~al.}(2020)\citenamefont {Ding}, \citenamefont {Pelcovits},\ and\ \citenamefont {Powers}}]{ding_shapes_2020}%
  \BibitemOpen
  \bibfield  {author} {\bibinfo {author} {\bibfnamefont {L.}~\bibnamefont {Ding}}, \bibinfo {author} {\bibfnamefont {R.~A.}\ \bibnamefont {Pelcovits}},\ and\ \bibinfo {author} {\bibfnamefont {T.~R.}\ \bibnamefont {Powers}},\ }\href {https://doi.org/10.1103/PhysRevE.102.032608} {\bibfield  {journal} {\bibinfo  {journal} {Phys. Rev. E}\ }\textbf {\bibinfo {volume} {102}},\ \bibinfo {pages} {032608} (\bibinfo {year} {2020})}\BibitemShut {NoStop}%
\bibitem [{\citenamefont {McMahon}\ and\ \citenamefont {Gallop}(2005)}]{mcmahon_membrane_2005}%
  \BibitemOpen
  \bibfield  {author} {\bibinfo {author} {\bibfnamefont {H.~T.}\ \bibnamefont {McMahon}}\ and\ \bibinfo {author} {\bibfnamefont {J.~L.}\ \bibnamefont {Gallop}},\ }\href {https://doi.org/10.1038/nature04396} {\bibfield  {journal} {\bibinfo  {journal} {Nature}\ }\textbf {\bibinfo {volume} {438}},\ \bibinfo {pages} {590} (\bibinfo {year} {2005})}\BibitemShut {NoStop}%
\bibitem [{\citenamefont {Y{\'a}\~{n}ez M{\'o}}\ \emph {et~al.}(2015)\citenamefont {Y{\'a}\~{n}ez M{\'o}}, \citenamefont {Siljander}, \citenamefont {Andreu}, \citenamefont {Bedina~Zavec}, \citenamefont {Borr{\`a}s}, \citenamefont {Buzas}, \citenamefont {Buzas}, \citenamefont {Casal}, \citenamefont {Cappello}, \citenamefont {Carvalho}, \citenamefont {Col{\'a}s}, \citenamefont {Cordeiro-da Silva}, \citenamefont {Fais}, \citenamefont {Falcon-Perez}, \citenamefont {Ghobrial}, \citenamefont {Giebel}, \citenamefont {Gimona}, \citenamefont {Graner}, \citenamefont {Gursel}, \citenamefont {Gursel}, \citenamefont {Heegaard}, \citenamefont {Hendrix}, \citenamefont {Kierulf}, \citenamefont {Kokubun}, \citenamefont {Kosanovic}, \citenamefont {Kralj-Iglic}, \citenamefont {Kr{\"a}mer-Albers}, \citenamefont {Laitinen}, \citenamefont {L{\"a}sser}, \citenamefont {Lener}, \citenamefont {Ligeti}, \citenamefont {Lin\={e}}, \citenamefont {Lipps}, \citenamefont {Llorente}, \citenamefont {Lötvall}, \citenamefont {Man{\v
  c}ek-Keber}, \citenamefont {Marcilla}, \citenamefont {Mittelbrunn}, \citenamefont {Nazarenko}, \citenamefont {Nolte-'t Hoen}, \citenamefont {Nyman}, \citenamefont {O'Driscoll}, \citenamefont {Olivan}, \citenamefont {Oliveira}, \citenamefont {Pállinger}, \citenamefont {del Portillo}, \citenamefont {Revent{\'o}s}, \citenamefont {Rigau}, \citenamefont {Rohde}, \citenamefont {Sammar}, \citenamefont {S{\'a}nchez-Madrid}, \citenamefont {Santar{\'e}m}, \citenamefont {Schallmoser}, \citenamefont {Stampe~Ostenfeld}, \citenamefont {Stoorvogel}, \citenamefont {Stukelj}, \citenamefont {Van~der Grein}, \citenamefont {Helena~Vasconcelos}, \citenamefont {Wauben},\ and\ \citenamefont {De~Wever}}]{yanez-mo_biological_2015}%
  \BibitemOpen
  \bibfield  {author} {\bibinfo {author} {\bibfnamefont {M.}~\bibnamefont {Y{\'a}\~{n}ez M{\'o}}}, \bibinfo {author} {\bibfnamefont {P.~R.-M.}\ \bibnamefont {Siljander}}, \bibinfo {author} {\bibfnamefont {Z.}~\bibnamefont {Andreu}}, \bibinfo {author} {\bibfnamefont {A.}~\bibnamefont {Bedina~Zavec}}, \bibinfo {author} {\bibfnamefont {F.~E.}\ \bibnamefont {Borr{\`a}s}}, \bibinfo {author} {\bibfnamefont {E.~I.}\ \bibnamefont {Buzas}}, \bibinfo {author} {\bibfnamefont {K.}~\bibnamefont {Buzas}}, \bibinfo {author} {\bibfnamefont {E.}~\bibnamefont {Casal}}, \bibinfo {author} {\bibfnamefont {F.}~\bibnamefont {Cappello}}, \bibinfo {author} {\bibfnamefont {J.}~\bibnamefont {Carvalho}}, \bibinfo {author} {\bibfnamefont {E.}~\bibnamefont {Col{\'a}s}}, \bibinfo {author} {\bibfnamefont {A.}~\bibnamefont {Cordeiro-da Silva}}, \bibinfo {author} {\bibfnamefont {S.}~\bibnamefont {Fais}}, \bibinfo {author} {\bibfnamefont {J.~M.}\ \bibnamefont {Falcon-Perez}}, \bibinfo {author} {\bibfnamefont {I.~M.}\ \bibnamefont {Ghobrial}},
  \bibinfo {author} {\bibfnamefont {B.}~\bibnamefont {Giebel}}, \bibinfo {author} {\bibfnamefont {M.}~\bibnamefont {Gimona}}, \bibinfo {author} {\bibfnamefont {M.}~\bibnamefont {Graner}}, \bibinfo {author} {\bibfnamefont {I.}~\bibnamefont {Gursel}}, \bibinfo {author} {\bibfnamefont {M.}~\bibnamefont {Gursel}}, \bibinfo {author} {\bibfnamefont {N.~H.~H.}\ \bibnamefont {Heegaard}}, \bibinfo {author} {\bibfnamefont {A.}~\bibnamefont {Hendrix}}, \bibinfo {author} {\bibfnamefont {P.}~\bibnamefont {Kierulf}}, \bibinfo {author} {\bibfnamefont {K.}~\bibnamefont {Kokubun}}, \bibinfo {author} {\bibfnamefont {M.}~\bibnamefont {Kosanovic}}, \bibinfo {author} {\bibfnamefont {V.}~\bibnamefont {Kralj-Iglic}}, \bibinfo {author} {\bibfnamefont {E.-M.}\ \bibnamefont {Kr{\"a}mer-Albers}}, \bibinfo {author} {\bibfnamefont {S.}~\bibnamefont {Laitinen}}, \bibinfo {author} {\bibfnamefont {C.}~\bibnamefont {L{\"a}sser}}, \bibinfo {author} {\bibfnamefont {T.}~\bibnamefont {Lener}}, \bibinfo {author} {\bibfnamefont {E.}~\bibnamefont
  {Ligeti}}, \bibinfo {author} {\bibfnamefont {A.}~\bibnamefont {Lin\={e}}}, \bibinfo {author} {\bibfnamefont {G.}~\bibnamefont {Lipps}}, \bibinfo {author} {\bibfnamefont {A.}~\bibnamefont {Llorente}}, \bibinfo {author} {\bibfnamefont {J.}~\bibnamefont {Lötvall}}, \bibinfo {author} {\bibfnamefont {M.}~\bibnamefont {Man{\v c}ek-Keber}}, \bibinfo {author} {\bibfnamefont {A.}~\bibnamefont {Marcilla}}, \bibinfo {author} {\bibfnamefont {M.}~\bibnamefont {Mittelbrunn}}, \bibinfo {author} {\bibfnamefont {I.}~\bibnamefont {Nazarenko}}, \bibinfo {author} {\bibfnamefont {E.~N.}\ \bibnamefont {Nolte-'t Hoen}}, \bibinfo {author} {\bibfnamefont {T.~A.}\ \bibnamefont {Nyman}}, \bibinfo {author} {\bibfnamefont {L.}~\bibnamefont {O'Driscoll}}, \bibinfo {author} {\bibfnamefont {M.}~\bibnamefont {Olivan}}, \bibinfo {author} {\bibfnamefont {C.}~\bibnamefont {Oliveira}}, \bibinfo {author} {\bibfnamefont {{\'E}.}~\bibnamefont {Pállinger}}, \bibinfo {author} {\bibfnamefont {H.~A.}\ \bibnamefont {del Portillo}}, \bibinfo {author}
  {\bibfnamefont {J.}~\bibnamefont {Revent{\'o}s}}, \bibinfo {author} {\bibfnamefont {M.}~\bibnamefont {Rigau}}, \bibinfo {author} {\bibfnamefont {E.}~\bibnamefont {Rohde}}, \bibinfo {author} {\bibfnamefont {M.}~\bibnamefont {Sammar}}, \bibinfo {author} {\bibfnamefont {F.}~\bibnamefont {S{\'a}nchez-Madrid}}, \bibinfo {author} {\bibfnamefont {N.}~\bibnamefont {Santar{\'e}m}}, \bibinfo {author} {\bibfnamefont {K.}~\bibnamefont {Schallmoser}}, \bibinfo {author} {\bibfnamefont {M.}~\bibnamefont {Stampe~Ostenfeld}}, \bibinfo {author} {\bibfnamefont {W.}~\bibnamefont {Stoorvogel}}, \bibinfo {author} {\bibfnamefont {R.}~\bibnamefont {Stukelj}}, \bibinfo {author} {\bibfnamefont {S.~G.}\ \bibnamefont {Van~der Grein}}, \bibinfo {author} {\bibfnamefont {M.}~\bibnamefont {Helena~Vasconcelos}}, \bibinfo {author} {\bibfnamefont {M.~H.~M.}\ \bibnamefont {Wauben}},\ and\ \bibinfo {author} {\bibfnamefont {O.}~\bibnamefont {De~Wever}},\ }\href {https://doi.org/https://doi.org/10.3402/jev.v4.27066} {\bibfield  {journal}
  {\bibinfo  {journal} {J. Extracell. Vesicles}\ }\textbf {\bibinfo {volume} {4}},\ \bibinfo {pages} {27066} (\bibinfo {year} {2015})}\BibitemShut {NoStop}%
\bibitem [{\citenamefont {Shin}\ \emph {et~al.}(2018)\citenamefont {Shin}, \citenamefont {Ge}, \citenamefont {Arpino}, \citenamefont {Villarreal}, \citenamefont {Hamid}, \citenamefont {Liu}, \citenamefont {Zhao}, \citenamefont {Wen}, \citenamefont {Chiang},\ and\ \citenamefont {Wu}}]{shin_visualization_2018}%
  \BibitemOpen
  \bibfield  {author} {\bibinfo {author} {\bibfnamefont {W.}~\bibnamefont {Shin}}, \bibinfo {author} {\bibfnamefont {L.}~\bibnamefont {Ge}}, \bibinfo {author} {\bibfnamefont {G.}~\bibnamefont {Arpino}}, \bibinfo {author} {\bibfnamefont {S.~A.}\ \bibnamefont {Villarreal}}, \bibinfo {author} {\bibfnamefont {E.}~\bibnamefont {Hamid}}, \bibinfo {author} {\bibfnamefont {H.}~\bibnamefont {Liu}}, \bibinfo {author} {\bibfnamefont {W.-D.}\ \bibnamefont {Zhao}}, \bibinfo {author} {\bibfnamefont {P.~J.}\ \bibnamefont {Wen}}, \bibinfo {author} {\bibfnamefont {H.-C.}\ \bibnamefont {Chiang}},\ and\ \bibinfo {author} {\bibfnamefont {L.-G.}\ \bibnamefont {Wu}},\ }\href {https://doi.org/10.1016/j.cell.2018.02.062} {\bibfield  {journal} {\bibinfo  {journal} {Cell}\ }\textbf {\bibinfo {volume} {173}},\ \bibinfo {pages} {934} (\bibinfo {year} {2018})}\BibitemShut {NoStop}%
\bibitem [{\citenamefont {Wiklander}\ \emph {et~al.}(2019)\citenamefont {Wiklander}, \citenamefont {Brennan}, \citenamefont {L\"otvall}, \citenamefont {Breakefield},\ and\ \citenamefont {EL~Andaloussi}}]{wiklander_advances_2019}%
  \BibitemOpen
  \bibfield  {author} {\bibinfo {author} {\bibfnamefont {O.~P.~B.}\ \bibnamefont {Wiklander}}, \bibinfo {author} {\bibfnamefont {M.~{\'A}.}\ \bibnamefont {Brennan}}, \bibinfo {author} {\bibfnamefont {J.}~\bibnamefont {L\"otvall}}, \bibinfo {author} {\bibfnamefont {X.~O.}\ \bibnamefont {Breakefield}},\ and\ \bibinfo {author} {\bibfnamefont {S.}~\bibnamefont {EL~Andaloussi}},\ }\href {https://doi.org/10.1126/scitranslmed.aav8521} {\bibfield  {journal} {\bibinfo  {journal} {Sci. Transl. Med.}\ }\textbf {\bibinfo {volume} {11}},\ \bibinfo {pages} {eaav8521} (\bibinfo {year} {2019})}\BibitemShut {NoStop}%
\bibitem [{\citenamefont {Tavakoli}\ \emph {et~al.}(2025)\citenamefont {Tavakoli}, \citenamefont {Hu}, \citenamefont {Ebrahim},\ and\ \citenamefont {Kachar}}]{tavakoli_hemifusomes_2025}%
  \BibitemOpen
  \bibfield  {author} {\bibinfo {author} {\bibfnamefont {A.}~\bibnamefont {Tavakoli}}, \bibinfo {author} {\bibfnamefont {S.}~\bibnamefont {Hu}}, \bibinfo {author} {\bibfnamefont {S.}~\bibnamefont {Ebrahim}},\ and\ \bibinfo {author} {\bibfnamefont {B.}~\bibnamefont {Kachar}},\ }\href {https://doi.org/10.1038/s41467-025-59887-9} {\bibfield  {journal} {\bibinfo  {journal} {Nat. Commun.}\ }\textbf {\bibinfo {volume} {16}},\ \bibinfo {pages} {4609} (\bibinfo {year} {2025})}\BibitemShut {NoStop}%
\bibitem [{\citenamefont {Discher}\ \emph {et~al.}(1999)\citenamefont {Discher}, \citenamefont {Won}, \citenamefont {Ege}, \citenamefont {Lee}, \citenamefont {Bates}, \citenamefont {Discher},\ and\ \citenamefont {Hammer}}]{discher_polymersomes_1999}%
  \BibitemOpen
  \bibfield  {author} {\bibinfo {author} {\bibfnamefont {B.~M.}\ \bibnamefont {Discher}}, \bibinfo {author} {\bibfnamefont {Y.-Y.}\ \bibnamefont {Won}}, \bibinfo {author} {\bibfnamefont {D.~S.}\ \bibnamefont {Ege}}, \bibinfo {author} {\bibfnamefont {J.~C.-M.}\ \bibnamefont {Lee}}, \bibinfo {author} {\bibfnamefont {F.~S.}\ \bibnamefont {Bates}}, \bibinfo {author} {\bibfnamefont {D.~E.}\ \bibnamefont {Discher}},\ and\ \bibinfo {author} {\bibfnamefont {D.~A.}\ \bibnamefont {Hammer}},\ }\href {https://doi.org/10.1126/science.284.5417.1143} {\bibfield  {journal} {\bibinfo  {journal} {Science}\ }\textbf {\bibinfo {volume} {284}},\ \bibinfo {pages} {1143} (\bibinfo {year} {1999})}\BibitemShut {NoStop}%
\bibitem [{\citenamefont {Dinsmore}\ \emph {et~al.}(2002)\citenamefont {Dinsmore}, \citenamefont {Hsu}, \citenamefont {Nikolaides}, \citenamefont {Marquez}, \citenamefont {Bausch},\ and\ \citenamefont {Weitz}}]{dinsmore_colloidosomes_2002}%
  \BibitemOpen
  \bibfield  {author} {\bibinfo {author} {\bibfnamefont {A.~D.}\ \bibnamefont {Dinsmore}}, \bibinfo {author} {\bibfnamefont {M.~F.}\ \bibnamefont {Hsu}}, \bibinfo {author} {\bibfnamefont {M.~G.}\ \bibnamefont {Nikolaides}}, \bibinfo {author} {\bibfnamefont {M.}~\bibnamefont {Marquez}}, \bibinfo {author} {\bibfnamefont {A.~R.}\ \bibnamefont {Bausch}},\ and\ \bibinfo {author} {\bibfnamefont {D.~A.}\ \bibnamefont {Weitz}},\ }\href {https://doi.org/10.1126/science.1074868} {\bibfield  {journal} {\bibinfo  {journal} {Science}\ }\textbf {\bibinfo {volume} {298}},\ \bibinfo {pages} {1006} (\bibinfo {year} {2002})}\BibitemShut {NoStop}%
\bibitem [{\citenamefont {Riske}\ and\ \citenamefont {Dimova}(2005)}]{riske_electro-deformation_2005}%
  \BibitemOpen
  \bibfield  {author} {\bibinfo {author} {\bibfnamefont {K.~A.}\ \bibnamefont {Riske}}\ and\ \bibinfo {author} {\bibfnamefont {R.}~\bibnamefont {Dimova}},\ }\href {https://doi.org/10.1529/biophysj.104.050310} {\bibfield  {journal} {\bibinfo  {journal} {Biophys. J.}\ }\textbf {\bibinfo {volume} {88}},\ \bibinfo {pages} {1143} (\bibinfo {year} {2005})}\BibitemShut {NoStop}%
\bibitem [{\citenamefont {Park}\ \emph {et~al.}(2004)\citenamefont {Park}, \citenamefont {Lim}, \citenamefont {Chung},\ and\ \citenamefont {Mirkin}}]{park_self-assembly_2004}%
  \BibitemOpen
  \bibfield  {author} {\bibinfo {author} {\bibfnamefont {S.}~\bibnamefont {Park}}, \bibinfo {author} {\bibfnamefont {J.-H.}\ \bibnamefont {Lim}}, \bibinfo {author} {\bibfnamefont {S.-W.}\ \bibnamefont {Chung}},\ and\ \bibinfo {author} {\bibfnamefont {C.~A.}\ \bibnamefont {Mirkin}},\ }\href {https://doi.org/10.1126/science.1093276} {\bibfield  {journal} {\bibinfo  {journal} {Science}\ }\textbf {\bibinfo {volume} {303}},\ \bibinfo {pages} {348} (\bibinfo {year} {2004})}\BibitemShut {NoStop}%
\bibitem [{\citenamefont {Baranov}\ \emph {et~al.}(2010)\citenamefont {Baranov}, \citenamefont {Fiore}, \citenamefont {van Huis}, \citenamefont {Giannini}, \citenamefont {Falqui}, \citenamefont {Lafont}, \citenamefont {Zandbergen}, \citenamefont {Zanella}, \citenamefont {Cingolani},\ and\ \citenamefont {Manna}}]{baranov_assembly_2010}%
  \BibitemOpen
  \bibfield  {author} {\bibinfo {author} {\bibfnamefont {D.}~\bibnamefont {Baranov}}, \bibinfo {author} {\bibfnamefont {A.}~\bibnamefont {Fiore}}, \bibinfo {author} {\bibfnamefont {M.}~\bibnamefont {van Huis}}, \bibinfo {author} {\bibfnamefont {C.}~\bibnamefont {Giannini}}, \bibinfo {author} {\bibfnamefont {A.}~\bibnamefont {Falqui}}, \bibinfo {author} {\bibfnamefont {U.}~\bibnamefont {Lafont}}, \bibinfo {author} {\bibfnamefont {H.}~\bibnamefont {Zandbergen}}, \bibinfo {author} {\bibfnamefont {M.}~\bibnamefont {Zanella}}, \bibinfo {author} {\bibfnamefont {R.}~\bibnamefont {Cingolani}},\ and\ \bibinfo {author} {\bibfnamefont {L.}~\bibnamefont {Manna}},\ }\href {https://doi.org/10.1021/nl903946n} {\bibfield  {journal} {\bibinfo  {journal} {Nano Letters}\ }\textbf {\bibinfo {volume} {10}},\ \bibinfo {pages} {743} (\bibinfo {year} {2010})}\BibitemShut {NoStop}%
\bibitem [{\citenamefont {Young}\ \emph {et~al.}(2013)\citenamefont {Young}, \citenamefont {Personick}, \citenamefont {Engel}, \citenamefont {Damasceno}, \citenamefont {Barnaby}, \citenamefont {Bleher}, \citenamefont {Li}, \citenamefont {Glotzer}, \citenamefont {Lee},\ and\ \citenamefont {Mirkin}}]{young_directional_2013}%
  \BibitemOpen
  \bibfield  {author} {\bibinfo {author} {\bibfnamefont {K.~L.}\ \bibnamefont {Young}}, \bibinfo {author} {\bibfnamefont {M.~L.}\ \bibnamefont {Personick}}, \bibinfo {author} {\bibfnamefont {M.}~\bibnamefont {Engel}}, \bibinfo {author} {\bibfnamefont {P.~F.}\ \bibnamefont {Damasceno}}, \bibinfo {author} {\bibfnamefont {S.~N.}\ \bibnamefont {Barnaby}}, \bibinfo {author} {\bibfnamefont {R.}~\bibnamefont {Bleher}}, \bibinfo {author} {\bibfnamefont {T.}~\bibnamefont {Li}}, \bibinfo {author} {\bibfnamefont {S.~C.}\ \bibnamefont {Glotzer}}, \bibinfo {author} {\bibfnamefont {B.}~\bibnamefont {Lee}},\ and\ \bibinfo {author} {\bibfnamefont {C.~A.}\ \bibnamefont {Mirkin}},\ }\href {https://doi.org/10.1002/anie.201306009} {\bibfield  {journal} {\bibinfo  {journal} {Angewandte Chemie International Edition}\ }\textbf {\bibinfo {volume} {52}},\ \bibinfo {pages} {13980} (\bibinfo {year} {2013})}\BibitemShut {NoStop}%
\bibitem [{\citenamefont {Zanella}\ \emph {et~al.}(2011)\citenamefont {Zanella}, \citenamefont {Bertoni}, \citenamefont {Franchini}, \citenamefont {Brescia}, \citenamefont {Baranov},\ and\ \citenamefont {Manna}}]{zanella_assembly_2010}%
  \BibitemOpen
  \bibfield  {author} {\bibinfo {author} {\bibfnamefont {M.}~\bibnamefont {Zanella}}, \bibinfo {author} {\bibfnamefont {G.}~\bibnamefont {Bertoni}}, \bibinfo {author} {\bibfnamefont {I.~R.}\ \bibnamefont {Franchini}}, \bibinfo {author} {\bibfnamefont {R.}~\bibnamefont {Brescia}}, \bibinfo {author} {\bibfnamefont {D.}~\bibnamefont {Baranov}},\ and\ \bibinfo {author} {\bibfnamefont {L.}~\bibnamefont {Manna}},\ }\href {https://doi.org/10.1039/C0CC02477E} {\bibfield  {journal} {\bibinfo  {journal} {Chem. Commun.}\ }\textbf {\bibinfo {volume} {47}},\ \bibinfo {pages} {203} (\bibinfo {year} {2011})}\BibitemShut {NoStop}%
\bibitem [{\citenamefont {Smith}\ \emph {et~al.}(2024)\citenamefont {Smith}, \citenamefont {Carnevale}, \citenamefont {Das},\ and\ \citenamefont {Chen}}]{smith_electron_2024}%
  \BibitemOpen
  \bibfield  {author} {\bibinfo {author} {\bibfnamefont {J.~W.}\ \bibnamefont {Smith}}, \bibinfo {author} {\bibfnamefont {L.~N.}\ \bibnamefont {Carnevale}}, \bibinfo {author} {\bibfnamefont {A.}~\bibnamefont {Das}},\ and\ \bibinfo {author} {\bibfnamefont {Q.}~\bibnamefont {Chen}},\ }\href {https://doi.org/10.1126/sciadv.adk0217} {\bibfield  {journal} {\bibinfo  {journal} {Sci. Adv.}\ }\textbf {\bibinfo {volume} {10}},\ \bibinfo {pages} {eadk0217} (\bibinfo {year} {2024})}\BibitemShut {NoStop}%
\bibitem [{\citenamefont {van Blaaderen}\ and\ \citenamefont {Wiltzius}(1995)}]{van_blaaderen_real-space_1995}%
  \BibitemOpen
  \bibfield  {author} {\bibinfo {author} {\bibfnamefont {A.}~\bibnamefont {van Blaaderen}}\ and\ \bibinfo {author} {\bibfnamefont {P.}~\bibnamefont {Wiltzius}},\ }\href {https://doi.org/10.1126/science.270.5239.1177} {\bibfield  {journal} {\bibinfo  {journal} {Science}\ }\textbf {\bibinfo {volume} {270}},\ \bibinfo {pages} {1177} (\bibinfo {year} {1995})}\BibitemShut {NoStop}%
\bibitem [{\citenamefont {Weeks}\ \emph {et~al.}(2000)\citenamefont {Weeks}, \citenamefont {Crocker}, \citenamefont {Levitt}, \citenamefont {Schofield},\ and\ \citenamefont {Weitz}}]{weeks_three-dimensional_2000}%
  \BibitemOpen
  \bibfield  {author} {\bibinfo {author} {\bibfnamefont {E.~R.}\ \bibnamefont {Weeks}}, \bibinfo {author} {\bibfnamefont {J.~C.}\ \bibnamefont {Crocker}}, \bibinfo {author} {\bibfnamefont {A.~C.}\ \bibnamefont {Levitt}}, \bibinfo {author} {\bibfnamefont {A.}~\bibnamefont {Schofield}},\ and\ \bibinfo {author} {\bibfnamefont {D.~A.}\ \bibnamefont {Weitz}},\ }\href {https://doi.org/10.1126/science.287.5453.627} {\bibfield  {journal} {\bibinfo  {journal} {Science}\ }\textbf {\bibinfo {volume} {287}},\ \bibinfo {pages} {627} (\bibinfo {year} {2000})}\BibitemShut {NoStop}%
\bibitem [{\citenamefont {Gasser}\ \emph {et~al.}(2001)\citenamefont {Gasser}, \citenamefont {Weeks}, \citenamefont {Schofield}, \citenamefont {Pusey},\ and\ \citenamefont {Weitz}}]{gasser_real-space_2001}%
  \BibitemOpen
  \bibfield  {author} {\bibinfo {author} {\bibfnamefont {U.}~\bibnamefont {Gasser}}, \bibinfo {author} {\bibfnamefont {E.~R.}\ \bibnamefont {Weeks}}, \bibinfo {author} {\bibfnamefont {A.}~\bibnamefont {Schofield}}, \bibinfo {author} {\bibfnamefont {P.~N.}\ \bibnamefont {Pusey}},\ and\ \bibinfo {author} {\bibfnamefont {D.~A.}\ \bibnamefont {Weitz}},\ }\href {https://doi.org/10.1126/science.1058457} {\bibfield  {journal} {\bibinfo  {journal} {Science}\ }\textbf {\bibinfo {volume} {292}},\ \bibinfo {pages} {258} (\bibinfo {year} {2001})}\BibitemShut {NoStop}%
\bibitem [{\citenamefont {Schall}\ \emph {et~al.}(2004)\citenamefont {Schall}, \citenamefont {Cohen}, \citenamefont {Weitz},\ and\ \citenamefont {Spaepen}}]{schall_visualization_2004}%
  \BibitemOpen
  \bibfield  {author} {\bibinfo {author} {\bibfnamefont {P.}~\bibnamefont {Schall}}, \bibinfo {author} {\bibfnamefont {I.}~\bibnamefont {Cohen}}, \bibinfo {author} {\bibfnamefont {D.~A.}\ \bibnamefont {Weitz}},\ and\ \bibinfo {author} {\bibfnamefont {F.}~\bibnamefont {Spaepen}},\ }\href {https://doi.org/10.1126/science.1102186} {\bibfield  {journal} {\bibinfo  {journal} {Science}\ }\textbf {\bibinfo {volume} {305}},\ \bibinfo {pages} {1944} (\bibinfo {year} {2004})}\BibitemShut {NoStop}%
\bibitem [{\citenamefont {Barry}\ and\ \citenamefont {Dogic}(2010)}]{barry_entropy_2010}%
  \BibitemOpen
  \bibfield  {author} {\bibinfo {author} {\bibfnamefont {E.}~\bibnamefont {Barry}}\ and\ \bibinfo {author} {\bibfnamefont {Z.}~\bibnamefont {Dogic}},\ }\href {https://doi.org/10.1073/pnas.1000406107} {\bibfield  {journal} {\bibinfo  {journal} {Proc. Natl. Acad. Sci. U.S.A.}\ }\textbf {\bibinfo {volume} {107}},\ \bibinfo {pages} {10348} (\bibinfo {year} {2010})}\BibitemShut {NoStop}%
\bibitem [{\citenamefont {Sharma}\ \emph {et~al.}(2014)\citenamefont {Sharma}, \citenamefont {Ward}, \citenamefont {Gibaud}, \citenamefont {Hagan},\ and\ \citenamefont {Dogic}}]{sharma_hierarchical_2014}%
  \BibitemOpen
  \bibfield  {author} {\bibinfo {author} {\bibfnamefont {P.}~\bibnamefont {Sharma}}, \bibinfo {author} {\bibfnamefont {A.}~\bibnamefont {Ward}}, \bibinfo {author} {\bibfnamefont {T.}~\bibnamefont {Gibaud}}, \bibinfo {author} {\bibfnamefont {M.~F.}\ \bibnamefont {Hagan}},\ and\ \bibinfo {author} {\bibfnamefont {Z.}~\bibnamefont {Dogic}},\ }\href {https://doi.org/10.1038/nature13694} {\bibfield  {journal} {\bibinfo  {journal} {Nature}\ }\textbf {\bibinfo {volume} {513}},\ \bibinfo {pages} {77} (\bibinfo {year} {2014})}\BibitemShut {NoStop}%
\bibitem [{\citenamefont {Khanra}\ \emph {et~al.}(2022)\citenamefont {Khanra}, \citenamefont {Jia}, \citenamefont {Mitchell}, \citenamefont {Balchunas}, \citenamefont {Pelcovits}, \citenamefont {Powers}, \citenamefont {Dogic},\ and\ \citenamefont {Sharma}}]{khanra_controlling_2022}%
  \BibitemOpen
  \bibfield  {author} {\bibinfo {author} {\bibfnamefont {A.}~\bibnamefont {Khanra}}, \bibinfo {author} {\bibfnamefont {L.~L.}\ \bibnamefont {Jia}}, \bibinfo {author} {\bibfnamefont {N.~P.}\ \bibnamefont {Mitchell}}, \bibinfo {author} {\bibfnamefont {A.}~\bibnamefont {Balchunas}}, \bibinfo {author} {\bibfnamefont {R.~A.}\ \bibnamefont {Pelcovits}}, \bibinfo {author} {\bibfnamefont {T.~R.}\ \bibnamefont {Powers}}, \bibinfo {author} {\bibfnamefont {Z.}~\bibnamefont {Dogic}},\ and\ \bibinfo {author} {\bibfnamefont {P.}~\bibnamefont {Sharma}},\ }\href {https://doi.org/10.1073/pnas.2204453119} {\bibfield  {journal} {\bibinfo  {journal} {Proc. Natl. Acad. Sci. U.S.A.}\ }\textbf {\bibinfo {volume} {119}},\ \bibinfo {pages} {e2204453119} (\bibinfo {year} {2022})}\BibitemShut {NoStop}%
\bibitem [{\citenamefont {Adkins}\ \emph {et~al.}(2025)\citenamefont {Adkins}, \citenamefont {Robaszewski}, \citenamefont {Shin}, \citenamefont {Brauns}, \citenamefont {Jia}, \citenamefont {Khanra}, \citenamefont {Sharma}, \citenamefont {Pelcovits}, \citenamefont {Powers},\ and\ \citenamefont {Dogic}}]{adkins_topology_2025}%
  \BibitemOpen
  \bibfield  {author} {\bibinfo {author} {\bibfnamefont {R.}~\bibnamefont {Adkins}}, \bibinfo {author} {\bibfnamefont {J.}~\bibnamefont {Robaszewski}}, \bibinfo {author} {\bibfnamefont {S.}~\bibnamefont {Shin}}, \bibinfo {author} {\bibfnamefont {F.}~\bibnamefont {Brauns}}, \bibinfo {author} {\bibfnamefont {L.}~\bibnamefont {Jia}}, \bibinfo {author} {\bibfnamefont {A.}~\bibnamefont {Khanra}}, \bibinfo {author} {\bibfnamefont {P.}~\bibnamefont {Sharma}}, \bibinfo {author} {\bibfnamefont {R.~A.}\ \bibnamefont {Pelcovits}}, \bibinfo {author} {\bibfnamefont {T.~R.}\ \bibnamefont {Powers}},\ and\ \bibinfo {author} {\bibfnamefont {Z.}~\bibnamefont {Dogic}},\ }\href {https://doi.org/10.1073/pnas.2427024122} {\bibfield  {journal} {\bibinfo  {journal} {Proc. Natl. Acad. Sci. U.S.A.}\ }\textbf {\bibinfo {volume} {122}},\ \bibinfo {pages} {e2427024122} (\bibinfo {year} {2025})}\BibitemShut {NoStop}%
\bibitem [{\citenamefont {Helfrich}(1973)}]{helfrich_elastic_1973}%
  \BibitemOpen
  \bibfield  {author} {\bibinfo {author} {\bibfnamefont {W.}~\bibnamefont {Helfrich}},\ }\href {https://doi.org/10.1515/znc-1973-11-1209} {\bibfield  {journal} {\bibinfo  {journal} {Z. Naturforsch. C}\ }\textbf {\bibinfo {volume} {28}},\ \bibinfo {pages} {693} (\bibinfo {year} {1973})}\BibitemShut {NoStop}%
\bibitem [{\citenamefont {Rawicz}\ \emph {et~al.}(2000)\citenamefont {Rawicz}, \citenamefont {Olbrich}, \citenamefont {McIntosh}, \citenamefont {Needham},\ and\ \citenamefont {Evans}}]{rawicz_effect_2000}%
  \BibitemOpen
  \bibfield  {author} {\bibinfo {author} {\bibfnamefont {W.}~\bibnamefont {Rawicz}}, \bibinfo {author} {\bibfnamefont {K.~C.}\ \bibnamefont {Olbrich}}, \bibinfo {author} {\bibfnamefont {T.}~\bibnamefont {McIntosh}}, \bibinfo {author} {\bibfnamefont {D.}~\bibnamefont {Needham}},\ and\ \bibinfo {author} {\bibfnamefont {E.}~\bibnamefont {Evans}},\ }\href {https://doi.org/10.1016/S0006-3495(00)76295-3} {\bibfield  {journal} {\bibinfo  {journal} {Biophys. J.}\ }\textbf {\bibinfo {volume} {79}},\ \bibinfo {pages} {328} (\bibinfo {year} {2000})}\BibitemShut {NoStop}%
\bibitem [{\citenamefont {Dimova}\ \emph {et~al.}(2002)\citenamefont {Dimova}, \citenamefont {Seifert}, \citenamefont {Pouligny}, \citenamefont {Förster},\ and\ \citenamefont {Döbereiner}}]{dimova_hyperviscous_2002}%
  \BibitemOpen
  \bibfield  {author} {\bibinfo {author} {\bibfnamefont {R.}~\bibnamefont {Dimova}}, \bibinfo {author} {\bibfnamefont {U.}~\bibnamefont {Seifert}}, \bibinfo {author} {\bibfnamefont {B.}~\bibnamefont {Pouligny}}, \bibinfo {author} {\bibfnamefont {S.}~\bibnamefont {Förster}},\ and\ \bibinfo {author} {\bibfnamefont {H.-G.}\ \bibnamefont {Döbereiner}},\ }\href {https://doi.org/10.1140/epje/i200101032} {\bibfield  {journal} {\bibinfo  {journal} {Eur. Phys. J. E}\ }\textbf {\bibinfo {volume} {7}},\ \bibinfo {pages} {241} (\bibinfo {year} {2002})}\BibitemShut {NoStop}%
\bibitem [{\citenamefont {Dimova}(2014)}]{dimova_recent_2014}%
  \BibitemOpen
  \bibfield  {author} {\bibinfo {author} {\bibfnamefont {R.}~\bibnamefont {Dimova}},\ }\href {https://doi.org/10.1016/j.cis.2014.03.003} {\bibfield  {journal} {\bibinfo  {journal} {Adv. Colloid Interface Sci.}\ }\bibinfo {series} {Special issue in honour of {Wolfgang} {Helfrich}},\ \textbf {\bibinfo {volume} {208}},\ \bibinfo {pages} {225} (\bibinfo {year} {2014})}\BibitemShut {NoStop}%
\bibitem [{\citenamefont {Tian}\ \emph {et~al.}(2007)\citenamefont {Tian}, \citenamefont {Johnson}, \citenamefont {Wang},\ and\ \citenamefont {Baumgart}}]{tian_line_2007}%
  \BibitemOpen
  \bibfield  {author} {\bibinfo {author} {\bibfnamefont {A.}~\bibnamefont {Tian}}, \bibinfo {author} {\bibfnamefont {C.}~\bibnamefont {Johnson}}, \bibinfo {author} {\bibfnamefont {W.}~\bibnamefont {Wang}},\ and\ \bibinfo {author} {\bibfnamefont {T.}~\bibnamefont {Baumgart}},\ }\href {https://doi.org/10.1103/PhysRevLett.98.208102} {\bibfield  {journal} {\bibinfo  {journal} {Phys. Rev. Lett.}\ }\textbf {\bibinfo {volume} {98}},\ \bibinfo {pages} {208102} (\bibinfo {year} {2007})}\BibitemShut {NoStop}%
\bibitem [{\citenamefont {Elson}\ \emph {et~al.}(2010)\citenamefont {Elson}, \citenamefont {Fried}, \citenamefont {Dolbow},\ and\ \citenamefont {Genin}}]{elson_phase_2010}%
  \BibitemOpen
  \bibfield  {author} {\bibinfo {author} {\bibfnamefont {E.~L.}\ \bibnamefont {Elson}}, \bibinfo {author} {\bibfnamefont {E.}~\bibnamefont {Fried}}, \bibinfo {author} {\bibfnamefont {J.~E.}\ \bibnamefont {Dolbow}},\ and\ \bibinfo {author} {\bibfnamefont {G.~M.}\ \bibnamefont {Genin}},\ }\href {https://doi.org/10.1146/annurev.biophys.093008.131238} {\bibfield  {journal} {\bibinfo  {journal} {Annu. Rev. Biophys.}\ }\textbf {\bibinfo {volume} {39}},\ \bibinfo {pages} {207} (\bibinfo {year} {2010})}\BibitemShut {NoStop}%
\bibitem [{\citenamefont {Asakura}\ and\ \citenamefont {Oosawa}(1954)}]{asakura_interaction_1954}%
  \BibitemOpen
  \bibfield  {author} {\bibinfo {author} {\bibfnamefont {S.}~\bibnamefont {Asakura}}\ and\ \bibinfo {author} {\bibfnamefont {F.}~\bibnamefont {Oosawa}},\ }\href {https://doi.org/10.1063/1.1740347} {\bibfield  {journal} {\bibinfo  {journal} {J. Chem. Phys.}\ }\textbf {\bibinfo {volume} {22}},\ \bibinfo {pages} {1255} (\bibinfo {year} {1954})}\BibitemShut {NoStop}%
\bibitem [{\citenamefont {Balchunas}\ \emph {et~al.}(2019)\citenamefont {Balchunas}, \citenamefont {Cabanas}, \citenamefont {Zakhary}, \citenamefont {Gibaud}, \citenamefont {Fraden}, \citenamefont {Sharma}, \citenamefont {Hagan},\ and\ \citenamefont {Dogic}}]{balchunas_equation_2019}%
  \BibitemOpen
  \bibfield  {author} {\bibinfo {author} {\bibfnamefont {A.~J.}\ \bibnamefont {Balchunas}}, \bibinfo {author} {\bibfnamefont {R.~A.}\ \bibnamefont {Cabanas}}, \bibinfo {author} {\bibfnamefont {M.~J.}\ \bibnamefont {Zakhary}}, \bibinfo {author} {\bibfnamefont {T.}~\bibnamefont {Gibaud}}, \bibinfo {author} {\bibfnamefont {S.}~\bibnamefont {Fraden}}, \bibinfo {author} {\bibfnamefont {P.}~\bibnamefont {Sharma}}, \bibinfo {author} {\bibfnamefont {M.~F.}\ \bibnamefont {Hagan}},\ and\ \bibinfo {author} {\bibfnamefont {Z.}~\bibnamefont {Dogic}},\ }\href {https://doi.org/10.1039/C9SM01054H} {\bibfield  {journal} {\bibinfo  {journal} {Soft Matter}\ }\textbf {\bibinfo {volume} {15}},\ \bibinfo {pages} {6791} (\bibinfo {year} {2019})}\BibitemShut {NoStop}%
\bibitem [{\citenamefont {Gibaud}\ \emph {et~al.}(2012)\citenamefont {Gibaud}, \citenamefont {Barry}, \citenamefont {Zakhary}, \citenamefont {Henglin}, \citenamefont {Ward}, \citenamefont {Yang}, \citenamefont {Berciu}, \citenamefont {Oldenbourg}, \citenamefont {Hagan}, \citenamefont {Nicastro}, \citenamefont {Meyer},\ and\ \citenamefont {Dogic}}]{gibaud_reconfigurable_2012}%
  \BibitemOpen
  \bibfield  {author} {\bibinfo {author} {\bibfnamefont {T.}~\bibnamefont {Gibaud}}, \bibinfo {author} {\bibfnamefont {E.}~\bibnamefont {Barry}}, \bibinfo {author} {\bibfnamefont {M.~J.}\ \bibnamefont {Zakhary}}, \bibinfo {author} {\bibfnamefont {M.}~\bibnamefont {Henglin}}, \bibinfo {author} {\bibfnamefont {A.}~\bibnamefont {Ward}}, \bibinfo {author} {\bibfnamefont {Y.}~\bibnamefont {Yang}}, \bibinfo {author} {\bibfnamefont {C.}~\bibnamefont {Berciu}}, \bibinfo {author} {\bibfnamefont {R.}~\bibnamefont {Oldenbourg}}, \bibinfo {author} {\bibfnamefont {M.~F.}\ \bibnamefont {Hagan}}, \bibinfo {author} {\bibfnamefont {D.}~\bibnamefont {Nicastro}}, \bibinfo {author} {\bibfnamefont {R.~B.}\ \bibnamefont {Meyer}},\ and\ \bibinfo {author} {\bibfnamefont {Z.}~\bibnamefont {Dogic}},\ }\href {https://doi.org/10.1038/nature10769} {\bibfield  {journal} {\bibinfo  {journal} {Nature}\ }\textbf {\bibinfo {volume} {481}},\ \bibinfo {pages} {348} (\bibinfo {year} {2012})}\BibitemShut {NoStop}%
\bibitem [{\citenamefont {Gibaud}\ \emph {et~al.}(2017)\citenamefont {Gibaud}, \citenamefont {Kaplan}, \citenamefont {Sharma}, \citenamefont {Zakhary}, \citenamefont {Ward}, \citenamefont {Oldenbourg}, \citenamefont {Meyer}, \citenamefont {Kamien}, \citenamefont {Powers},\ and\ \citenamefont {Dogic}}]{gibaud_achiral_2017}%
  \BibitemOpen
  \bibfield  {author} {\bibinfo {author} {\bibfnamefont {T.}~\bibnamefont {Gibaud}}, \bibinfo {author} {\bibfnamefont {C.~N.}\ \bibnamefont {Kaplan}}, \bibinfo {author} {\bibfnamefont {P.}~\bibnamefont {Sharma}}, \bibinfo {author} {\bibfnamefont {M.~J.}\ \bibnamefont {Zakhary}}, \bibinfo {author} {\bibfnamefont {A.}~\bibnamefont {Ward}}, \bibinfo {author} {\bibfnamefont {R.}~\bibnamefont {Oldenbourg}}, \bibinfo {author} {\bibfnamefont {R.~B.}\ \bibnamefont {Meyer}}, \bibinfo {author} {\bibfnamefont {R.~D.}\ \bibnamefont {Kamien}}, \bibinfo {author} {\bibfnamefont {T.~R.}\ \bibnamefont {Powers}},\ and\ \bibinfo {author} {\bibfnamefont {Z.}~\bibnamefont {Dogic}},\ }\href {https://doi.org/10.1073/pnas.1617043114} {\bibfield  {journal} {\bibinfo  {journal} {Proc. Natl. Acad. Sci. U.S.A.}\ }\textbf {\bibinfo {volume} {114}},\ \bibinfo {pages} {E3376} (\bibinfo {year} {2017})}\BibitemShut {NoStop}%
\bibitem [{\citenamefont {Szleifer}(1988)}]{szleifer_curvature_1988}%
  \BibitemOpen
  \bibfield  {author} {\bibinfo {author} {\bibfnamefont {I.}~\bibnamefont {Szleifer}},\ }\href {https://doi.org/10.1103/PhysRevLett.60.1966} {\bibfield  {journal} {\bibinfo  {journal} {Phys. Rev. Lett.}\ }\textbf {\bibinfo {volume} {60}},\ \bibinfo {pages} {1966} (\bibinfo {year} {1988})}\BibitemShut {NoStop}%
\bibitem [{\citenamefont {Nafisi}\ \emph {et~al.}(2018)\citenamefont {Nafisi}, \citenamefont {Aksel},\ and\ \citenamefont {Douglas}}]{nafisi_construction_2018}%
  \BibitemOpen
  \bibfield  {author} {\bibinfo {author} {\bibfnamefont {P.~M.}\ \bibnamefont {Nafisi}}, \bibinfo {author} {\bibfnamefont {T.}~\bibnamefont {Aksel}},\ and\ \bibinfo {author} {\bibfnamefont {S.~M.}\ \bibnamefont {Douglas}},\ }\href {https://doi.org/10.1093/synbio/ysy015} {\bibfield  {journal} {\bibinfo  {journal} {Synth. Biol.}\ }\textbf {\bibinfo {volume} {3}},\ \bibinfo {pages} {ysy015} (\bibinfo {year} {2018})}\BibitemShut {NoStop}%
\bibitem [{\citenamefont {Dogic}\ and\ \citenamefont {Fraden}(2001)}]{dogic_development_2001}%
  \BibitemOpen
  \bibfield  {author} {\bibinfo {author} {\bibfnamefont {Z.}~\bibnamefont {Dogic}}\ and\ \bibinfo {author} {\bibfnamefont {S.}~\bibnamefont {Fraden}},\ }\href {https://doi.org/10.1098/rsta.2000.0814} {\bibfield  {journal} {\bibinfo  {journal} {Phil. Trans. R. Soc. A.}\ }\textbf {\bibinfo {volume} {359}},\ \bibinfo {pages} {997} (\bibinfo {year} {2001})}\BibitemShut {NoStop}%
\bibitem [{\citenamefont {Schneider}\ \emph {et~al.}(1984)\citenamefont {Schneider}, \citenamefont {Jenkins},\ and\ \citenamefont {Webb}}]{schneider_thermal_1984}%
  \BibitemOpen
  \bibfield  {author} {\bibinfo {author} {\bibfnamefont {M.~B.}\ \bibnamefont {Schneider}}, \bibinfo {author} {\bibfnamefont {J.~T.}\ \bibnamefont {Jenkins}},\ and\ \bibinfo {author} {\bibfnamefont {W.~W.}\ \bibnamefont {Webb}},\ }\href {https://doi.org/10.1051/jphys:019840045090145700} {\bibfield  {journal} {\bibinfo  {journal} {J. de Physique}\ }\textbf {\bibinfo {volume} {45}},\ \bibinfo {pages} {1457} (\bibinfo {year} {1984})}\BibitemShut {NoStop}%
\bibitem [{\citenamefont {Gracià}\ \emph {et~al.}(2010)\citenamefont {Gracià}, \citenamefont {Bezlyepkina}, \citenamefont {Knorr}, \citenamefont {Lipowsky},\ and\ \citenamefont {Dimova}}]{gracia_effect_2010}%
  \BibitemOpen
  \bibfield  {author} {\bibinfo {author} {\bibfnamefont {R.~S.}\ \bibnamefont {Gracià}}, \bibinfo {author} {\bibfnamefont {N.}~\bibnamefont {Bezlyepkina}}, \bibinfo {author} {\bibfnamefont {R.~L.}\ \bibnamefont {Knorr}}, \bibinfo {author} {\bibfnamefont {R.}~\bibnamefont {Lipowsky}},\ and\ \bibinfo {author} {\bibfnamefont {R.}~\bibnamefont {Dimova}},\ }\href {https://doi.org/10.1039/B920629A} {\bibfield  {journal} {\bibinfo  {journal} {Soft Matter}\ }\textbf {\bibinfo {volume} {6}},\ \bibinfo {pages} {1472} (\bibinfo {year} {2010})}\BibitemShut {NoStop}%
\bibitem [{\citenamefont {Jia}\ \emph {et~al.}(2017)\citenamefont {Jia}, \citenamefont {Zakhary}, \citenamefont {Dogic}, \citenamefont {Pelcovits},\ and\ \citenamefont {Powers}}]{jia_chiral_2017}%
  \BibitemOpen
  \bibfield  {author} {\bibinfo {author} {\bibfnamefont {L.~L.}\ \bibnamefont {Jia}}, \bibinfo {author} {\bibfnamefont {M.~J.}\ \bibnamefont {Zakhary}}, \bibinfo {author} {\bibfnamefont {Z.}~\bibnamefont {Dogic}}, \bibinfo {author} {\bibfnamefont {R.~A.}\ \bibnamefont {Pelcovits}},\ and\ \bibinfo {author} {\bibfnamefont {T.~R.}\ \bibnamefont {Powers}},\ }\href {https://doi.org/10.1103/PhysRevE.95.060701} {\bibfield  {journal} {\bibinfo  {journal} {Phys. Rev. E}\ }\textbf {\bibinfo {volume} {95}},\ \bibinfo {pages} {060701} (\bibinfo {year} {2017})}\BibitemShut {NoStop}%
\bibitem [{\citenamefont {J\"ulicher}\ and\ \citenamefont {Seifert}(1994)}]{Julicher1994}%
  \BibitemOpen
  \bibfield  {author} {\bibinfo {author} {\bibfnamefont {F.}~\bibnamefont {J\"ulicher}}\ and\ \bibinfo {author} {\bibfnamefont {U.}~\bibnamefont {Seifert}},\ }\href {https://doi.org/10.1103/PhysRevE.49.4728} {\bibfield  {journal} {\bibinfo  {journal} {Phys. Rev. E}\ }\textbf {\bibinfo {volume} {49}},\ \bibinfo {pages} {4728} (\bibinfo {year} {1994})}\BibitemShut {NoStop}%
\bibitem [{\citenamefont {Portet}\ and\ \citenamefont {Dimova}(2010)}]{portet_new_2010}%
  \BibitemOpen
  \bibfield  {author} {\bibinfo {author} {\bibfnamefont {T.}~\bibnamefont {Portet}}\ and\ \bibinfo {author} {\bibfnamefont {R.}~\bibnamefont {Dimova}},\ }\href {https://doi.org/10.1016/j.bpj.2010.09.032} {\bibfield  {journal} {\bibinfo  {journal} {Biophys. J.}\ }\textbf {\bibinfo {volume} {99}},\ \bibinfo {pages} {3264} (\bibinfo {year} {2010})}\BibitemShut {NoStop}%
\bibitem [{\citenamefont {Yang}\ \emph {et~al.}(2011)\citenamefont {Yang}, \citenamefont {Barry}, \citenamefont {Dogic},\ and\ \citenamefont {Hagan}}]{yang_self-assembly_2011}%
  \BibitemOpen
  \bibfield  {author} {\bibinfo {author} {\bibfnamefont {Y.}~\bibnamefont {Yang}}, \bibinfo {author} {\bibfnamefont {E.}~\bibnamefont {Barry}}, \bibinfo {author} {\bibfnamefont {Z.}~\bibnamefont {Dogic}},\ and\ \bibinfo {author} {\bibfnamefont {M.~F.}\ \bibnamefont {Hagan}},\ }\href {https://doi.org/10.1039/C1SM06201H} {\bibfield  {journal} {\bibinfo  {journal} {Soft Matter}\ }\textbf {\bibinfo {volume} {8}},\ \bibinfo {pages} {707} (\bibinfo {year} {2011})}\BibitemShut {NoStop}%
\bibitem [{\citenamefont {Zakhary}\ \emph {et~al.}(2014)\citenamefont {Zakhary}, \citenamefont {Gibaud}, \citenamefont {Nadir~Kaplan}, \citenamefont {Barry}, \citenamefont {Oldenbourg}, \citenamefont {Meyer},\ and\ \citenamefont {Dogic}}]{zakhary_imprintable_2014}%
  \BibitemOpen
  \bibfield  {author} {\bibinfo {author} {\bibfnamefont {M.~J.}\ \bibnamefont {Zakhary}}, \bibinfo {author} {\bibfnamefont {T.}~\bibnamefont {Gibaud}}, \bibinfo {author} {\bibfnamefont {C.}~\bibnamefont {Nadir~Kaplan}}, \bibinfo {author} {\bibfnamefont {E.}~\bibnamefont {Barry}}, \bibinfo {author} {\bibfnamefont {R.}~\bibnamefont {Oldenbourg}}, \bibinfo {author} {\bibfnamefont {R.~B.}\ \bibnamefont {Meyer}},\ and\ \bibinfo {author} {\bibfnamefont {Z.}~\bibnamefont {Dogic}},\ }\href {https://doi.org/10.1038/ncomms4063} {\bibfield  {journal} {\bibinfo  {journal} {Nat. Commun.}\ }\textbf {\bibinfo {volume} {5}},\ \bibinfo {pages} {3063} (\bibinfo {year} {2014})}\BibitemShut {NoStop}%
\bibitem [{\citenamefont {Hallett}\ \emph {et~al.}(1991)\citenamefont {Hallett}, \citenamefont {Nickel}, \citenamefont {Samuels},\ and\ \citenamefont {Krygsman}}]{hallett_determination_1991}%
  \BibitemOpen
  \bibfield  {author} {\bibinfo {author} {\bibfnamefont {F.~R.}\ \bibnamefont {Hallett}}, \bibinfo {author} {\bibfnamefont {B.}~\bibnamefont {Nickel}}, \bibinfo {author} {\bibfnamefont {C.}~\bibnamefont {Samuels}},\ and\ \bibinfo {author} {\bibfnamefont {P.~H.}\ \bibnamefont {Krygsman}},\ }\href {https://doi.org/10.1002/jemt.1060170409} {\bibfield  {journal} {\bibinfo  {journal} {J. Electron Microsc. Tech}\ }\textbf {\bibinfo {volume} {17}},\ \bibinfo {pages} {459} (\bibinfo {year} {1991})}\BibitemShut {NoStop}%
\bibitem [{\citenamefont {Lee}\ \emph {et~al.}(2001)\citenamefont {Lee}, \citenamefont {Bermudez}, \citenamefont {Discher}, \citenamefont {Sheehan}, \citenamefont {Won}, \citenamefont {Bates},\ and\ \citenamefont {Discher}}]{lee_preparation_2001}%
  \BibitemOpen
  \bibfield  {author} {\bibinfo {author} {\bibfnamefont {J.~C.-M.}\ \bibnamefont {Lee}}, \bibinfo {author} {\bibfnamefont {H.}~\bibnamefont {Bermudez}}, \bibinfo {author} {\bibfnamefont {B.~M.}\ \bibnamefont {Discher}}, \bibinfo {author} {\bibfnamefont {M.~A.}\ \bibnamefont {Sheehan}}, \bibinfo {author} {\bibfnamefont {Y.-Y.}\ \bibnamefont {Won}}, \bibinfo {author} {\bibfnamefont {F.~S.}\ \bibnamefont {Bates}},\ and\ \bibinfo {author} {\bibfnamefont {D.~E.}\ \bibnamefont {Discher}},\ }\href {https://doi.org/10.1002/bit.1045} {\bibfield  {journal} {\bibinfo  {journal} {Biotechnol Bioeng}\ }\textbf {\bibinfo {volume} {73}},\ \bibinfo {pages} {135} (\bibinfo {year} {2001})}\BibitemShut {NoStop}%
\bibitem [{\citenamefont {Maulucci}\ \emph {et~al.}(2005)\citenamefont {Maulucci}, \citenamefont {De~Spirito}, \citenamefont {Arcovito}, \citenamefont {Boffi}, \citenamefont {Castellano},\ and\ \citenamefont {Briganti}}]{maulucci_particle_2005}%
  \BibitemOpen
  \bibfield  {author} {\bibinfo {author} {\bibfnamefont {G.}~\bibnamefont {Maulucci}}, \bibinfo {author} {\bibfnamefont {M.}~\bibnamefont {De~Spirito}}, \bibinfo {author} {\bibfnamefont {G.}~\bibnamefont {Arcovito}}, \bibinfo {author} {\bibfnamefont {F.}~\bibnamefont {Boffi}}, \bibinfo {author} {\bibfnamefont {A.~C.}\ \bibnamefont {Castellano}},\ and\ \bibinfo {author} {\bibfnamefont {G.}~\bibnamefont {Briganti}},\ }\href {https://doi.org/10.1529/biophysj.104.048876} {\bibfield  {journal} {\bibinfo  {journal} {Biophys. J.}\ }\textbf {\bibinfo {volume} {88}},\ \bibinfo {pages} {3545} (\bibinfo {year} {2005})}\BibitemShut {NoStop}%
\bibitem [{\citenamefont {Yewle}\ \emph {et~al.}(2016)\citenamefont {Yewle}, \citenamefont {Wattamwar}, \citenamefont {Tao}, \citenamefont {Ostertag},\ and\ \citenamefont {Ghoroghchian}}]{yewle_progressive_2016}%
  \BibitemOpen
  \bibfield  {author} {\bibinfo {author} {\bibfnamefont {J.}~\bibnamefont {Yewle}}, \bibinfo {author} {\bibfnamefont {P.}~\bibnamefont {Wattamwar}}, \bibinfo {author} {\bibfnamefont {Z.}~\bibnamefont {Tao}}, \bibinfo {author} {\bibfnamefont {E.~M.}\ \bibnamefont {Ostertag}},\ and\ \bibinfo {author} {\bibfnamefont {P.~P.}\ \bibnamefont {Ghoroghchian}},\ }\href {https://doi.org/10.1007/s11095-015-1809-9} {\bibfield  {journal} {\bibinfo  {journal} {Pharm. Res}\ }\textbf {\bibinfo {volume} {33}},\ \bibinfo {pages} {573} (\bibinfo {year} {2016})}\BibitemShut {NoStop}%
\bibitem [{\citenamefont {Huang}\ \emph {et~al.}(2017)\citenamefont {Huang}, \citenamefont {Quinn}, \citenamefont {Sadovsky}, \citenamefont {Suresh},\ and\ \citenamefont {Hsia}}]{huang_formation_2017}%
  \BibitemOpen
  \bibfield  {author} {\bibinfo {author} {\bibfnamefont {C.}~\bibnamefont {Huang}}, \bibinfo {author} {\bibfnamefont {D.}~\bibnamefont {Quinn}}, \bibinfo {author} {\bibfnamefont {Y.}~\bibnamefont {Sadovsky}}, \bibinfo {author} {\bibfnamefont {S.}~\bibnamefont {Suresh}},\ and\ \bibinfo {author} {\bibfnamefont {K.~J.}\ \bibnamefont {Hsia}},\ }\href {https://doi.org/10.1073/pnas.1702065114} {\bibfield  {journal} {\bibinfo  {journal} {Proc. Natl. Acad. Sci. U.S.A.}\ }\textbf {\bibinfo {volume} {114}},\ \bibinfo {pages} {2910} (\bibinfo {year} {2017})}\BibitemShut {NoStop}%
\bibitem [{\citenamefont {Utada}\ \emph {et~al.}(2005)\citenamefont {Utada}, \citenamefont {Lorenceau}, \citenamefont {Link}, \citenamefont {Kaplan}, \citenamefont {Stone},\ and\ \citenamefont {Weitz}}]{utada_monodisperse_2005}%
  \BibitemOpen
  \bibfield  {author} {\bibinfo {author} {\bibfnamefont {A.~S.}\ \bibnamefont {Utada}}, \bibinfo {author} {\bibfnamefont {E.}~\bibnamefont {Lorenceau}}, \bibinfo {author} {\bibfnamefont {D.~R.}\ \bibnamefont {Link}}, \bibinfo {author} {\bibfnamefont {P.~D.}\ \bibnamefont {Kaplan}}, \bibinfo {author} {\bibfnamefont {H.~A.}\ \bibnamefont {Stone}},\ and\ \bibinfo {author} {\bibfnamefont {D.~A.}\ \bibnamefont {Weitz}},\ }\href {https://doi.org/10.1126/science.1109164} {\bibfield  {journal} {\bibinfo  {journal} {Science}\ }\textbf {\bibinfo {volume} {308}},\ \bibinfo {pages} {537} (\bibinfo {year} {2005})}\BibitemShut {NoStop}%
\bibitem [{\citenamefont {Shum}\ \emph {et~al.}(2008)\citenamefont {Shum}, \citenamefont {Kim},\ and\ \citenamefont {Weitz}}]{shum_microfluidic_2008}%
  \BibitemOpen
  \bibfield  {author} {\bibinfo {author} {\bibfnamefont {H.~C.}\ \bibnamefont {Shum}}, \bibinfo {author} {\bibfnamefont {J.-W.}\ \bibnamefont {Kim}},\ and\ \bibinfo {author} {\bibfnamefont {D.~A.}\ \bibnamefont {Weitz}},\ }\href {https://doi.org/10.1021/ja802157y} {\bibfield  {journal} {\bibinfo  {journal} {J. Am. Chem. Soc.}\ }\textbf {\bibinfo {volume} {130}},\ \bibinfo {pages} {9543} (\bibinfo {year} {2008})}\BibitemShut {NoStop}%
\bibitem [{\citenamefont {van Blaaderen}\ and\ \citenamefont {Vrij}(1993)}]{van_blaaderen_synthesis_1993}%
  \BibitemOpen
  \bibfield  {author} {\bibinfo {author} {\bibfnamefont {A.}~\bibnamefont {van Blaaderen}}\ and\ \bibinfo {author} {\bibfnamefont {A.}~\bibnamefont {Vrij}},\ }\href {https://doi.org/10.1006/jcis.1993.1073} {\bibfield  {journal} {\bibinfo  {journal} {J. Colloid Interface Sci.}\ }\textbf {\bibinfo {volume} {156}},\ \bibinfo {pages} {1} (\bibinfo {year} {1993})}\BibitemShut {NoStop}%
\bibitem [{\citenamefont {Hagan}(2021)}]{hagan_equilibrium_2021}%
  \BibitemOpen
  \bibfield  {author} {\bibinfo {author} {\bibfnamefont {M.~F.}\ \bibnamefont {Hagan}},\ }\bibfield  {journal} {\bibinfo  {journal} {Rev. Mod. Phys.}\ }\textbf {\bibinfo {volume} {93}},\ \href {https://doi.org/10.1103/RevModPhys.93.025008} {10.1103/RevModPhys.93.025008} (\bibinfo {year} {2021})\BibitemShut {NoStop}%
\bibitem [{\citenamefont {Oglęcka}\ \emph {et~al.}(2014)\citenamefont {Oglęcka}, \citenamefont {Rangamani}, \citenamefont {Liedberg}, \citenamefont {Kraut},\ and\ \citenamefont {Parikh}}]{oglecka_oscillatory_2014}%
  \BibitemOpen
  \bibfield  {author} {\bibinfo {author} {\bibfnamefont {K.}~\bibnamefont {Oglęcka}}, \bibinfo {author} {\bibfnamefont {P.}~\bibnamefont {Rangamani}}, \bibinfo {author} {\bibfnamefont {B.}~\bibnamefont {Liedberg}}, \bibinfo {author} {\bibfnamefont {R.~S.}\ \bibnamefont {Kraut}},\ and\ \bibinfo {author} {\bibfnamefont {A.~N.}\ \bibnamefont {Parikh}},\ }\href {https://doi.org/10.7554/eLife.03695} {\bibfield  {journal} {\bibinfo  {journal} {eLife}\ }\textbf {\bibinfo {volume} {3}},\ \bibinfo {pages} {e03695} (\bibinfo {year} {2014})}\BibitemShut {NoStop}%
\bibitem [{\citenamefont {Reynwar}\ \emph {et~al.}(2007)\citenamefont {Reynwar}, \citenamefont {Illya}, \citenamefont {Harmandaris}, \citenamefont {Müller}, \citenamefont {Kremer},\ and\ \citenamefont {Deserno}}]{reynwar_aggregation_2007}%
  \BibitemOpen
  \bibfield  {author} {\bibinfo {author} {\bibfnamefont {B.~J.}\ \bibnamefont {Reynwar}}, \bibinfo {author} {\bibfnamefont {G.}~\bibnamefont {Illya}}, \bibinfo {author} {\bibfnamefont {V.~A.}\ \bibnamefont {Harmandaris}}, \bibinfo {author} {\bibfnamefont {M.~M.}\ \bibnamefont {Müller}}, \bibinfo {author} {\bibfnamefont {K.}~\bibnamefont {Kremer}},\ and\ \bibinfo {author} {\bibfnamefont {M.}~\bibnamefont {Deserno}},\ }\href {https://doi.org/10.1038/nature05840} {\bibfield  {journal} {\bibinfo  {journal} {Nature}\ }\textbf {\bibinfo {volume} {447}},\ \bibinfo {pages} {461} (\bibinfo {year} {2007})}\BibitemShut {NoStop}%
\bibitem [{\citenamefont {Kozlovsky}\ and\ \citenamefont {Kozlov}(2002)}]{kozlovsky_stalk_2002}%
  \BibitemOpen
  \bibfield  {author} {\bibinfo {author} {\bibfnamefont {Y.}~\bibnamefont {Kozlovsky}}\ and\ \bibinfo {author} {\bibfnamefont {M.~M.}\ \bibnamefont {Kozlov}},\ }\href {https://doi.org/10.1016/S0006-3495(02)75450-7} {\bibfield  {journal} {\bibinfo  {journal} {Biophys. J.}\ }\textbf {\bibinfo {volume} {82}},\ \bibinfo {pages} {882} (\bibinfo {year} {2002})}\BibitemShut {NoStop}%
\bibitem [{\citenamefont {Lettinga}\ \emph {et~al.}(2005)\citenamefont {Lettinga}, \citenamefont {Barry},\ and\ \citenamefont {Dogic}}]{lettinga_self-diffusion_2005}%
  \BibitemOpen
  \bibfield  {author} {\bibinfo {author} {\bibfnamefont {M.~P.}\ \bibnamefont {Lettinga}}, \bibinfo {author} {\bibfnamefont {E.}~\bibnamefont {Barry}},\ and\ \bibinfo {author} {\bibfnamefont {Z.}~\bibnamefont {Dogic}},\ }\href {https://doi.org/10.1209/epl/i2005-10127-x} {\bibfield  {journal} {\bibinfo  {journal} {EPL}\ }\textbf {\bibinfo {volume} {71}},\ \bibinfo {pages} {692} (\bibinfo {year} {2005})}\BibitemShut {NoStop}%
\bibitem [{\citenamefont {Lau}\ \emph {et~al.}(2009)\citenamefont {Lau}, \citenamefont {Prasad},\ and\ \citenamefont {Dogic}}]{lau_condensation_2009}%
  \BibitemOpen
  \bibfield  {author} {\bibinfo {author} {\bibfnamefont {A.~W.~C.}\ \bibnamefont {Lau}}, \bibinfo {author} {\bibfnamefont {A.}~\bibnamefont {Prasad}},\ and\ \bibinfo {author} {\bibfnamefont {Z.}~\bibnamefont {Dogic}},\ }\href {https://doi.org/10.1209/0295-5075/87/48006} {\bibfield  {journal} {\bibinfo  {journal} {EPL}\ }\textbf {\bibinfo {volume} {87}},\ \bibinfo {pages} {48006} (\bibinfo {year} {2009})}\BibitemShut {NoStop}%
\end{thebibliography}%


\begin{thebibliography}{15}%
\makeatletter
\providecommand \@ifxundefined [1]{%
 \@ifx{#1\undefined}
}%
\providecommand \@ifnum [1]{%
 \ifnum #1\expandafter \@firstoftwo
 \else \expandafter \@secondoftwo
 \fi
}%
\providecommand \@ifx [1]{%
 \ifx #1\expandafter \@firstoftwo
 \else \expandafter \@secondoftwo
 \fi
}%
\providecommand \natexlab [1]{#1}%
\providecommand \enquote  [1]{``#1''}%
\providecommand \bibnamefont  [1]{#1}%
\providecommand \bibfnamefont [1]{#1}%
\providecommand \citenamefont [1]{#1}%
\providecommand \href@noop [0]{\@secondoftwo}%
\providecommand \href [0]{\begingroup \@sanitize@url \@href}%
\providecommand \@href[1]{\@@startlink{#1}\@@href}%
\providecommand \@@href[1]{\endgroup#1\@@endlink}%
\providecommand \@sanitize@url [0]{\catcode `\\12\catcode `\$12\catcode `\&12\catcode `\#12\catcode `\^12\catcode `\_12\catcode `\%12\relax}%
\providecommand \@@startlink[1]{}%
\providecommand \@@endlink[0]{}%
\providecommand \url  [0]{\begingroup\@sanitize@url \@url }%
\providecommand \@url [1]{\endgroup\@href {#1}{\urlprefix }}%
\providecommand \urlprefix  [0]{URL }%
\providecommand \Eprint [0]{\href }%
\providecommand \doibase [0]{https://doi.org/}%
\providecommand \selectlanguage [0]{\@gobble}%
\providecommand \bibinfo  [0]{\@secondoftwo}%
\providecommand \bibfield  [0]{\@secondoftwo}%
\providecommand \translation [1]{[#1]}%
\providecommand \BibitemOpen [0]{}%
\providecommand \bibitemStop [0]{}%
\providecommand \bibitemNoStop [0]{.\EOS\space}%
\providecommand \EOS [0]{\spacefactor3000\relax}%
\providecommand \BibitemShut  [1]{\csname bibitem#1\endcsname}%
\let\auto@bib@innerbib\@empty
\bibitem [{\citenamefont {Mutz}\ and\ \citenamefont {Helfrich}(1990)}]{mutz_bending_1990}%
  \BibitemOpen
  \bibfield  {author} {\bibinfo {author} {\bibfnamefont {M.}~\bibnamefont {Mutz}}\ and\ \bibinfo {author} {\bibfnamefont {W.}~\bibnamefont {Helfrich}},\ }\href {https://doi.org/10.1051/jphys:019900051010099100} {\bibfield  {journal} {\bibinfo  {journal} {J. Phys.}\ }\textbf {\bibinfo {volume} {51}},\ \bibinfo {pages} {991} (\bibinfo {year} {1990})}\BibitemShut {NoStop}%
\bibitem [{\citenamefont {Faizi}\ \emph {et~al.}(2024)\citenamefont {Faizi}, \citenamefont {Granek},\ and\ \citenamefont {Vlahovska}}]{faizi_curvature_2024}%
  \BibitemOpen
  \bibfield  {author} {\bibinfo {author} {\bibfnamefont {H.~A.}\ \bibnamefont {Faizi}}, \bibinfo {author} {\bibfnamefont {R.}~\bibnamefont {Granek}},\ and\ \bibinfo {author} {\bibfnamefont {P.~M.}\ \bibnamefont {Vlahovska}},\ }\href {https://doi.org/10.1073/pnas.2413557121} {\bibfield  {journal} {\bibinfo  {journal} {Proc. Natl. Acad. Sci. U.S.A.}\ }\textbf {\bibinfo {volume} {121}},\ \bibinfo {pages} {e2413557121} (\bibinfo {year} {2024})}\BibitemShut {NoStop}%
\bibitem [{\citenamefont {Peliti}\ and\ \citenamefont {Leibler}(1985)}]{peliti_effects_1985}%
  \BibitemOpen
  \bibfield  {author} {\bibinfo {author} {\bibfnamefont {L.}~\bibnamefont {Peliti}}\ and\ \bibinfo {author} {\bibfnamefont {S.}~\bibnamefont {Leibler}},\ }\href {https://doi.org/10.1103/PhysRevLett.54.1690} {\bibfield  {journal} {\bibinfo  {journal} {Phys. Rev. Lett.}\ }\textbf {\bibinfo {volume} {54}},\ \bibinfo {pages} {1690} (\bibinfo {year} {1985})}\BibitemShut {NoStop}%
\bibitem [{\citenamefont {Evans}\ and\ \citenamefont {Rawicz}(1990)}]{evans_entropy-driven_1990}%
  \BibitemOpen
  \bibfield  {author} {\bibinfo {author} {\bibfnamefont {E.}~\bibnamefont {Evans}}\ and\ \bibinfo {author} {\bibfnamefont {W.}~\bibnamefont {Rawicz}},\ }\href {https://doi.org/10.1103/PhysRevLett.64.2094} {\bibfield  {journal} {\bibinfo  {journal} {Phys. Rev. Lett.}\ }\textbf {\bibinfo {volume} {64}},\ \bibinfo {pages} {2094} (\bibinfo {year} {1990})}\BibitemShut {NoStop}%
\bibitem [{\citenamefont {Lipowsky}(1991)}]{lipowsky_conformation_1991}%
  \BibitemOpen
  \bibfield  {author} {\bibinfo {author} {\bibfnamefont {R.}~\bibnamefont {Lipowsky}},\ }\href {https://doi.org/10.1038/349475a0} {\bibfield  {journal} {\bibinfo  {journal} {Nature}\ }\textbf {\bibinfo {volume} {349}},\ \bibinfo {pages} {475} (\bibinfo {year} {1991})}\BibitemShut {NoStop}%
\bibitem [{\citenamefont {Gibaud}\ \emph {et~al.}(2012)\citenamefont {Gibaud}, \citenamefont {Barry}, \citenamefont {Zakhary}, \citenamefont {Henglin}, \citenamefont {Ward}, \citenamefont {Yang}, \citenamefont {Berciu}, \citenamefont {Oldenbourg}, \citenamefont {Hagan}, \citenamefont {Nicastro}, \citenamefont {Meyer},\ and\ \citenamefont {Dogic}}]{gibaud_reconfigurable_2012}%
  \BibitemOpen
  \bibfield  {author} {\bibinfo {author} {\bibfnamefont {T.}~\bibnamefont {Gibaud}}, \bibinfo {author} {\bibfnamefont {E.}~\bibnamefont {Barry}}, \bibinfo {author} {\bibfnamefont {M.~J.}\ \bibnamefont {Zakhary}}, \bibinfo {author} {\bibfnamefont {M.}~\bibnamefont {Henglin}}, \bibinfo {author} {\bibfnamefont {A.}~\bibnamefont {Ward}}, \bibinfo {author} {\bibfnamefont {Y.}~\bibnamefont {Yang}}, \bibinfo {author} {\bibfnamefont {C.}~\bibnamefont {Berciu}}, \bibinfo {author} {\bibfnamefont {R.}~\bibnamefont {Oldenbourg}}, \bibinfo {author} {\bibfnamefont {M.~F.}\ \bibnamefont {Hagan}}, \bibinfo {author} {\bibfnamefont {D.}~\bibnamefont {Nicastro}}, \bibinfo {author} {\bibfnamefont {R.~B.}\ \bibnamefont {Meyer}},\ and\ \bibinfo {author} {\bibfnamefont {Z.}~\bibnamefont {Dogic}},\ }\href {https://doi.org/10.1038/nature10769} {\bibfield  {journal} {\bibinfo  {journal} {Nature}\ }\textbf {\bibinfo {volume} {481}},\ \bibinfo {pages} {348} (\bibinfo {year} {2012})}\BibitemShut {NoStop}%
\bibitem [{\citenamefont {Jia}\ \emph {et~al.}(2017)\citenamefont {Jia}, \citenamefont {Zakhary}, \citenamefont {Dogic}, \citenamefont {Pelcovits},\ and\ \citenamefont {Powers}}]{jia_chiral_2017}%
  \BibitemOpen
  \bibfield  {author} {\bibinfo {author} {\bibfnamefont {L.~L.}\ \bibnamefont {Jia}}, \bibinfo {author} {\bibfnamefont {M.~J.}\ \bibnamefont {Zakhary}}, \bibinfo {author} {\bibfnamefont {Z.}~\bibnamefont {Dogic}}, \bibinfo {author} {\bibfnamefont {R.~A.}\ \bibnamefont {Pelcovits}},\ and\ \bibinfo {author} {\bibfnamefont {T.~R.}\ \bibnamefont {Powers}},\ }\href {https://doi.org/10.1103/PhysRevE.95.060701} {\bibfield  {journal} {\bibinfo  {journal} {Phys. Rev. E}\ }\textbf {\bibinfo {volume} {95}},\ \bibinfo {pages} {060701} (\bibinfo {year} {2017})}\BibitemShut {NoStop}%
\bibitem [{\citenamefont {Gibaud}\ \emph {et~al.}(2017)\citenamefont {Gibaud}, \citenamefont {Kaplan}, \citenamefont {Sharma}, \citenamefont {Zakhary}, \citenamefont {Ward}, \citenamefont {Oldenbourg}, \citenamefont {Meyer}, \citenamefont {Kamien}, \citenamefont {Powers},\ and\ \citenamefont {Dogic}}]{gibaud_achiral_2017}%
  \BibitemOpen
  \bibfield  {author} {\bibinfo {author} {\bibfnamefont {T.}~\bibnamefont {Gibaud}}, \bibinfo {author} {\bibfnamefont {C.~N.}\ \bibnamefont {Kaplan}}, \bibinfo {author} {\bibfnamefont {P.}~\bibnamefont {Sharma}}, \bibinfo {author} {\bibfnamefont {M.~J.}\ \bibnamefont {Zakhary}}, \bibinfo {author} {\bibfnamefont {A.}~\bibnamefont {Ward}}, \bibinfo {author} {\bibfnamefont {R.}~\bibnamefont {Oldenbourg}}, \bibinfo {author} {\bibfnamefont {R.~B.}\ \bibnamefont {Meyer}}, \bibinfo {author} {\bibfnamefont {R.~D.}\ \bibnamefont {Kamien}}, \bibinfo {author} {\bibfnamefont {T.~R.}\ \bibnamefont {Powers}},\ and\ \bibinfo {author} {\bibfnamefont {Z.}~\bibnamefont {Dogic}},\ }\href {https://doi.org/10.1073/pnas.1617043114} {\bibfield  {journal} {\bibinfo  {journal} {Proc. Natl. Acad. Sci. U.S.A.}\ }\textbf {\bibinfo {volume} {114}},\ \bibinfo {pages} {E3376} (\bibinfo {year} {2017})}\BibitemShut {NoStop}%
\bibitem [{\citenamefont {Senti}\ \emph {et~al.}(1955)\citenamefont {Senti}, \citenamefont {Hellman}, \citenamefont {Ludwig}, \citenamefont {Babcock}, \citenamefont {Tobin}, \citenamefont {Glass},\ and\ \citenamefont {Lamberts}}]{senti_viscosity_1955}%
  \BibitemOpen
  \bibfield  {author} {\bibinfo {author} {\bibfnamefont {F.~R.}\ \bibnamefont {Senti}}, \bibinfo {author} {\bibfnamefont {N.~N.}\ \bibnamefont {Hellman}}, \bibinfo {author} {\bibfnamefont {N.~H.}\ \bibnamefont {Ludwig}}, \bibinfo {author} {\bibfnamefont {G.~E.}\ \bibnamefont {Babcock}}, \bibinfo {author} {\bibfnamefont {R.}~\bibnamefont {Tobin}}, \bibinfo {author} {\bibfnamefont {C.~A.}\ \bibnamefont {Glass}},\ and\ \bibinfo {author} {\bibfnamefont {B.~L.}\ \bibnamefont {Lamberts}},\ }\href {https://doi.org/10.1002/pol.1955.120178605} {\bibfield  {journal} {\bibinfo  {journal} {J. Polym. Sci.}\ }\textbf {\bibinfo {volume} {17}},\ \bibinfo {pages} {527} (\bibinfo {year} {1955})}\BibitemShut {NoStop}%
\bibitem [{\citenamefont {Rapp}(2017)}]{RAPP2017243}%
  \BibitemOpen
  \bibfield  {author} {\bibinfo {author} {\bibfnamefont {B.~E.}\ \bibnamefont {Rapp}},\ }in\ \href {https://doi.org/https://doi.org/10.1016/B978-1-4557-3141-1.50009-5} {\emph {\bibinfo {booktitle} {Microfluidics: Modelling, Mechanics and Mathematics}}},\ \bibinfo {series and number} {Micro and Nano Technologies},\ \bibinfo {editor} {edited by\ \bibinfo {editor} {\bibfnamefont {B.~E.}\ \bibnamefont {Rapp}}}\ (\bibinfo  {publisher} {Elsevier},\ \bibinfo {address} {Oxford},\ \bibinfo {year} {2017})\ pp.\ \bibinfo {pages} {243--263}\BibitemShut {NoStop}%
\bibitem [{\citenamefont {J\"ulicher}\ and\ \citenamefont {Seifert}(1994)}]{Julicher1994}%
  \BibitemOpen
  \bibfield  {author} {\bibinfo {author} {\bibfnamefont {F.}~\bibnamefont {J\"ulicher}}\ and\ \bibinfo {author} {\bibfnamefont {U.}~\bibnamefont {Seifert}},\ }\href {https://doi.org/10.1103/PhysRevE.49.4728} {\bibfield  {journal} {\bibinfo  {journal} {Phys. Rev. E}\ }\textbf {\bibinfo {volume} {49}},\ \bibinfo {pages} {4728} (\bibinfo {year} {1994})}\BibitemShut {NoStop}%
\bibitem [{\citenamefont {Adkins}\ \emph {et~al.}(2025)\citenamefont {Adkins}, \citenamefont {Robaszewski}, \citenamefont {Shin}, \citenamefont {Brauns}, \citenamefont {Jia}, \citenamefont {Khanra}, \citenamefont {Sharma}, \citenamefont {Pelcovits}, \citenamefont {Powers},\ and\ \citenamefont {Dogic}}]{adkins_topology_2025}%
  \BibitemOpen
  \bibfield  {author} {\bibinfo {author} {\bibfnamefont {R.}~\bibnamefont {Adkins}}, \bibinfo {author} {\bibfnamefont {J.}~\bibnamefont {Robaszewski}}, \bibinfo {author} {\bibfnamefont {S.}~\bibnamefont {Shin}}, \bibinfo {author} {\bibfnamefont {F.}~\bibnamefont {Brauns}}, \bibinfo {author} {\bibfnamefont {L.}~\bibnamefont {Jia}}, \bibinfo {author} {\bibfnamefont {A.}~\bibnamefont {Khanra}}, \bibinfo {author} {\bibfnamefont {P.}~\bibnamefont {Sharma}}, \bibinfo {author} {\bibfnamefont {R.~A.}\ \bibnamefont {Pelcovits}}, \bibinfo {author} {\bibfnamefont {T.~R.}\ \bibnamefont {Powers}},\ and\ \bibinfo {author} {\bibfnamefont {Z.}~\bibnamefont {Dogic}},\ }\href {https://doi.org/10.1073/pnas.2427024122} {\bibfield  {journal} {\bibinfo  {journal} {Proc. Natl. Acad. Sci. U.S.A.}\ }\textbf {\bibinfo {volume} {122}},\ \bibinfo {pages} {e2427024122} (\bibinfo {year} {2025})}\BibitemShut {NoStop}%
\bibitem [{\citenamefont {Balchunas}\ \emph {et~al.}(2019)\citenamefont {Balchunas}, \citenamefont {Cabanas}, \citenamefont {Zakhary}, \citenamefont {Gibaud}, \citenamefont {Fraden}, \citenamefont {Sharma}, \citenamefont {Hagan},\ and\ \citenamefont {Dogic}}]{balchunas_equation_2019}%
  \BibitemOpen
  \bibfield  {author} {\bibinfo {author} {\bibfnamefont {A.~J.}\ \bibnamefont {Balchunas}}, \bibinfo {author} {\bibfnamefont {R.~A.}\ \bibnamefont {Cabanas}}, \bibinfo {author} {\bibfnamefont {M.~J.}\ \bibnamefont {Zakhary}}, \bibinfo {author} {\bibfnamefont {T.}~\bibnamefont {Gibaud}}, \bibinfo {author} {\bibfnamefont {S.}~\bibnamefont {Fraden}}, \bibinfo {author} {\bibfnamefont {P.}~\bibnamefont {Sharma}}, \bibinfo {author} {\bibfnamefont {M.~F.}\ \bibnamefont {Hagan}},\ and\ \bibinfo {author} {\bibfnamefont {Z.}~\bibnamefont {Dogic}},\ }\href {https://doi.org/10.1039/C9SM01054H} {\bibfield  {journal} {\bibinfo  {journal} {Soft Matter}\ }\textbf {\bibinfo {volume} {15}},\ \bibinfo {pages} {6791} (\bibinfo {year} {2019})}\BibitemShut {NoStop}%
\bibitem [{\citenamefont {Szleifer}(1988)}]{szleifer_curvature_1988}%
  \BibitemOpen
  \bibfield  {author} {\bibinfo {author} {\bibfnamefont {I.}~\bibnamefont {Szleifer}},\ }\href {https://doi.org/10.1103/PhysRevLett.60.1966} {\bibfield  {journal} {\bibinfo  {journal} {Phys. Rev. Lett.}\ }\textbf {\bibinfo {volume} {60}},\ \bibinfo {pages} {1966} (\bibinfo {year} {1988})}\BibitemShut {NoStop}%
\bibitem [{\citenamefont {Rawicz}\ \emph {et~al.}(2000)\citenamefont {Rawicz}, \citenamefont {Olbrich}, \citenamefont {McIntosh}, \citenamefont {Needham},\ and\ \citenamefont {Evans}}]{rawicz_effect_2000}%
  \BibitemOpen
  \bibfield  {author} {\bibinfo {author} {\bibfnamefont {W.}~\bibnamefont {Rawicz}}, \bibinfo {author} {\bibfnamefont {K.~C.}\ \bibnamefont {Olbrich}}, \bibinfo {author} {\bibfnamefont {T.}~\bibnamefont {McIntosh}}, \bibinfo {author} {\bibfnamefont {D.}~\bibnamefont {Needham}},\ and\ \bibinfo {author} {\bibfnamefont {E.}~\bibnamefont {Evans}},\ }\href {https://doi.org/10.1016/S0006-3495(00)76295-3} {\bibfield  {journal} {\bibinfo  {journal} {Biophys. J.}\ }\textbf {\bibinfo {volume} {79}},\ \bibinfo {pages} {328} (\bibinfo {year} {2000})}\BibitemShut {NoStop}%
\end{thebibliography}%
\end{document}